\documentclass[12pt,a4paper]{article}

\usepackage{amsmath}
\usepackage{mathtools}
\usepackage{amssymb}
\usepackage{graphicx}
\usepackage{cite}
\usepackage{hyperref}

\begin{document}
\title{Thin and thick bubble walls III: wall energy}
\author{
	Ariel M\'{e}gevand\thanks{Member of CONICET, Argentina. E-mail address:
		megevand@mdp.edu.ar}~ 
	and Federico Agust\'{\i}n Membiela\thanks{Member of CONICET, Argentina. E-mail
		address: membiela@mdp.edu.ar} \\[0.5cm]
	\normalsize \it IFIMAR (CONICET-UNMdP)\\
	\normalsize \it Departamento de F\'{\i}sica, Facultad de Ciencias Exactas
	y Naturales, \\
	\normalsize \it UNMdP, De\'{a}n Funes 3350, (7600) Mar del Plata, Argentina }
\date{}
\maketitle

\begin{abstract}
We study the energy-momentum tensor of a bubble wall beyond the approximation of an infinitely thin wall.
To this end, we discuss the proper decomposition into wall and bulk contributions, and we 
use a systematic method to calculate the energy-momentum tensor at any order in the wall width.
We consider the specific examples of spherical bubbles with different initial configurations,
and we compare our approximations with a numerical computation.
\end{abstract}

\section{Introduction}

A cosmological phase transition is a general prediction of particle physics models.
In particular, a first-order electroweak phase transition is possible in extensions of the Standard Model,
where the expansion and collision of bubbles and the bulk fluid motions 
may lead to phenomena such as
electroweak baryogenesis \cite{krs85,ckn93} or
the generation of a stochastic background of gravitational waves
\cite{tw90,ktw92,kkt94}.
A bubble is generally described by a configuration of a scalar field $\phi(x^\mu)$ which varies between the values $\phi_\pm$ corresponding to the minima of the potential $V(\phi)$.
To study the evolution of bubbles and some of their consequences, a useful approximation is that their walls are infinitely thin.
This approximation allows analytic calculations, thus giving an important 
alternative to a lattice computation.

It is worth noting that the question of whether the wall is thin or thick depends on the problem at hand.
For instance, for the interaction of the wall with plasma particles
(e.g., in calculations of the wall velocity \cite{mp95} or electroweak baryogenesis  \cite{ckn93}),
the relevant length scale will be given by the mean free paths of particles 
or the inverse of particle momenta, which are typically of order $T^{-1}$.
This scale is comparable to the wall width $l$, which is roughly given by the inverse energy scale of the theory, $l\sim v^{-1}$.
In contrast, for the generation of gravitational waves, 
the relevant size scale will be, e.g., the bubble radius at  percolation \cite{kkt94}
or the wavelength of wall deformations \cite{mms15,mm21a}.
These can be much larger than $l$ since the bubbles typically grow to sizes of cosmological order 
$H^{-1}\sim M_P/v^2$, where $M_P$ is the Planck mass.

In addition, each of these problems requires knowledge of the wall evolution.
For example, the estimation of friction and calculations of electroweak baryogenesis require the field profile,
and a tanh ansatz is often used.
For hydrodynamics considerations (where the wall can be assumed to be infinitely thin), the wall velocity is often assumed to be constant for simplicity.
In gravitational wave generation, a relevant quantity is the energy stored in the walls,                 
and a realistic treatment of the wall dynamics becomes particularly important.
The thin-wall approximation greatly simplifies the field equation \cite{c77}. 
As a result, it provides an analytical approximation for the field profile across the wall as well as an effective equation of motion (EOM) for the wall as a surface.

Besides the thin-wall approximation, bubbles are often assumed to be spherical.
In particular, to track the progress of the phase transition analytically, the phase transition is generally
modeled in terms of spherical bubbles that expand at a fixed velocity and simply overlap until they fill the space
(for a recent review, see \cite{abf+24cosmological}).
Such an approximation is also used in some models of gravitational wave generation, both in analytical and numerical computations (see, e.g., \cite{kt93,hk08,jt17,mm21b}).
However, bubbles can deviate from the spherical shape for several reasons.
For instance, fluctuations of the surface can be amplified due to different kinds of instabilities
\cite{l92,hkllm93,mm14,bbm1}.
Also, the wall dynamics is affected by collisions \cite{jkt19} and merging \cite{w84}.

The use of the thin-wall approximation for the dynamics of a domain wall is valid when $l\ll L$, where $L$ is the 
radius of curvature of the wall hypersurface \cite{w89thick,ghg90}.
Note that this condition refers to the spacetime curvature of the worldvolume, so the thin-wall approximation breaks down if the surface is very curved in space (e.g., for a small spherical wall), but also if it has a high enough acceleration.
For a bubble wall, this condition translates into a condition for the pressure difference, 
so the potential must have a small energy density difference between the vacua \cite{c77}
(for a detailed discussion of the validity of the various assumptions involved in the thin-wall approximation, see Ref.~\cite{mm23}).

Recently, we discussed the EOM for a general (not necessarily spherical) wall \cite{mm23}
and derived an extension of the thin-wall approximation that gives the field profile and the wall EOM as an expansion in powers of $l/L$ \cite{mm24}.
This is important for thick-walled bubbles which can occur in very supercooled phase transitions \cite{jkt19,cghw20}.
(In such a case, neglecting the fluid is a good approximation, and we use this simplification).
In the present paper, we will use these results to calculate the energy-momentum tensor of the wall.
Obtaining this quantity beyond the thin-wall approximation is particularly important for improving analytical estimates of the production of gravitational waves.
Since the wall is not infinitely thin, there are some ambiguities about its extension that need to be resolved in order to obtain a reliable result for the wall energy.
A proper identification of the wall is also useful for numerical treatments, where
it is of interest to analyze the results in terms of the wall dynamics.
This is sometimes attempted by setting fixed values of the field to define the wall boundaries \cite{jkt19,cghw20},
which is ambiguous since such values are arbitrary.
An alternative is to define the wall energy density as the kinetic and gradient components
\cite{kt93,elnv19}. 
However, this omits the energy of the potential barrier, which is also concentrated at the wall.
We propose an alternative definition that solves all these problems.

The plan of the paper is the following.
In the next section, we discuss the identification of the wall for a general field configuration 
and the appropriate definition of the energy-momentum tensor of the wall.
In Sec.~\ref{sec:O31} we illustrate the decomposition into wall and bulk contributions with specific examples.
In Sec.~\ref{sec:pert_method}, 
we briefly review the results of Ref.~\cite{mm24} and apply them to calculate the wall energy-momentum tensor
to the next-to-next-to-leading order in the wall width.
In Sec.~\ref{sec:O3} we compare the approximation with numerical calculations for the case of spherical bubbles.
We consider realistic initial conditions for the bubble evolution, which we discuss further in App.~\ref{app:instanton}.
In Sec.~\ref{sec:conclu} we summarize our conclusions.

\section{Wall contribution to the energy-momentum tensor}

\label{sec:Tmunu}

We will describe the first-order phase transition with a scalar potential $V(\phi)$
with two minima $\phi_+$ and $\phi_-$ corresponding to the false and true vacuum, respectively.
Whether we want to describe a bubble of any shape, or a set of overlapping bubbles,
the simplest approximation is that we have
infinitely thin walls separating false-vacuum and true-vacuum domains.
In each domain, the potential takes one of the two values $V_-=V(\phi_-)$ or $V_+=V(\phi_+)$,
and we only need to follow the evolution of the network of interfaces.
Near an interface $S$, the stress-energy tensor takes the form
\begin{equation}
	T_{\mu\nu}= S_{\mu\nu}\delta(n) +g_{\mu\nu}\left[V_{+}\Theta(n)+V_{-}\Theta(-n)\right],
	\label{eq:Tmunudelta}
\end{equation}
where $n$ is the distance from the worldvolume $\Sigma$ of the surface.
The quantity $S_{\mu\nu}$ is called the surface stress-energy tensor
and is formally defined as (see, e.g., \cite{misner,bkt87})
\begin{equation}
	S_{\mu\nu}=\lim_{\epsilon\to0}\int_{-\epsilon}^{+\epsilon}T_{\mu\nu}dn . \label{eq:Smunuthin}
\end{equation}
Although Eq.~(\ref{eq:Tmunudelta}) will be a reasonable approximation for many applications,
in order to implement this approximation for a real bubble,
the surface stress energy tensor $S_{\mu\nu}$ must be calculated from the actual quantity $T_{\mu\nu}$, which is not a delta function.
The formal expression (\ref{eq:Smunuthin}) is not very useful for this purpose.
In particular, the infinitesimal integration range should be replaced by a suitable finite range
associated to the wall region.
Furthermore, a surface $S$ representing the wall must be defined within this region
in order to consider the distance $n$ to the corresponding hypersurface $\Sigma$ (see \cite{mm23} for a recent discussion).
We need to limit the ambiguity in the choice of the surface and the wall range as much as possible.

\subsection{Wall surface and wall region}
\label{subsec:wallregion}

The evolution of one or more bubbles of arbitrary shape is described by the field equation,
which for a vacuum phase transition is given by
\begin{equation}
	\nabla_{\mu}\nabla^{\mu}\phi+V'(\phi)=0. \label{eq:eccampo}
\end{equation}
Given a solution $\phi(x^\mu)$, the problem of identifying the bubble walls is greatly simplified when
the field in each domain is in a minimum of the potential.
We can then define the walls as the regions where the field varies between these two values.
We can choose a surface representing the wall position as the points where $\phi$ takes some intermediate value $\phi_w$.
Unfortunately, the field will not always approach the values $\phi_\pm$ on each side of the interface.
For example, we may have bubbles where $\phi$ is (initially) closer to the metastable minimum $\phi_+$ than to the stable minimum $\phi_-$ (see below).
This is because the barrier between the minima is not always in the middle between $\phi_+$ and $\phi_-$.
The top of the barrier provides a natural separation between field values we can associate with false and true vacuum, 
and we can always define the wall position by taking $\phi_w$ as this maximum point $\phi_{\max}$
(the value $\phi_w=\phi_{\max}$  is also motivated by gauge invariance%
\footnote{It is well known that the effective potential is generally gauge dependent \cite{jackiw74functional,dj74gauge},
	but its values at the stationary points are not \cite{nielsen75gauge,fk76gauge}
	(however, inconsistent truncation at a given order in perturbation theory can introduce artificial gauge dependence; see, e.g., \cite{laine95gauge,pr11baryon}).
	Therefore, defining the wall position as the points where $V$ takes the physical value $V_{\max}$
	gives a gauge independent definition of the wall surface.	
	It is beyond the scope of this paper to check the gauge invariance of the results, and 
	we will consider a toy-model potential for our specific examples.
	However, even if the surface representing the wall position is gauge dependent, this surface is only a reference
	for the calculations, and the final results should not depend on it, at least at sufficiently high orders in the wall width.}).
In any case, the choice of the specific surface representing the wall position should not matter if we develop an appropriate method to calculate (perturbatively in the wall width) physical quantities associated with the whole wall region (such as the wall energy).
Alternative definitions that do not involve a fixed value $\phi=\phi_w$ may be convenient in some cases,
and we will discuss some of them below.
The independence on the specific choice of the surface can be checked by comparing the results of our approximations with a numerical calculation, which we do in Sec.~\ref{sec:O3}.

The boundaries of the wall region could be defined as surfaces where $\phi$ takes certain values $\phi_{\mathrm{in}},\phi_{\mathrm{out}}$
(see, e.g., \cite{cghw20}).
This is useful when the values $\phi_\pm$ are approached asymptotically as we move away from the surface,
where one can choose $\phi_{\mathrm{in}}$ and $\phi_{\mathrm{out}}$ close to $\phi_-$ and $\phi_+$, respectively.
However, this prescription becomes difficult to implement in cases where the field varies within a domain
(see examples below).
In any case, the ambiguity in the choice of the specific values $\phi_{\mathrm{in}},\phi_{\mathrm{out}}$ leads to an inaccuracy in quantities such as the energy of the wall.
One way to avoid this problem is to assign the kinetic and gradient energy to the wall and the potential energy to the domains
(see, e.g., \cite{kt93,elnv19}).
Since the kinetic and gradient terms generally vanish away from the wall, there is no need to specify boundaries for their integration.
However, field variations within a domain will contribute to the kinetic and gradient energy.
Conversely, the potential energy density peaks at the wall due to the barrier between minima.
A reasonable definition of the wall energy should at least include this contribution.

We will look for a decomposition of the energy-momentum tensor
\begin{equation}
	T_{\mu\nu}=\partial_{\mu}\phi\partial_{\nu}\phi
	-\frac{1}{2}g_{\mu\nu}g^{\alpha\beta}\partial_{\alpha}\phi\partial_{\beta}\phi
	+g_{\mu\nu}V(\phi) 
	\label{eq:Tmunufield}
\end{equation}
into wall and bulk components,
\begin{equation}
	T_{\mu\nu}(x^\mu)= T_{\mu\nu}^{w}(x^\mu) + T_{\mu\nu}^{b}(x^\mu) ,
	\label{Tmunudecomp}
\end{equation}
where $T^w_{\mu\nu}$ peaks at the wall and falls rapidly to zero away from it,
while $T_{\mu\nu}^{b}$ does not have a peak at the wall but only interpolates between the behavior in each domain.
Such a decomposition generalizes the approximation (\ref{eq:Tmunudelta}).
Thus, using the component $T^w_{\mu\nu}$ in Eq.~(\ref{eq:Smunuthin}), we can safely drop the formal limit $\epsilon\to 0$ 
and replace the integration limits $\pm\epsilon$ with $\pm\infty$ or any convenient values.
This approach requires writing $T_{\mu\nu}(x^\mu)$ as a function of the variable $n$.
In flat space, $n$ is the distance along a straight line orthogonal to the hypersurface $\Sigma$.
More generally, $n$ is the proper distance along a normal geodesic.
Such a change of variables can be cumbersome, but it is the usual approach for a very thin wall,
where $\phi$ only depends on $n$ \cite{vs00}.

\begin{figure}[tb]
	\centering
	\hspace*{\fill}\includegraphics[width=0.4\textwidth]{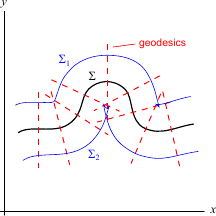}%
	\hfill%
	\includegraphics[width=0.4\textwidth]{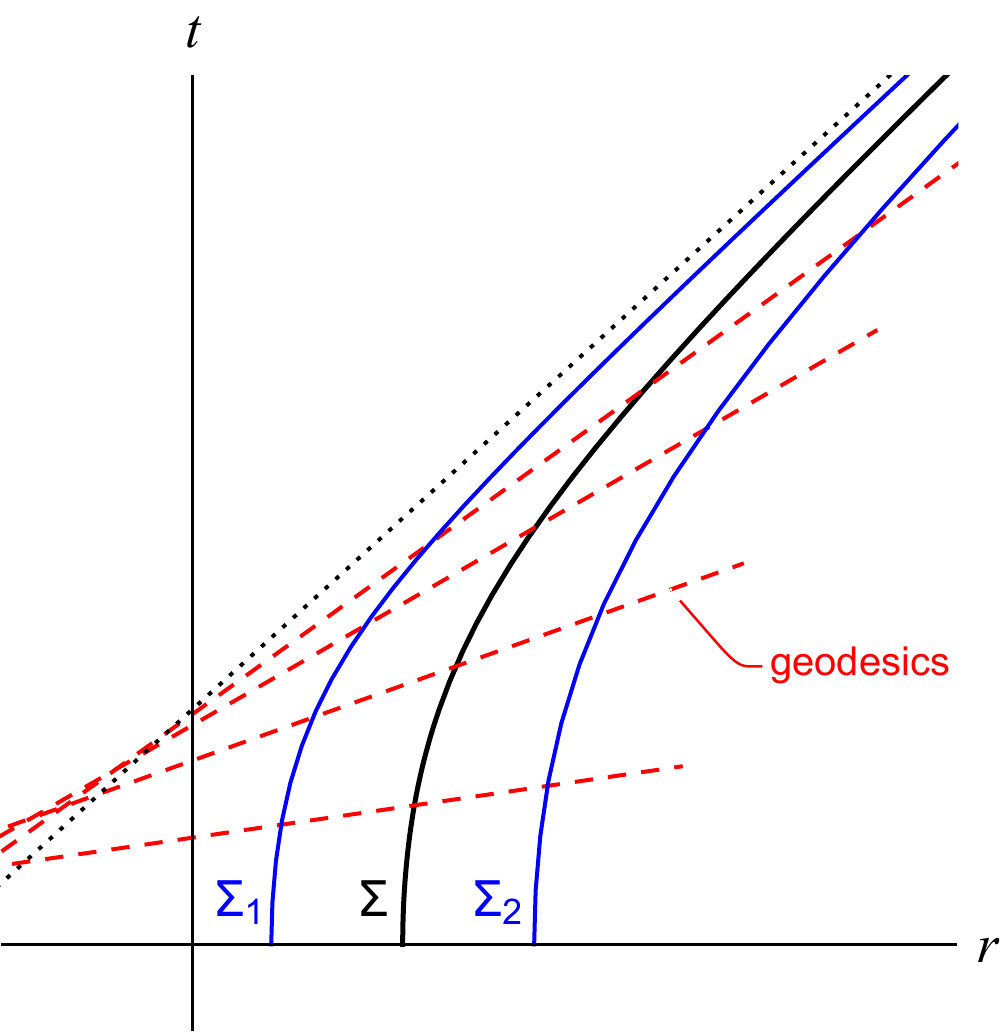}\hspace*{\fill}
	\caption{The left panel shows schematically a wall surface that is curved in space.
		The right panel shows a spacetime section of the worldvolume for the realistic case of an accelerated spherical wall. 
		The two hypersurfaces $\Sigma_1$ and $\Sigma_2$ are at a fixed distance $n$ from $\Sigma$ along the normal geodesics.
		\label{fig:curved}}
\end{figure}

For a curved wall, $n$ will be well defined as long as the geodesics do not cross,
which is always valid in a neighborhood of $\Sigma$,
as sketched in the left panel of Fig.~\ref{fig:curved}.
Beyond this neighborhood, the description of the interface in terms of a field profile $\phi(n)$ breaks down.
In such a case, the points where the geodesics cross define a boundary for the wall region.
Notice that this boundary is more natural than delimiting the wall with arbitrary field values $\phi_{\mathrm{in}},\phi_{\mathrm{out}}$.
The origin of the restriction on $n$ is that the wall overlaps itself where it is wider than its local radius of curvature.
Indeed, consider two hypersurfaces $\Sigma_{1}$, $\Sigma_{2}$ 
representing wall boundaries at a fixed distance $n$ from $\Sigma$.
In the left panel of Fig.~\ref{fig:curved} we see that these surfaces
intersect themselves if the geodesics cross within this distance.

If the wall is accelerated, the worldvolume is curved even if the wall is spatially planar.
The right panel of Fig.~\ref{fig:curved} illustrates this spacetime curvature for the case of a spherical bubble considered below.
In this case, the normal lines intersect outside the physical region.
However, due to the light-speed asymptote of the bubble radius,
these geodesics never penetrate the lightcone indicated by a dotted line.
The orthogonal distance $n$ is well defined only outside this lightcone.
As a consequence, a field profile of the form $\phi(n)$ will represent only a part of 
the bubble profile $\phi(r)$ at a given time $t$.
In a way, this is a shortcoming compared to other methods of describing the wall.
Nevertheless, as we will see, 
the region inside the lightcone can be interpreted as the bubble interior.
In particular, the field may have oscillations just beyond the dotted line
and the energy density may not fall to zero as we move away from $\Sigma$.
In such a case, the valid range of $n$ gives a physical determination of the wall boundaries.

\subsection{The distance to the surface}
\label{CGN}

Let us assume that we have conveniently chosen a wall surface by some condition such as $\phi(x^\mu)=\phi_w$.
This condition defines a hypersurface  $\Sigma$ that describes the evolution of the two-dimensional surface $S$.
The hypersurface has a parametric representation of the form $x^{\mu}=X^{\mu}(\xi^{a})$ with $\mu=0,1,2,3$ and $a=0,1,2$.
The surface $S$ is determined by the intersection of $\Sigma$ with the fixed-time hypersurface  $x^{0}=t$.
At a given time, the surface $S$ can be parametrized as $x^{i}=Y^{i}(\zeta^{A})$ with $i=1,2,3$ and $A=1,2$.
The function $Y^{i}(\zeta^{A})$ depends on $t$ and can be obtained from $X^\mu(\xi^a)$. 
Indeed, from the condition $X^{0}(\xi^{a})=t$ we can solve for $\xi^{0}=\Xi(\xi^{1},\xi^{2})$ and choose $\zeta^{A}=\xi^{A}$, so we have $Y^{i}(\zeta^A)=X^{i}(\Xi(\zeta^{1},\zeta^{2}),\zeta^{1},\zeta^{2})$.
We will also consider implicit equations $F(x^{\mu})=0$, $G(x^i)=0$
for $\Sigma$ and $S$, respectively.
A possible choice for the latter is $G(x^i)=F(t,x^{i})$.
The obvious choice for $F$ would be $F(x^\mu)=\phi(x^\mu)-\phi_w$, but it is convenient to consider a function that does not have a rapid variation across the interface.
A useful choice is an explicit representation, where we write one of the coordinates as a function of the others, e.g.,
\begin{align}
	F &=x^{3}-x_{w}^{3}(x^{0},x^{1},x^{2}), 
	&
	X^\mu(\xi^a) &=\left(\xi^{0},\xi^{1},\xi^{2},x^3_w(\xi^{a})\right),
	\label{eq:Monge}
	\\
	G &=x^{3}-x_{w}^{3}(t,x^{1},x^{2}),
	&
	Y^i(\zeta^A) &=\left(\zeta^{1},\zeta^{2},x^3_w(t,\zeta^1,\zeta^2)\right).
	\label{MongeS}
\end{align}

The parameters $\xi^a$ can be used as coordinates on $\Sigma$, and
the induced metric is given by $\gamma_{ab} = g_{\mu\nu}\partial_a X^\mu \partial_b X^\nu$.
The surface element of the hypersurface is given by
$d\Sigma=d^{3}\xi\sqrt{\gamma}$, with $\gamma \equiv \det\gamma_{ab}$.
Similarly, on $S$ we can use the coordinates $\zeta^A$, where
the induced metric is given by $\sigma_{AB} = g_{ij}\partial_A Y^i \partial_B Y^j$
and the surface element is given by $dS=d^{2}\zeta\sqrt{\det \sigma_{AB}}$.

The normal vectors to $\Sigma$ and $S$ are given by%
\footnote{In 4D spacetime, $S$ has two independent normal vectors.
	One of them is $N^\mu$, and we could take the other one tangent to $\Sigma$ \cite{poisson}.
	However, we will be interested in
	the embedding in 3D space.}
\begin{align}
	N_{\mu} &=-\partial_{\mu}F/s, \quad \text{with} \quad s ={|F_{,\mu}F^{,\mu}|^{1/2}} ,
	\label{eq:Nmu}
	\\
	M_{i} &=-\partial_{i}F/q, \quad  \text{with} \quad q ={|F_{,i}F^{,i}|}^{1/2} .
	\label{eq:Mi}
\end{align}
Indeed, we have, e.g., $N_{\mu} \partial_a X^\mu=0$ and $N_{\mu}N^{\mu}=-1$.
We can always choose $F$ such that these vectors point towards the false vacuum.
The tensor
\begin{equation}
	h^{\mu\nu} = \gamma^{ab}\partial_a X^\mu \partial_b X^\nu = g^{\mu\nu}+N^\mu N^{\nu} 
	\label{eq:proyector}
\end{equation}
gives the orthogonal projection operator ${h^\mu}_\nu$ which projects any tensor onto a tensor tangent to $\Sigma$
(see, e.g., \cite{poisson,carroll}). 
For $S$, the projection tensor is given by $k_{ij}=g_{ij}+M_i M_j$.

The geodesics orthogonal to the hypersurface $\Sigma$ are given by the equations
\begin{equation}
	\frac{d^{2}x^{\mu}}{dn^{2}}+\Gamma_{\nu\rho}^{\mu}\frac{dx^{\nu}}{dn}\frac{dx^{\rho}}{dn} =0 ,
	\quad x^{\mu}(0)=X^{\mu}(\xi^{a}), \quad
	\frac{dx^{\mu}}{dn}(0)=N^{\mu}(X^{\nu}(\xi^{a})) .
	\label{geod}
\end{equation}
The solution can be written as an expansion in powers of $n$, where the first few terms follow immediately from Eq.~(\ref{geod}),
\begin{equation}
	x^{\mu}=X^{\mu}(\xi^{a})+N^{\mu}|_{X^\nu(\xi^{a})} \, n 
	-\frac{1}{2}\left(\Gamma^\mu_{\nu\rho}N^\nu N^\rho\right)_{X^\nu(\xi^{a})} n^2 + \cdots .
	\label{eq:Gaussian}
\end{equation}
Due to the normalization of $N^\mu$, the parameter $n$ gives the proper distance from $x^\mu$ to $\Sigma$ along the geodesic.
In a certain neighborhood of $\Sigma$, any point $x^\mu$ will belong to one and only one of these geodesics (see Fig.~\ref{fig:curved}).
This point is univocally determined by the distance $n$ to $\Sigma$ and the point $X^{\mu}(\xi^{a})$ where the geodesic crosses the hypersurface.
Thus, $\xi^a$ and $n$ can be used as new coordinates $\bar{x}^\mu$, which are called Gaussian normal coordinates \cite{carroll}.
The change of coordinates between $x^\mu$ and $\bar{x}^\mu$ is given by Eq.~(\ref{eq:Gaussian}).

We will be interested in the inverse transformation.
In particular, we want to obtain the quantity $n$ at a given point.
Near the hypersurface we have the expansion $F=\partial_n F|_0 n +\frac{1}{2}\partial_n^2F|_0 n^2+\cdots$
(for brevity, we will often use the notation $|_0$ or $|_{n=0}$ instead of $|_{X^\nu (\xi^a)}$).
Inverting this expansion, we obtain
\begin{equation}
	n=\left.\frac{1}{\partial_nF}\right|_{X^\nu(\xi^{a})} F
	-\left.\frac{\partial_n^2F}{2(\partial_nF)^{3}}\right|_{X^\nu(\xi^{a})} F^{2}
	+\left[\frac{(\partial_n^2F)^{2}}{2(\partial_nF)^{5}}
	-\frac{\partial_n^3F}{6(\partial_nF)^{4}}\right]_{X^\nu(\xi^{a})} F^{3}+\cdots .
	\label{eq:nsegundoorden}
\end{equation}
The derivatives of the function $F(x^\mu)$ with respect to $n$ are readily obtained as%
\footnote{Since the tangent vector $n^\mu$ is parallel transported along the geodesic, we have, e.g., 
	$\partial_n^2 = (n^\nu \nabla_\nu) (n^\mu\nabla_\mu)  = n^\mu n^\nu \nabla_\mu\nabla_\nu $, 
	and $n^\mu |_{n=0}=N^\mu$.}
\begin{equation}
	\partial_n F|_0=N^\mu\partial_\mu F=s,\quad
	\partial_n^2 F|_0=N^\mu\partial_\mu s=\partial_n s,\quad
	\partial_n^3 F|_0=N^{\rho}N^{\nu}N^{\mu}\nabla_{\rho}\nabla_{\nu}\partial_{\mu}F.
	\label{dnF}
\end{equation}
The coefficients of the expansion (\ref{eq:nsegundoorden}) are evaluated at the point $(\xi^a,n=0)$ on $\Sigma$. 
For small $n$, we can relate them to their values at $(\xi^a,n)$,
and we obtain \cite{mm24}
\begin{equation}
	n=\frac{F}{s} + \frac{N^{\mu}\partial_{\mu}s}{2s} \left(\frac{F}{s}\right)^{2} + \left[\frac{(N^{\mu}\partial_{\mu}s)^{2}}{2s^{2}}
	-\frac{N^{\rho}N^{\nu}N^{\mu}\nabla_{\rho}\nabla_{\nu}\partial_{\mu}F}{6s}\right] \left(\frac{F}{s}\right)^{3}+\cdots .
	\label{eq:n_orden2}
\end{equation}
where all the quantities are evaluated at  $x^\mu$.
Thus, Eqs.~(\ref{eq:nsegundoorden}) and (\ref{eq:n_orden2}) give two alternative expressions for $n$.
In Eq.~(\ref{eq:n_orden2}) we just have the expression for $n(x^\mu)$ in any coordinates, 
while in Eq.~(\ref{eq:nsegundoorden}) the coefficients are a function of the corresponding point $X^\mu(\xi^a)$ on $\Sigma$. 
The latter is useful to consider the variation of $n$ along a normal geodesic (i.e., at fixed $\xi^a$).
Notice that the expressions for the coefficients in Eq.~(\ref{eq:n_orden2}) are similar to those of the expansion
(\ref{eq:nsegundoorden})-(\ref{dnF}), but there is a change of sign in the quadratic term.

Similarly, we construct Gaussian normal coordinates  $\tilde{x}^i= (\zeta^{A},m)$ for $S$ at a given time through the change of coordinates
\begin{equation}
	x^{i}=Y^{i}(\zeta^{A})+M^{i}|_{Y^i(\zeta^{A})}m+\cdots ,
	\label{eq:GaussianS}
\end{equation}
where $m$ is the distance from $S$ along a normal geodesic in the slice of constant $t$.
We have an expression analogous to Eq.~(\ref{eq:nsegundoorden}), which
gives $m$ as a function of $G$ and $\zeta^A$, 
and an expression analogous to Eq.~(\ref{eq:n_orden2}), which gives $m(x^i)$.
Although the obvious choice for the function $G$ is $G(x^i)=F(t,x^{i})$, once we obtain $n(x^\mu)$
from Eq.~(\ref{eq:n_orden2}) we can also define $S$ through $G'(x^i)=n(t,x^i)=0$.
With this choice, the normal to $S$ is given by 
\begin{equation}
	M'_{i}=-\partial_{i}n/q' \quad \text{with} \quad q'={|n_{,i}n^{,i}|^{1/2}} .
\end{equation}
The vector fields $M^i$ and $M^{\prime i}$ coincide on $S$.
The advantage of using $G'=n$ is that the analogous of Eq.~(\ref{eq:nsegundoorden}) gives in this case
a direct relation between $m$ and $n$ along the geodesic of constant $\zeta^A$ and $t$, which we will need later,
\begin{equation}
	m= \left.\frac{1}{q'}\right|_{Y^i(\zeta^A)} n
	-\left.\frac{\partial_mq'}{2q^{\prime 3}}\right|_{Y^i(\zeta^A)} n^{2} +
	\left[\frac{(\partial_mq')^{2}}{2q^{\prime 5}}
	-\frac{\partial_m^3n}{6q^{\prime 4}}\right]_{Y^i(\zeta^A)} n^{3}+\cdots .
	\label{eq:nprin}
\end{equation}
We have%
\footnote{Here, $\nabla_i$ is the covariant derivative in 3D space.
	In flat space and using a metric $g_{\mu\nu}$ with $g_{00}=1$ and $g_{0i}=0$
	(e.g., in Cartesian or spherical coordinates), 
	the components $g_{ij}$ just give the induced metric on the spacelike hypersurface $x^0=t$.
	Thus, the intrinsic covariant derivative in 3D space
	is just given by the component $i$ of $\nabla_\mu$.}
$\partial_mq'|_{m=0}={M}^{i}\partial_{i}q'=M^{i}M^{j}\nabla_{i}\partial_{j}n$ and 
$\partial_m^3n|_{m=0}={M}^{i}{M}^{j}{M}^{k} \nabla_{i}\nabla_{j}\partial_{k}n$.
Using Eq.~(\ref{eq:n_orden2}), the derivatives of $n$ can be obtained from the function $F(x^\mu)$. 
For instance, we have $\partial_{i}n|_{m=0}=s^{-1}{\partial_{i}F}=-N_{i}$.

Fixing $\xi^a$ in Eq.~(\ref{eq:Gaussian}) we obtain the parametrization of a normal geodesic of $\Sigma$.
According to Eq.~(\ref{geod}), its tangent vector $n^{\mu}=\partial_n x^\mu$ fulfills $n^\sigma \nabla_\sigma n^\mu=0$ and $n^\mu|_{n=0}=N^\mu$, which means that $n^\mu$ is parallel transported from $N^\mu$.
It should be noted that Eq.~(\ref{eq:Nmu}) defines a vector field $N_\mu$ which is generally different from $n_\mu$, except on $\Sigma$.
Parallel transport implies that we have $n_\mu n^\mu=-1$ everywhere.
On the other hand, fixing $n$ in Eq.~(\ref{eq:Gaussian})
we obtain the parametrization of a hypersurface $\Sigma_n$ which is at a fixed distance $n$ from $\Sigma$.
It is not difficult to see that $n_\mu \partial_a x^\mu=0$ (see, e.g., \cite{carroll}),
so $n_\mu$ is the normal to $\Sigma_n$.
Since $\Sigma_n$ is defined by the condition $n=$ constant, we also have ${n}_{\mu}=-\partial_\mu n$.
The vectors $\partial_a x^\mu$ and $n^\mu$ make up the transformation matrix $\partial x^\mu/\partial\bar{x}^\alpha$, 
since the coordinate transformation is given by Eq.~(\ref{eq:Gaussian}).
As a consequence, the metric tensor in Gaussian normal coordinates
has components $\bar{g}_{nn}=-1 $ and $ \bar{g}_{an}=0$,
$\bar{g}_{ab} = g_{\mu\nu} \partial_a x^\mu \partial_b x^\nu$. 
The latter gives the induced metric on $\Sigma_n$
(in particular, we have $\bar{g}_{ab}|_{n=0} = \gamma_{ab}$).
The volume element is $d^{4}x\sqrt{-g}=dnd^{3}\xi\sqrt{\bar{g}}$, where $\bar{g}=\det\bar{g}_{ab}$.
Besides, $d^3\xi \sqrt{\bar{g}}$ is the surface element of $\Sigma_n$.
The projection tensor orthogonal to $n_{\mu}$ and tangent to $\Sigma_n$ (which coincides with $h^{\mu\nu}$ at $n=0$) is 
\begin{equation}
	P^{\mu\nu} = \bar{g}^{ab}\partial_a x^\mu \partial_b x^\nu = g^{\mu\nu}+n^\mu n^{\nu}.
	\label{eq:proyectorn}
\end{equation}

Similarly, in the Gaussian normal coordinates $\tilde{x}^i$ associated to $S$, 
the metric tensor  $\tilde{g}_{ij}$ fulfills $\tilde{g}_{mm}=-1$ and $\tilde{g}_{mA}=0$,
and $\tilde{g}_{AB}|_{m=0}$ gives the induced metric $\sigma_{AB}$.
The 3D volume element is given by $d^3x=dm\,d^{2}\zeta\sqrt{\tilde{g}}$, where $\tilde{g}=\det \tilde{g}_{AB}$
and $d^{2}\zeta\sqrt{\tilde{g}}$ is the surface element on the surface $S_m$ which is at a distance $m$ from $S$.
In particular, at $m=0$ we have $\sqrt{\tilde{g}}=\sqrt{\det \sigma_{AB}}$ and
we obtain the surface element on $S$.
The projection tensor on $S_m$ is given by $k_{ij}=g_{ij}+m_i m_j$, where $m^i$ is the geodesic's tangent vector,
which fulfills $m^i|_{m=0}=M^i$.

The tensor $K_{\mu\nu}=-\nabla_{\mu}n_{\nu}$ is the extrinsic curvature tensor of the hypersurface $\Sigma_n$
 \cite{carroll}.
In Gaussian normal coordinates we have $\bar{n}^{\mu}=(0,0,0,1)$, so $\bar{K}_{n\mu}=0 $ and $\bar{K}_{ab}=-\bar{\Gamma}_{ab}^{n}$.
The Christoffel symbol is given by $\bar{\Gamma}_{ab}^{n}=\frac{1}{2}\partial_{n}\bar{g}_{ab}$,
so we have
$\partial_{n}\bar{g}_{ab}=-2\bar{K}_{ab}$.
It can be shown that 
$n^\rho \nabla_\rho K_{\mu\nu} = {K_\mu}^\rho K_{\rho\nu}$
(see, e.g., Ref.~\cite{mm24}).
Using recursively this equation we obtain 
$n^{\sigma}n^{\mu}\nabla_{\sigma}\nabla_{\mu}K_{\rho\nu}=2{K_{\rho}}^{\sigma}{K_{\sigma}}^{\mu}K_{\mu\nu}$
and so on.
In Gaussian normal coordinates, these expressions give
the successive derivatives of $\bar{g}_{ab}$ with respect to $n$, 
and we obtain the expansion
\begin{equation}
	\bar{g}_{ab}=\gamma_{ab}-2\bar{K}_{ab}|_{n=0}\,n
	+{\bar{K}_{a}}^{c}\bar{K}_{cb}|_{n=0}n^{2} + \cdots .
	\label{gab}
\end{equation}
Using the well-known result $\delta g=gg^{\mu\nu}\delta g_{\mu\nu}$ (see,
e.g., \cite{poisson}), we obtain
\begin{equation}
	\sqrt{\bar{g}}=\sqrt{\gamma}
	\left[1-K|_{n=0}\, n+({1}/{2}) (K^{2}- K^{\mu\nu}K_{\mu\nu})_{n=0}\, n^{2}+ \cdots  \right],
	\label{expang}
\end{equation}
where we have defined the mean curvature $K$,
\begin{equation}
	K =g^{\mu\nu}K_{\mu\nu} =\bar{g}^{ab}\bar{K}_{ab}
	=-\bar{g}^{ab}\bar{\Gamma}_{ab}^{n}.
	\label{eq:K}
\end{equation}
We have $\partial_{n}K=K^{\mu\nu}K_{\mu\nu}$, $\partial_{n}^{2}K=2{K^{\mu}}_{\nu}{K^{\nu}}_{\rho}{K^{\rho}}_{\mu}$, and so on.
At $n=0$, we can write $K_{\mu\nu}=-{h_{\mu}}^{\rho}\nabla_{\rho}N_{\nu}$, and we obtain
\begin{equation}
	K|_{n=0}=-N_{;\mu}^{\mu},
	\quad
	\partial_{n}K|_{n=0}=N_{;\nu}^{\mu}N_{;\mu}^{\nu}
	\quad
	\partial_{n}^{2}K|_{n=0}=-2N_{;\nu}^{\mu}N_{;\rho}^{\nu}N_{;\mu}^{\rho}.
	\label{eq:Kcovar}
\end{equation}

We have analogous expressions for the extrinsic curvature of the surfaces $S_m$ in 3D space,
$\kappa_{ij}$, and related quantities.
For example, we have $\kappa_{ij}|_{m=0}=-{k_i}^k M_{j;k}$, 
the extrinsic curvature is given by $\kappa=g^{ij}\kappa_{ij}=-M^i_{;i}$,
and the quantities $\tilde{g}_{AB}$ and $\sqrt{\tilde{g}}$ have expansions
in powers of $m$ analogous to (\ref{gab})-(\ref{expang}), 
e.g.,
\begin{equation}
	\sqrt{\tilde{g}/\tilde{g}|_{m=0}} = 1-\kappa|_{m=0}m+ ({1}/{2})  (\kappa^2-\kappa^{ij} \kappa_{ij})_{m=0} m^{2}+\cdots .
	\label{expangtil}
\end{equation}

\subsection{Energy decomposition}

Given a reference hypersurface $\Sigma$ for the wall, we establish Gaussian normal coordinates 
$\bar{x}^\mu=(\xi^a,n)$
such that $\Sigma$ is located at $n=0$ and the false-vacuum region is at $n>0$.
In these coordinates, the field equation,  Eq.~(\ref{eq:eccampo}), takes the form
\begin{equation}
	\partial_{n}^{2}\phi-K\partial_{n}\phi-D_{a}D^{a}\phi=V'(\phi), \label{ecfi}
\end{equation}
where we have used Eq.~(\ref{eq:K}) and we have defined%
\footnote{The operator $D_a$ is the intrinsic covariant derivative, which can be defined as the projection
	$D_bA_a= \partial_b X^\beta \partial_a X^\alpha \nabla_\beta A_\alpha $ \cite{poisson} 
	or $D_\nu A_\mu= {P^\beta}_\nu {P^\alpha}_\mu \nabla_\beta A_\alpha$ \cite{carroll}.
	In Gaussian normal coordinates, we have $D_b A_a =\partial_bA_a-\bar{\Gamma}_{ba}^c A_c$.}
$
D_{a}D^{a}\phi\equiv\bar{g}^{ab}\left(\partial_{a}\partial_{b}\phi-\bar{\Gamma}_{ab}^{c}\partial_{c}\phi\right)
$.
The first integral of this EOM gives a relation between $V(\phi(\bar{x}^\mu))$ and the derivatives of $\phi(\bar{x}^\mu)$.
As we shall see, this relation can be used to determine what part of the potential energy density to associate with the wall. 
Assuming that we have the undisturbed false vacuum on one side of the interface, 
we impose the asymptotic boundary condition $\phi\to\phi_+$ for $n\to\infty$.
Hence, multiplying Eq.~(\ref{ecfi}) by $\partial_n\phi$ and integrating with respect to $n$, we obtain 
\begin{equation}
	V(\phi(\xi^a,n)) = \frac{1}{2}\left(\partial_{n}\phi\right)^{2}  
	+ \int_{n}^{\infty}d n' \left[K(\partial_{n}\phi)^{2}+\partial_{n}\phi D_{a}D^{a}\phi\right] +  V_{+}.
	 \label{eq:primeraint}
\end{equation}
This equation provides a decomposition of the potential energy density at the point $\bar{x}^\mu$ into two functions,
\begin{equation}
	V^w(\xi^a,n)=\frac12(\partial_n\phi)^2 , \quad
	V^b(\xi^a,n) = V_{+}
	+ \int_{n}^{\infty}d n' \left[K(\partial_{n}\phi)^{2}+\partial_{n}\phi D_{a}D^{a}\phi\right] .
	\label{defVbulk}
\end{equation}
The function $V^w(\bar{x}^\mu)$ has a peak at the interface and vanishes where $\phi$ is constant. 
In Eq.~(\ref{eq:primeraint}), this peak of $V^w(\bar{x}^\mu)$ describes the field crossing the potential barrier of $V(\phi(\bar{x}^\mu))$.
Therefore, the gradient energy density $\frac12(\partial_n\phi)^2$ and the barrier potential energy density $V^w(\bar{x}^\mu)$ contribute equally to the wall energy density (see below).
On the other hand, the function $V^b(\bar{x}^\mu)$ gives a smooth transition between the potential densities in the two domains.
In particular, if we assume that the field takes the values $\phi_\pm$ at $n=\pm\infty$, Eq.~(\ref{eq:primeraint}) implies that $V^b$
varies between $V_-$ and  $V_+$.

Writing the energy-momentum tensor, Eq.~(\ref{eq:Tmunufield}), in Gaussian normal coordinates and 
using Eq.~(\ref{eq:primeraint}), we obtain
\begin{align}
	\bar{T}_{ab} & =\bar{g}_{ab} 
	\left[(\partial_{n}\phi)^{2} + V^b\right]
	+\left(\bar{g}_{a}^{c}\bar{g}_{b}^{d}-{\textstyle\frac{1}{2}}\bar{g}_{ab}\bar{g}^{cd}\right)
	\partial_{c}\phi\partial_{d}\phi, 
	\label{TabNGC2}
	\\
	\bar{T}_{an} & = \partial_{a}\phi\partial_{n}\phi , \qquad
	\bar{T}_{nn} = - V^b
	+{\textstyle\frac{1}{2}}\bar{g}^{ab}\partial_{a}\phi\partial_{b}\phi .
	\label{TnnTanNGC2}
\end{align}
If the wall is thin enough, $\phi$ only depends on $n$
and we have 
$\bar{T}_{ab} =\bar{g}_{ab} [\phi^{\prime 2} + V^b]$, $\bar{T}_{an}=0$, and $\bar{T}_{nn} = - V^b$,
which we can write covariantly as
\begin{equation}
	T_{\mu\nu}^{\mathrm{thin}}= [\phi'(n)]^{2}P_{\mu\nu} + V^b(n){g}_{\mu\nu}.
	\label{eq:Tmunuthin}
\end{equation}
We recognize the terms in this expression as smooth versions of those in Eq.~(\ref{eq:Tmunudelta}).
In the limit of an infinitely thin wall, $[\phi'(n)]^{2}$ becomes a delta function, while $V^b(n)$ gives the potential jump at the interface.
Therefore, in this case it is clear that we have to define 
$T_{\mu\nu}^{w} =\phi^{\prime 2}P_{\mu\nu}$ and
${T}_{\mu\nu}^{b}=V^b{g}_{\mu\nu}$.
To obtain the decomposition (\ref{Tmunudecomp}) 
in the general case, we notice that the only term in Eqs.~(\ref{TabNGC2})-(\ref{TnnTanNGC2}) with the desired properties for a wall contribution is still the one proportional to
$(\partial_{n}\phi)^{2}$, which only occurs in the components $\bar{T}_{ab}$,
so we define
\begin{equation}
	T_{\mu\nu}^{w} =(\partial_{n}\phi)^{2}P_{\mu\nu}
				 = (n^\mu\partial_\mu\phi)^2 P_{\mu\nu} .
	\label{eq:Tmunuw}
\end{equation}
Besides being concentrated at the wall, this tensor is tangent to the family of hypersurfaces $\Sigma_n$,
indicating that no momentum associated to the wall flows out of the wall region.
The term $\partial_{a}\phi\partial_{n}\phi$ has a (lower) peak inside the wall,
but this peak occurs in the components $\bar{T}_{an}$ and
represents momentum flowing orthogonal to the wall.
We may interpret this term as an interaction between the wall and the bulk.
The rest of the terms can be assigned to the bulk part of $T_{\mu\nu}$.
As we will see, nonvanishing contributions to $\bar{T}_{an}$ occur only beyond the next-to-next-to-leading order in the wall width.

We will be interested in particular in the energy density of the wall,
$T_{00}^{w}=P_{00}(\partial_{n}\phi)^{2}$, with $P_{00}=g_{00}+n_0n_0$. 
In flat space and in a coordinate system with $g_{00}=1$ and $g_{0i}=0$, 
we have $P_{00}=-n_{i}n^{i}=-n_{,i}n^{,i}=q^{\prime2}$, and
\begin{equation}
	T_{00}^{w} =  q^{\prime2} (\partial_n\phi)^2 =(\nabla n)^2 (n^\mu \partial_\mu \phi)^2.
	\label{Tw00}
\end{equation}
If $\phi$ only depends on $n$ (which is true only in some specific cases or as an approximation for thin walls), 
we have $T_{00}^{w}=-g^{ij}\partial_{i}\phi\partial_{j}\phi=(\nabla\phi)^{2}$.
The dependence with $n$ disappears and the quantity $T_{00}^{w}$ 
can be continued beyond the range of validity of the Gaussian normal coordinates.

Having isolated the wall contribution, we can now compute the surface stress-energy tensor $S_{\mu\nu}$
using $T_{\mu\nu}^w$ instead of $T_{\mu\nu}$ in Eq.~(\ref{eq:Smunuthin}).
We can extend the integration interval as far as we like, as long as the Gaussian normal coordinates are valid.
However, there are still some issues regarding the volume element and the covariant integration.
Let us consider first a related but simpler quantity, namely, the surface energy density $\varepsilon_w$.

The energy of the wall is given by $E_w  =\int d^{3}x\,T_{00}^{w} $.
In the Gaussian normal coordinates associated to the two-dimensional surface $S$, 
we have
\begin{equation}
		E_w  = \int d^{2}\zeta dm\sqrt{\tilde{g}}\,T_{00}^{w} 
	\equiv \int_{S}d^{2}\zeta\sqrt{\tilde{g}|_{m=0}} \,\varepsilon_{w},
\end{equation}
i.e., we define $\varepsilon_w$ by the condition that its surface integral gives the total wall energy.
Therefore, this quantity is given by
\begin{equation}
		\varepsilon_{w}=\int dm\sqrt{\tilde{g}/\tilde{g}|_{m=0}}\,T_{00}^{w}.
	\label{eq:Ewdef}
\end{equation}
Note that Eq.~(\ref{eq:Ewdef}) is a line integral along a geodesic that intersects $S$ at a given point $\zeta^A$
at a given time $t$.
The curve is given by Eq.~(\ref{eq:GaussianS}).
In Cartesian coordinates, we just have $x^{i}=Y^{i}(\zeta^{A})+M^{i}(\zeta^{A})m$.
On the other hand, the wall energy density is given by Eq.~(\ref{Tw00}) as a function of the Gaussian normal coordinates $\xi^a,n$ associated to the hypersurface $\Sigma$.
The integration requires the change of variables to the coordinates $\zeta^A,m$.

The definition of $\varepsilon_w$, Eq.~(\ref{eq:Ewdef}), is similar to that of $S_{\mu\nu}$ in Eq.~(\ref{eq:Smunuthin}),
except that here 
we have taken into account the variation of the surface element with $m$.
If the wall is really thin, then the factors of $\sqrt{\tilde{g}}$ cancel.
Similarly, we may define  $S_{\mu\nu}$ as
\begin{equation}
	S_{\mu\nu}  = \int dn\sqrt{\bar{g}/\bar{g}|_{n=0}}\,T^{w}_{\mu\nu} 
	=\gamma^{-1/2}\int dn\sqrt{\bar{g}}\,(\partial_{n}\phi)^{2} P_{\mu\nu} .
	\label{Smunudef}
\end{equation}
The caveat with this definition is 
that integrating the components of a tensor will generally not give a tensor, unless we are in flat space and in a Lorentz frame \cite{misner}.
Indeed, under a general coordinate transformation, 
we will have the transformation matrix in the integrand of Eq.~(\ref{Smunudef}),
while $S_{\mu\nu}$ should transform with the transformation matrix evaluated at $n=0$.
Similarly, raising or lowering indices of $T_{\mu\nu}$ will not be equivalent to raising or lowering indices of $S_{\mu\nu}$ if $g_{\mu\nu}$ is not constant.
These are not problems for $\varepsilon_w$ because the covariance was already lost 
when we decided to integrate the energy density $T_{00}$ in a given frame.
If the specific application favors some definite coordinates that are appropriate for the problem at hand,
the definition (\ref{Smunudef}) will still be useful.
In this work, we will consider the Gaussian normal coordinates to be such preferred coordinates. 
We have $\bar{P}_{\mu n}=0$ and $\bar{P}_{ab}=\bar{g}_{ab}$, so we obtain
\begin{equation}
	\bar{S}_{\mu n}=0 , \quad \bar{S}_{ab}=\gamma^{-1/2}\int dn\sqrt{\bar{g}}\,(\partial_{n}\phi)^{2} \bar{g}_{ab} .
	\label{Sabdef}
\end{equation}

The definition (\ref{Smunudef}) is approximately covariant if
the coordinate transformation is approximately constant in the wall range, so that it can be removed from the integral.
Similarly, raising and lowering indices of $T^w_{\mu\nu}$ is equivalent to raising and lowering indices of $S_{\mu\nu}$
if $g_{\mu\nu}$ is approximately constant within the wall.
In this case, the projection tensor $P_{\mu\nu}$ in Eq.~(\ref{eq:Tmunuw}) is approximately constant,
and $S_{\mu\nu}$ takes the surface-layer form
\begin{equation}
	S_{\mu\nu} = \sigma h_{\mu\nu},
	\label{Smunuthin}
\end{equation}
where $\sigma$ is the surface tension,
\begin{equation}
	\sigma=\int(\partial_{n}\phi)^{2}dn.
	\label{eq:defsigma}
\end{equation}
In Eq.~(\ref{eq:defsigma}), the integral is along the normal geodesic at a given position $\xi^a$ on $\Sigma$.
In general, $\phi$ is not a function of $n$ alone, and $\sigma$ may depend on $\xi^a$.

To finish this section, let us discuss alternative definitions of the wall surface.
If we are interested in the evolution of the energy density, 
we can choose the surface $S$ at a given time to be located at the point of maximum $T_{00}^w$.
Alternatively, we can choose the wall position as the average of $m$ using the energy density as a weight.
If the energy is concentrated in a small range of $m$, the two points will be very close.
Since we set the origin of the coordinate $m$ at $S$,
any of these definitions gives an equation $m_w(t,\zeta^A)=0$ which defines the surface.
Notice that this approach is somewhat recursive, as it involves using the field $\phi(t,\zeta^A,m)$ as a function of the Gaussian normal coordinates associated to $S$ in the definition of this surface.
Its utility depends on the problem.

Similarly, we can choose the hypersurface $\Sigma$ to be located at the maximum of $(\partial_{n}\phi)^{2}$
or at the mean position $n$ weighted by this quantity.
These options are particularly useful for cases where $\phi$ depends only on $n$
and when using the thin-wall approximation.
In this case, we obtain a condition for $\Sigma$ of the form $n_w(\xi^a)=0$.
For example, using the mean value, we have
\begin{equation}
	\int(\partial_{n}\phi)^{2}\,n\,dn=0 , \label{eq:cero_n}
\end{equation}
where the integration extends over the entire range where the variable $n$ is well defined.
If the field only depends on $n$, these definitions of the wall position are equivalent to the condition $\phi(x^\mu)=\phi_w$ discussed above, with the specific value $\phi_w$ corresponding to $n_w=0$.
The use of an average with an appropriate weight function can be 
generalized to define other quantities characterizing the wall.
Here we will define the wall width $l$ by
\begin{equation}
	l^2 = \frac{\int(\partial_{n}\phi)^{2}n^{2}dn}{\int(\partial_{n}\phi)^{2}dn}
	\equiv \frac{\mu}{\sigma} .
	\label{eq:defl}
\end{equation}

\subsection{Spherical bubbles}

For the particular case of a spherical bubble in flat space, 
Eq.~(\ref{eq:eccampo}) takes the form%
\footnote{In spherical coordinates, 
	the non-vanishing Christoffel symbols are $\Gamma_{\theta\theta}^{r}=-r$, $\Gamma_{\varphi\varphi}^{r}=-r\sin^{2}\theta$, $\Gamma_{\varphi\varphi}^{\theta}=-\sin\theta\cos\theta$, $\Gamma_{r\theta}^{\theta}=r^{-1}$, $\Gamma_{r\varphi}^{\varphi}=r^{-1}$, and $\Gamma_{\theta\varphi}^{\varphi}=\cot\theta$.}
\begin{equation}
	-\partial_t^2\phi + \partial_r^2\phi + {2}{r^{-1}}\partial_r\phi = V'(\phi) 
	\label{ecfiO3}
\end{equation}
and the boundary conditions are $\partial_r\phi=0$ at $r=0$ and $\phi\to \phi_+$ for $r\to\infty$.
The specific solution depends on the initial configuration $\phi(t=0,r)$.
Using any of the definitions of the wall surface discussed above, e.g., $\phi(t,r)=\phi_w$, we obtain the wall position $r_w(t)$.
Thus the hypersurface $\Sigma$ can be defined as in Eq.~(\ref{eq:Monge}), either
by $F=r-r_w(t)=0$ or by writing $x^\mu = X^{\mu}(\xi^{a})$ in spherical coordinates,
\begin{equation}
	(t,r,\theta,\varphi)=\left(\xi^{0},r_w(\xi^{0}),\xi^{1},\xi^{2}\right) .
	\label{eq:paramSigmaesf}
\end{equation}
From Eq.~(\ref{eq:Nmu}), we have $s = \sqrt{1-\dot{r}_w^2}\equiv \gamma_w^{-1}$ and
\begin{equation}
	N^\mu=\gamma_w(\dot{r}_w,1,0,0) .
	\label{NmuO3}
\end{equation}
The quantities associated with the extrinsic curvature of the hypersurface that appear in Eqs.~(\ref{expang})-(\ref{eq:Kcovar}) are given by
\begin{equation}
	N^{\mu}_{;\mu} = \gamma_{w}^{3}\ddot{r}_{w}+2\frac{\gamma_{w}}{r_{w}},
	\quad
	N^{\mu}_{;\nu}N^{\nu}_{;\mu}=
	(\gamma_{w}^{3}\ddot{r}_{w})^{2}+2\frac{\gamma_{w}^{2}}{r_{w}^2},
	\quad
	N^{\mu}_{;\nu}N^{\nu}_{;\rho}N^{\rho}_{;\mu}=
	(\gamma_{w}^{3}\ddot{r}_{w})^{3}+2\frac{\gamma_{w}^{3}}{r_{w}^{3}} .
	\label{d2nKesf}
\end{equation}

For the surface $S$, 
we can use $G=r-r_w(t)=0$ as in Eq.~(\ref{MongeS}) (with the additional constraint that $t$ is fixed).
Setting $X^0=\xi^0=t$ and $\xi^A=\zeta^A$ in Eq.~(\ref{eq:paramSigmaesf}),
the parametric representation $x^i=Y^i(\zeta^A)$ takes the form
\begin{equation}
	(r,\theta,\varphi)=\left(r_w(t),\zeta^1,\zeta^2\right).
\end{equation}
The normal vector, given by Eq.~(\ref{eq:Mi}), is the unit radial vector $M^i=(1,0,0)$,
and the quantities associated with the extrinsic curvature that appear in Eq.~(\ref{expangtil}) are given by 
$\kappa=-2r^{-1}$, $\kappa^{ij}\kappa_{ij}=2r^{-2}$.

In flat space, the geodesics are straight lines, and the ones tangent to $N^\mu$ at the point $X^\mu(\xi^a)$ are given by
\begin{equation}
	(t,r,\theta,\varphi)=
	\big(\xi^{0},r_{w}(\xi^{0}),\xi^{1},\xi^{2}\big)+\gamma_{w}(\xi^{0})\big(\dot{r}_{w}(\xi^{0}),1,0,0\big)n ,
	\label{geodesicaO3}
\end{equation}
i.e., the expansion (\ref{eq:Gaussian}) is truncated at the linear term.
This equation also gives the transformation to Gaussian normal coordinates,
namely, $\theta=\xi^{1}$, $\varphi=\xi^{2}$, 
\begin{equation}
	t=\xi^{0}+\gamma_{w}(\xi^{0})\dot{r}_{w}(\xi^{0})n,\quad r=r_{w}(\xi^{0})+\gamma_{w}(\xi^{0})n.
	\label{gaussianasO3}
\end{equation}
To obtain the inverse transformation, we must solve the equations
\begin{equation}
	\frac{t-\xi^{0}}{r-r_{w}(\xi^{0})}=\dot{r}_{w}(\xi^{0}),\quad n=\frac{r-r_{w}(\xi^{0})}{\gamma_{w}(\xi^{0})} .
	\label{gaussianasO3inv}
\end{equation}
We can solve the first one for $\xi^0(t,r)$ and then use the result in the second one.
We can readily obtain the vector tangent to the normal geodesics, either using $n^\mu = \partial_n x^\mu$ in Eq.~(\ref{gaussianasO3}) or $n_{\mu}=-\partial_{\mu}n$ in Eq.~(\ref{gaussianasO3inv}). 
We have (in spherical coordinates)
\begin{equation}
	n^\mu=\gamma_w(\xi^0)(\dot{r}_w(\xi^0),1,0,0) .
	\label{nmuesf}
\end{equation}
As expected, $n^\mu$ is constant along the geodesics, which are lines of fixed $\xi^0$ in the $rt$ plane.
To obtain $n^\mu(t,r)$, we must use the function
$\xi^0(t,r)$ obtained from Eq.~(\ref{gaussianasO3inv}).

The normal geodesics to $S$ at constant $t$ are radial lines
\begin{equation}
	(r,\theta,\varphi)=\left(r_w(t),\zeta^{1},\zeta^{2} \right) + \left(1,0,0\right) m ,
	\label{eq:GaussianSphere}
\end{equation}
so the coordinate transformation (\ref{eq:GaussianS})
is in this case truncated at the linear term. 
Thus, the Gaussian normal coordinates are essentially the spherical coordinates,
with the radial coordinate shifted to the surface $S$, i.e., $\theta=\zeta^1$, $\varphi=\zeta^2$, $r = r_w(t) + m$, 
and the metric tensor in these coordinates is given by 
$\tilde{g}_{ij}=-\mathrm{diag} (1,r^{2},r^{2}\sin^{2}\theta)$.

The energy-momentum tensor of the wall is given by Eq.~(\ref{eq:Tmunuw}) and, in particular, the wall energy density is given by Eq.~(\ref{Tw00}). We have
\begin{equation}
	T_{00}^w=(\partial_r n)^2(\partial_n\phi)^2.
	\label{T00wesf}
\end{equation}
If $\phi$ depends only on $n$, we just have $T_{00}^w=(\partial_r\phi)^2$.
However, this is not the general case.
On the other hand, we will generally have a numerical solution of Eq.~(\ref{ecfiO3}) 
as a function of $r$ and $t$ rather than $n$ and $\xi^0$.
To obtain $T_{00}^w(t,r)$, 
we can write  $\partial_n\phi = n^0 \partial_t \phi+n^r \partial_r \phi$, 
where we have, from Eqs.~(\ref{gaussianasO3})-(\ref{nmuesf}),
$ n^0= \dot{r}_{w}(\xi^0)\gamma_{w}(\xi^0)$ and $n^r=\partial_r n=\gamma_w(\xi^0)$.
We still need to obtain $\xi^0(t,r)$ by solving the first of Eqs.~(\ref{gaussianasO3inv}).
This procedure is tedious. 
Fortunately, as we will see, in most cases the dependence of $\phi(\xi^0,n)$ on $\xi^0$ is negligible and we can use the approximation 
$T_{00}^w \simeq (\partial_r\phi)^2$.

The surface energy density $\varepsilon_w$ is given by Eq.~(\ref{eq:Ewdef}).
In this case, we have $dm=dr$ and the metric determinant is straightforward.
Assuming that $T_{00}^w$ is well defined in all the range of the variable $r$, we obtain
\begin{equation}
	E_w = 4\pi r_{w}^{2} \varepsilon_w = 4\pi\int_{0}^{\infty}dr T_{00}^w.
	\label{Ewesf}
\end{equation}

\section{An illustrative example: O(3,1)-symmetric bubble}
\label{sec:O31}

To illustrate the concepts developed in the previous section, 
let us consider a simple case where the field depends only on the normal coordinate $n$.
To this end, we will further simplify the O(3)-symmetric problem by considering an O(3,1)-invariant solution,
where $\phi$ only depends on the variable $\rho=\sqrt{r^2-t^2}$ \cite{c77}.
Thus, we write  $\phi(x^\mu)=\bar\phi(\rho)$, and the field equation becomes
\begin{equation}
	\bar{\phi}''(\rho)+3\rho^{-1}\bar{\phi}'(\rho)=V'(\bar{\phi}),
	\label{eq:ecfiColeman}
\end{equation}
with the boundary conditions $\bar{\phi}'(0)=0$, $\bar{\phi}(\rho)\to\phi_{+}$ for $\rho\to\infty$.
This problem is easily solved by the shooting method, 
where we replace the condition at infinity by $\bar{\phi}(0)=\phi_i$ to solve the equation numerically, 
and then iteratively search for the correct value of $\phi_i$ such that $\bar{\phi}$ approaches $\phi_+=0$ for large $\rho$.
The solution thus obtained is only valid for $r\geq t$.
To obtain the solution inside the lightcone numerically, 
we can consider the analytic continuation
$\phi(x^\mu)=\bar{\phi}(i\tau)=\tilde{\phi}(\tau)$, with $\tau=\sqrt{t^2-r^2}$, for which we have the equation
\begin{equation}
	\tilde{\phi}''(\tau)+3\tau^{-1}\tilde{\phi}'(\tau)=-V'(\tilde{\phi})
	\label{eq:ecfiColemanin}
\end{equation}
with boundary conditions $\tilde\phi (0)=\phi_i$, $\tilde\phi'(0)=0$ 
(see, e.g., \cite{mm23,bgm16,jkt19}).

\subsection{Bubble profile and wall position}

Since the field depends only on $\rho$, it makes sense to set the wall position to a fixed value $\rho=\rho_w$,
which is equivalent to the condition $\phi=\phi_w$.
Therefore, the bubble radius is given by
\begin{equation}
	r_w(t)= \sqrt{\rho_{w}^{2}+t^{2}} ,
	\label{rwColeman}
\end{equation}
and we have $\dot{r}_w=t/r_w$ and $\gamma_w=r_w/\rho_w$.
Hence, the normal geodesic crossing $\Sigma$ at a point $t=\xi^0$, $r=r_w(\xi^0)$ is given by 
\begin{equation}
	t=\xi^{0}(1+n/\rho_{w}), \quad r=r_{w}(\xi^{0})(1+n/\rho_{w}) .
	\label{eq:GaussianColeman}
\end{equation}
This line goes through the origin and has slope  $t/r=\xi^{0}/r_{w}(\xi^{0})$.
These relations also give the transformation to the Gaussian normal coordinates.
The inverse transformation is
$\xi^{0}=t\rho_{w}/\rho$, $n=\rho-\rho_{w}$,
so $n$ is just the variable $\rho$ shifted to $\Sigma$.
If we define $\Sigma$ by the condition (\ref{eq:cero_n})
and the wall width by Eq.~(\ref{eq:defl}), we have
\begin{equation}
	\rho_{w} = \sigma^{-1}\int_{0}^{\infty}\left[\bar{\phi}'(\rho)\right]^{2} \rho\, d\rho ,
	\quad
	l^2 = \sigma^{-1}\int_0^\infty\left[\bar{\phi}'(\rho)\right]^{2} (\rho-\rho_w)^{2}d\rho ,
	\label{rhow}
\end{equation}
with $\sigma=\int_{0}^{\infty} \left[\bar{\phi}'(\rho)\right]^{2} d\rho$.

For concrete computations, we will consider a toy-model with a quartic potential
\begin{equation}
	V(\phi)=\frac{1}{2}m^{2}\phi^{2}-\frac{e}{3}\phi^{3}+\frac{\lambda}{4}\phi^{4} 
	+ V_+ .
	\label{eq:pot}
\end{equation}
This potential has a minimum at $\phi_{+}=0$, another minimum at
\begin{equation}
	\phi_{-}=(e/2\lambda)(1+\sqrt{1-4\lambda m^{2}/e^{2}}),
\end{equation}
and a maximum between them, at $\phi_{\max}=e/\lambda-\phi_{-}$.
In the parameter region $e^{2}/\lambda m^{2} > 9/2 $ we have $V(\phi_+)>V(\phi_-)$.
The height of the barrier, $V_{\max} \equiv V(\phi_{\max})-V(0)$,
is given by $V_{\max} =\frac{1}{4}\phi_{\max}^{2}(m^{2}-\frac{e}{3}\phi_{\max})$.
For $e^{2}/\lambda m^{2} = 9/2 $, the two minima are degenerate, while  for $e^{2}/\lambda m^{2}\to \infty$
the barrier disappears.
The constant term in Eq.~(\ref{eq:pot}) is given by
$V_{+} = \frac{1}{4}\phi_{-}^{2}\left(\frac{e}{3}\phi_{-}-m^{2}\right)$
so that the vacuum energy density vanishes in the true vacuum, 
i.e., we have $V_{-}=0$ and
$\Delta V\equiv V_{+}-V_- =V_+$.

\begin{figure}[tb]
	\centering
	\includegraphics[width=0.45\textwidth]{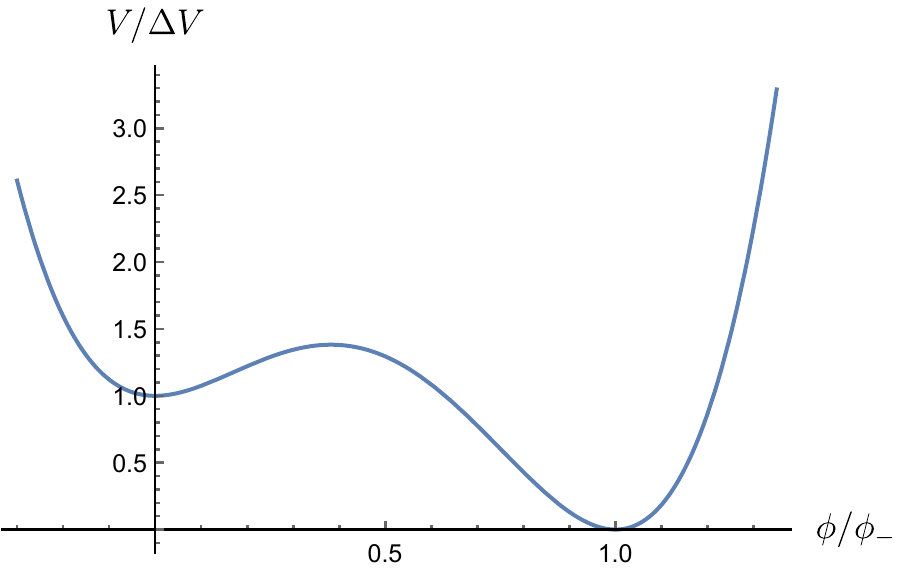}\hfill\includegraphics[width=0.5\textwidth]{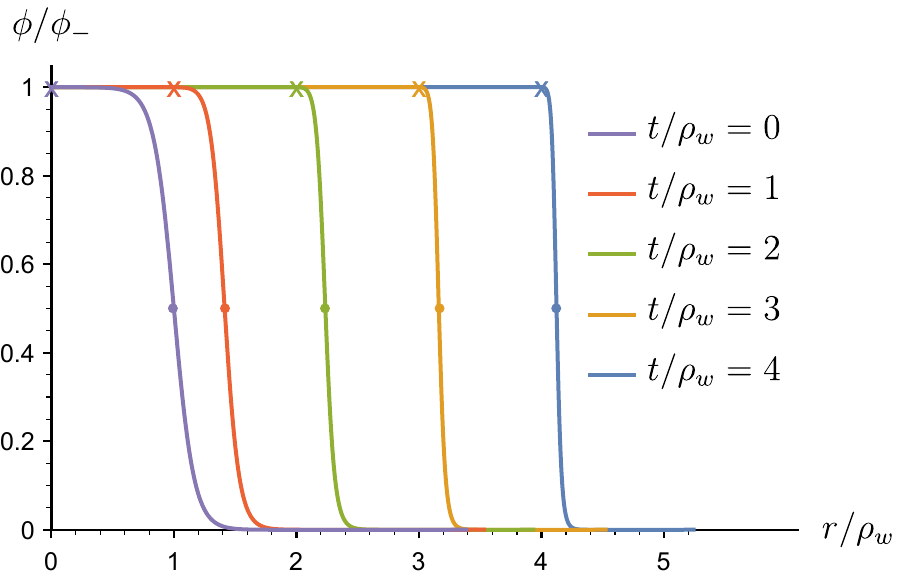}
	\caption{Evolution of the bubble profile for the potential (\ref{eq:pot}) with
		$e^2/\lambda m^2=5$, which gives $\Delta V/V_{\max}\simeq 2.6$.
		\label{fig:potperf1}}
\end{figure}

It is well known that if the potential is nearly degenerate (i.e., if $\Delta V \ll V_{\max}$),
the solution $\bar{\phi}(\rho)$ has a thin wall interpolating
between constant values $\phi_-$ and $\phi_+$ \cite{c77}.
Since we want to go beyond the thin-wall approximation,
let us consider a potential shape with $\Delta V\gtrsim V_{\max}$ (see Fig.~\ref{fig:potperf1}).
The right panel shows the solution $\phi(r,t)= \bar{\phi}(\sqrt{r^2-t^2})$ at different times.
The leftmost curve gives the graph of the function $\bar{\phi}(\rho)$, since at $t=0$ we have $\rho=r$.
In this curve, $\phi$ varies appreciably in a range of $\rho$ of order $\rho_w$.
However, using Eq.~(\ref{rhow}) as a measure of the wall width $l$, we have $l/\rho_w\simeq 0.093$.
At later times, the wall gets thinner due to the Lorentz contraction.
The value $\bar\phi(0)=\phi_i$ (which in this case is very close to $\phi_-$) is indicated by a cross on the curves.
The corresponding point $\rho=0$ ($r=t$) separates $\bar\phi(\rho)$ from the 
innermost solution $\tilde{\phi}(\tau)$ (in this case, $\tilde{\phi}\simeq \phi_-$).
The dots indicate the wall position $r=r_w(t)$ given by Eqs.~(\ref{rwColeman}) and (\ref{rhow}).

\begin{figure}[bt]
	\centering
	\includegraphics[width=0.45\textwidth]{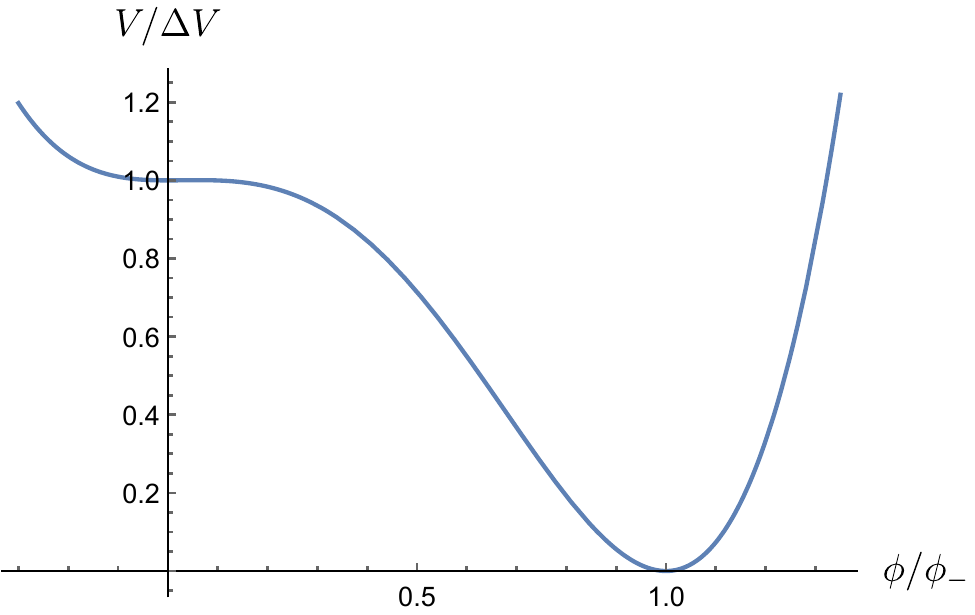}\hfill\includegraphics[width=0.5\textwidth]{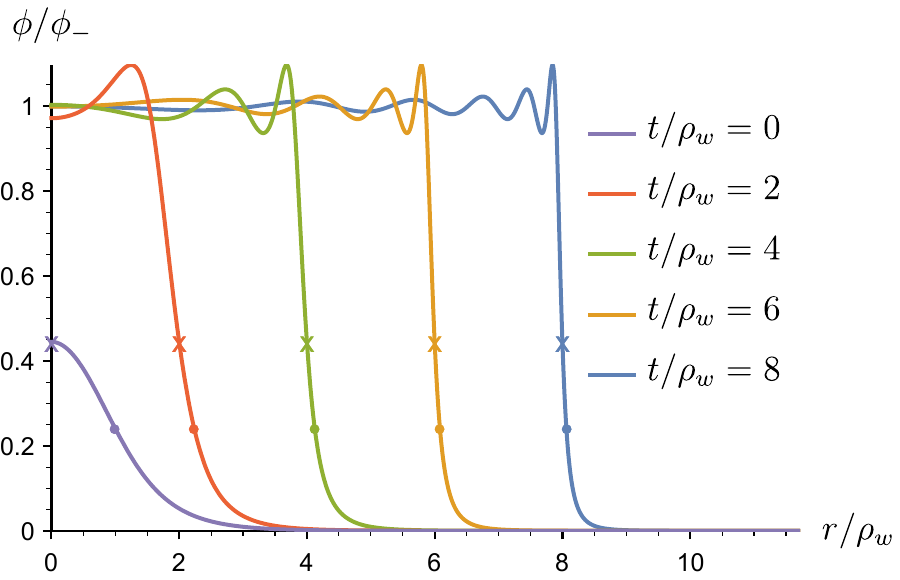}
	\caption{Evolution of the bubble profile for the potential (\ref{eq:pot})
		with $e^2/\lambda m^2 = 18$, which gives $\Delta V/V_{\max}\simeq 1800$.
		\label{fig:potperf2}}
\end{figure}

\begin{figure}[tb]
	\centering
	\includegraphics[height=5.6cm]{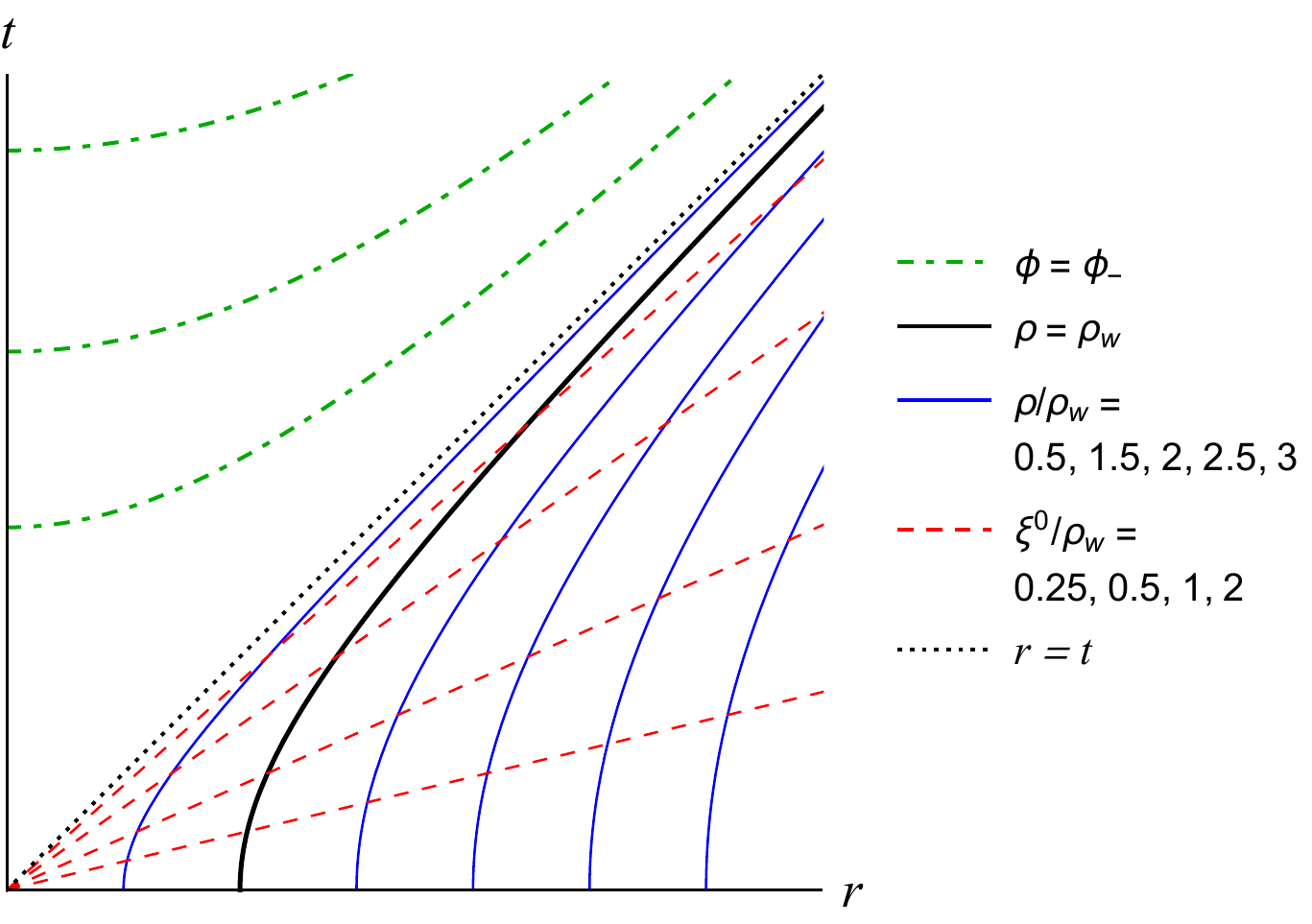}
	\hfill%
	\includegraphics[height=4.5cm]{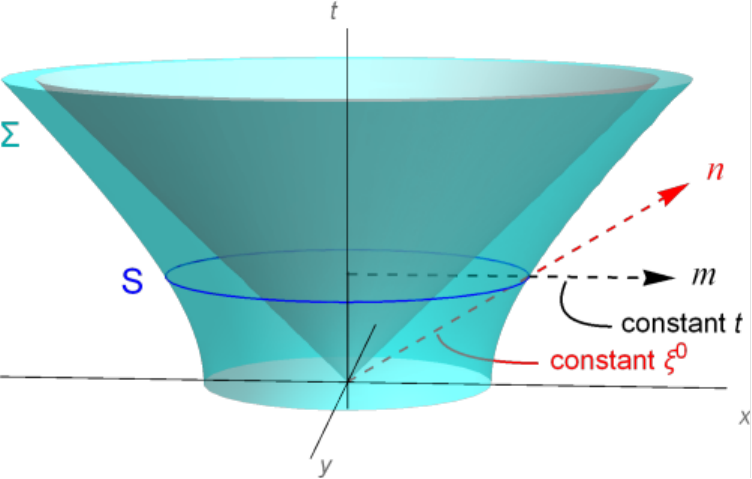}%
	\caption{Left: In the $rt$ plane, lines of constant $\phi=\phi(\rho_w)$ for several values of $\rho_w$, 
		lines of constant $\phi=\phi_-$, and normal geodesics for different values of $\xi^0$, for the potential of Fig.~\ref{fig:potperf2}.
		Right: Schematically, the hypersurface $\Sigma$, the surface $S$, and the lightcone $r=t$. 
		The variable $n=\rho-\rho_w$ measures the proper distance from a point of $\Sigma$ along a geodesic tangent to $N^\mu$,
		which is a curve of constant $\xi^0$.
		The variable $m=r-r_w(t)$ measures the distance from a point of $S$ along a geodesic at constant $t$ tangent to $M^{i}$.}\label{fig:cono}
\end{figure}

To illustrate the scope and limitations of this treatment, we consider in Fig.~\ref{fig:potperf2} a potential with a very small barrier compared with the potential difference $\Delta V$.
In this case, the initial profile does not even have a domain with $\phi=\phi_-$.
If we consider the wall as the region where $\phi$ varies, then the bubble is initially all wall.
Formally, Eq.~(\ref{rhow}) gives $l/\rho_w \simeq 0.44$.
Since the initial value of the field at the bubble center is quite far from the minimum $\phi_-$,
the field evolves inside the bubble and oscillates around the minimum.
Due to the oscillations, it is not evident, looking at field gradients, where 
the wall ends and the bubble interior begins.
The point $r=t$, $\phi=\phi_i$ (indicated by the crosses) gives a natural boundary as it separates the solutions $\tilde{\phi}(\tau)$ and $\bar{\phi}(\rho)$.
Indeed, for any $\rho>0$, a fixed point of the profile given by $\phi=\bar{\phi}(\rho)$ describes a hyperbola $r^2-t^2=\rho^2$ (see the left panel of Fig.~\ref{fig:cono}).
Therefore, all these points move with constant acceleration, which is in agreement with the interpretation 
that this part of the profile makes up the bubble wall.
In contrast, for $r<t$, a given value of $\phi$ (say, $\phi=\phi_-$) corresponds to several spacelike hypersurfaces
(due to the field oscillations).

The hypersurfaces $\Sigma_n$ are those of constant $\rho$ shown in the left panel of Fig.~\ref{fig:cono}.
The normal geodesics (dashed lines) only cover the region $r>t$, and the Gaussian normal coordinates can only be used there%
\footnote{This region is the so-called Rindler wedge, where one can introduce Rindler coordinates $\rho,\tau$, where the latter is given by $\sinh \tau = t/\rho = \xi^0/\rho_w $.
	These coordinates are adapted to accelerated observers sitting at fixed points in the wall profile.}.
This confirms that $r=t$ is a good boundary for the wall region.
This boundary is present in the definition of $\rho_w$ by 
Eq.~(\ref{rhow}).
With this definition, the value $\phi_w$ is halfway between $\phi_i$ and $\phi_+$
and the wall position $r_w$ is in the middle of wall region, as can be seen in Fig.~\ref{fig:potperf2}.
A 2-dimensional representation of the hypersurface, including the normal geodesics to $\Sigma$ and $S$ and the lightcone, is shown in the right panel of Fig.~\ref{fig:cono}.

\subsection{Energy-momentum tensor}

The energy-momentum tensor, Eq.~(\ref{eq:Tmunufield}), is given in this case by
\begin{equation}
	T_{\mu\nu}=\begin{cases}
		\left[\partial_{\mu}\rho\partial_{\nu}\rho+\frac{1}{2}g_{\mu\nu}\right]\bar{\phi}'(\rho)^{2}+g_{\mu\nu}V(\bar{\phi}) & \text{for }r>t,\\
		\left[\partial_{\mu}\tau\partial_{\nu}\tau-\frac{1}{2}g_{\mu\nu}\right]\tilde{\phi}'(\tau)^{2}+g_{\mu\nu}V(\tilde{\phi}) & \text{for }r\leq t.
	\end{cases}
	\label{eq:TmunuColeman}
\end{equation}
The wall contribution is defined in the domain of the Gaussian normal coordinates, $r>t$, and is given by Eq.~(\ref{eq:Tmunuw}), $T_{\mu\nu}^{w} = \bar{\phi}^{\prime2} P_{\mu\nu}$.
Furthermore, for this symmetric solution we have $\phi = \bar{\phi}(\rho_w+n)$, $\partial_a\phi=0$, and most of the terms in Eqs.~(\ref{TabNGC2})-(\ref{TnnTanNGC2}) vanish. 
Therefore, we have the simple decomposition 
\begin{equation}
	T_{\mu\nu}= P_{\mu\nu} \bar{\phi}'(\rho)^{2}
	+ g_{\mu\nu} V^b(\rho) ,
	\label{eq:TmunuColemansep}
\end{equation}
which looks like Eq.~(\ref{eq:Tmunuthin}), but in this case it is not an approximation.
The tensor $P_{\mu\nu}$ is defined in Eq.~(\ref{eq:proyectorn}).
In this case, the vector $n_\mu$ is given by $n_\mu= -\partial_\mu n= -\partial_{\mu}\rho$, and we have
$P_{\mu\nu}=g_{\mu\nu}+\partial_{\mu}\rho\partial_{\nu}\rho$.
In spherical coordinates, we have
\begin{equation}
	P_{\mu\nu} =
	\begin{pmatrix}r^{2}/\rho^{2} & -tr/\rho^{2} & 0 & 0\\
		-tr/\rho^{2} & t^{2}/\rho^{2} & 0 & 0\\
		0 & 0 & -r^{2} & 0\\
		0 & 0 & 0 & -r^{2}\sin^{2}\theta
	\end{pmatrix} .
	\label{PmunuColeman}
\end{equation}
In particular, we have $T^w_{00} = \bar{\phi}^{\prime}(\rho)^2 \, r^{2}/\rho^{2} = (\partial_r\phi)^2$.

The bulk potential energy density $V^b$ is given by Eq.~(\ref{defVbulk}), where the term with the derivatives $D_a$ vanishes in this case.
It is worth discussing this quantity in more detail for this simple case.
In the first place, the decomposition of the potential energy comes from the first integral of the field equation in Gaussian normal coordinates, Eq.~(\ref{ecfi}).
In this case we have%
\footnote{The extrinsic curvature tensor is given by $K_{\mu\nu}=-\nabla_{\mu}n_{\nu} = -\rho^{-1}P_{\mu\nu}$.}
$K=-3\rho^{-1}$ and Eq.~(\ref{ecfi}) is the same as Eq.~(\ref{eq:ecfiColeman}).
The first integral of Eq.~(\ref{eq:ecfiColeman}) gives two equivalent equations
using either the boundary condition at $\rho=\infty$ or at $\rho=0$,
\begin{equation}
	V\left(\bar{\phi}(\rho)\right) 
	=\frac{1}{2}\bar{\phi}'(\rho)^{2}+V_{+}-3\int_{\rho}^{\infty}\bar{\phi}'(\eta)^{2}\frac{d\eta}{\eta}
	=\frac{1}{2}\bar{\phi}'(\rho)^{2}+V_{i}+3\int_{0}^{\rho}\bar{\phi}'(\eta)^{2}\frac{d\eta}{\eta}.
\end{equation}
In both expressions, the term $\frac{1}{2}\bar{\phi}'(\rho)^{2}$ 
gives the peak at the wall
due to the potential barrier between the minima,
while the function
\begin{equation}
	V^b(\rho)=V_{+}-3\int_{\rho}^{\infty}\bar{\phi}'(\eta)^{2}\frac{d\eta}{\eta}
	=V_{i}+3\int_{0}^{\rho}\bar{\phi}'(\eta)^{2}\frac{d\eta}{\eta}
	\label{Vbulk}
\end{equation}
is monotonic and varies between the value $V_i$ at $\rho=0$ and the value $V_+$ at $\rho=\infty$.

\begin{figure}[tb]
	\centering
	\includegraphics[width=0.48\textwidth]{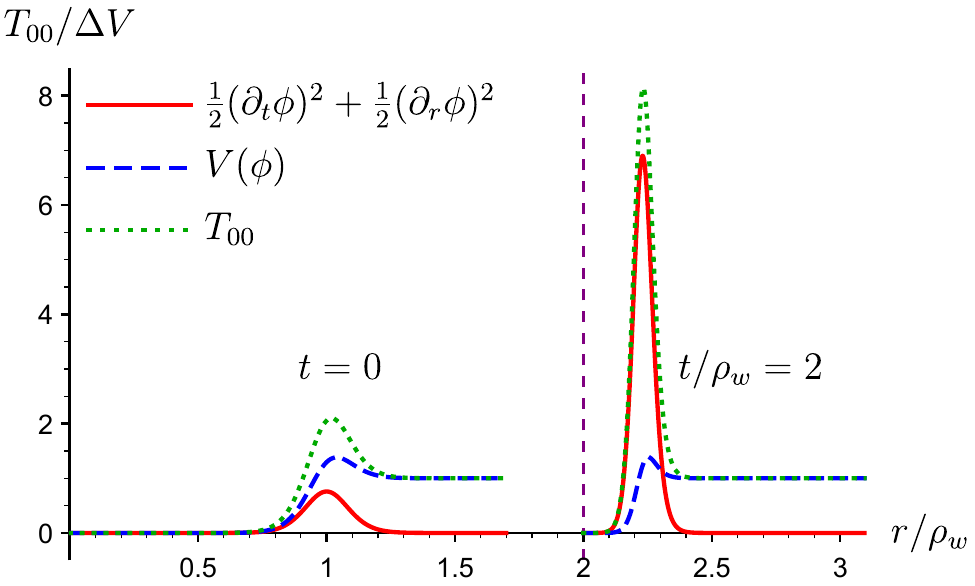}\hfill\includegraphics[width=0.48\textwidth]{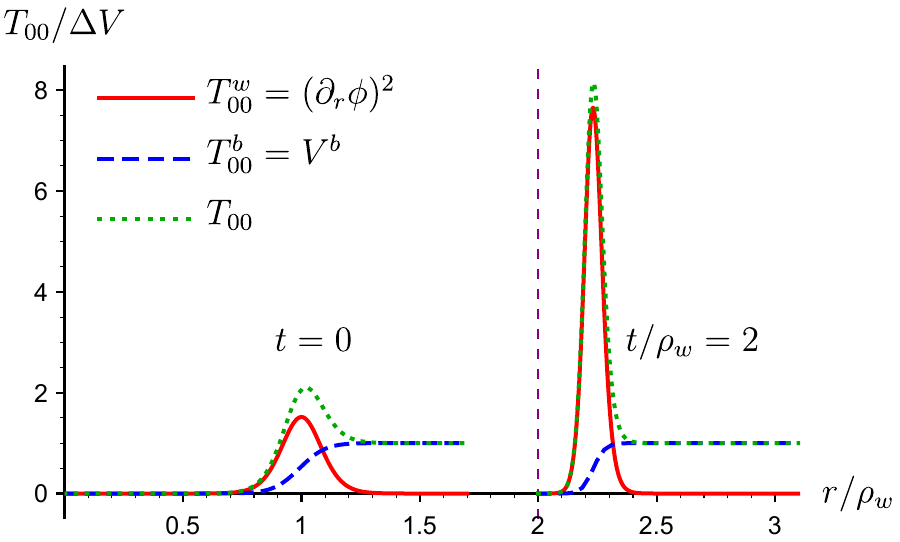}
	\caption{Alternative decompositions of the energy density for the case of Fig.~\ref{fig:potperf1} at two different times.
		The vertical line indicates the point $r=t$.\label{fig:T001}}
\end{figure}

As discussed in Sec.~\ref{subsec:wallregion}, an alternative decomposition often used in the literature consists in 
assigning the kinetic and gradient energy to the wall and all the potential energy to the domains.
In Fig.~\ref{fig:T001} we compare these decompositions for the case of Fig.~\ref{fig:potperf1}.
The potential energy density has a peak at the wall, which initially has the same amplitude as the kinetic and gradient energy densities.
In our decomposition, this peak is included in the wall energy density.
As the wall gets thinner due to the Lorentz contraction, this contribution becomes unimportant.

For the case of Fig.~\ref{fig:potperf1}, where we have $\phi\simeq \phi_-$ for $r\leq t$, we do not need to consider the solution $\tilde{\phi}$ at all.
In contrast, for the case of Fig.~\ref{fig:potperf2}, the function $\bar{\phi}(\rho)$ only describes the part of the profile to the right of the crosses.
This is not a problem if we consider that the wall region is limited to the domain of the Gaussian normal coordinates, $r>t$.
As a matter of fact, the quantities associated with the wall, such as $T^w_{\mu\nu}$, 
are defined in terms of the variable $n$ and depend on the solution $\bar{\phi}(\rho)$.
However, if we are interested in such a quantity as a function of the variable $m=r-r_w$, 
we may try to extend its definition to $r\leq t$, where $m$ is well defined (see the right panel of Fig.~\ref{fig:cono}).
For example, the wall energy density in this case is given by $T_{00}^w=(\partial_r\phi)^2$, and this expression is well defined inside the lightcone.
To verify that this extension makes sense, let us consider the first integral of Eq.~(\ref{eq:ecfiColemanin}),
\begin{equation}
	V(\tilde{\phi})=
	-\frac{1}{2}\tilde{\phi}'(\tau)^{2}+V_{i}-3\int_{0}^{\tau}\tilde{\phi}'(\eta)^{2}d\eta/\eta .
	\label{eq:sepVtil}
\end{equation}
The quantity $V^b(\tau)=V_{i}-3\int_{0}^{\tau}\tilde{\phi}'(\eta)^{2}d\eta/\eta$ is a natural continuation of $V^b(\rho)$.
Indeed, it is a monotonic function that takes the value $V_i$ at $\tau=\rho=0$ and approaches 
the value $V_-$ as we penetrate deeper into the bubble%
\footnote{To see this, notice that $\tilde{\phi}(\tau)$ does damped oscillations around $\phi_-$, initially with amplitude $\phi_{-}-\phi_{i}$ but decaying with $\tau$ \cite{mm23}, so we have $\tilde{\phi}\to\phi_{-}$ for $\tau\to\infty$.
The value $V_-$ is only reached asymptotically, so we will not have exactly $V=V_-$ at the bubble center.}.
Inserting Eq.~(\ref{eq:sepVtil}) into Eq.~(\ref{eq:TmunuColeman}), we obtain a decomposition similar to Eq.~(\ref{eq:TmunuColemansep}),
$T_{\mu\nu}=\tilde{P}_{\mu\nu}\tilde{\phi}^{\prime2} + g_{\mu\nu}V^b(\tau)$,
where $\tilde{P}_{\mu\nu}=\partial_{\mu}\tau\partial_{\nu}\tau - g_{\mu\nu}$.
Since $V^b(\tau)$ is the natural continuation of $V^b(\rho)$, the first term in this decomposition gives a natural continuation for the definition of $T_{\mu\nu}^{w}$.
In spherical coordinates, we have $\partial_{\mu}\tau=\tau^{-1}(t,-r,0,0)$, and we obtain,
in particular, $T^{w}_{00}=\tilde{\phi}^{\prime2}r^{2}/\tau^{2}= (\partial_{r}\phi)^{2}$.

\begin{figure}[tb]
	\centering
	\includegraphics[width=0.49\textwidth]{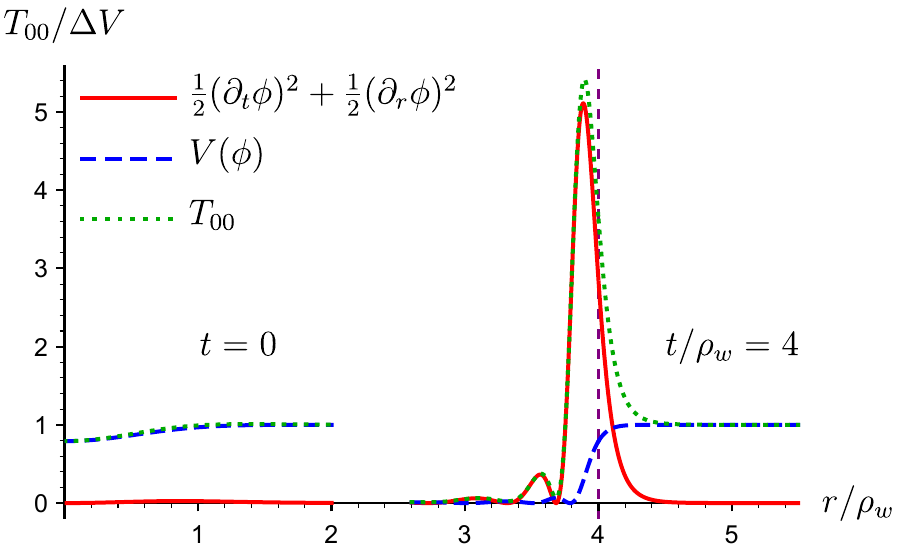}\hfill\includegraphics[width=0.49\textwidth]{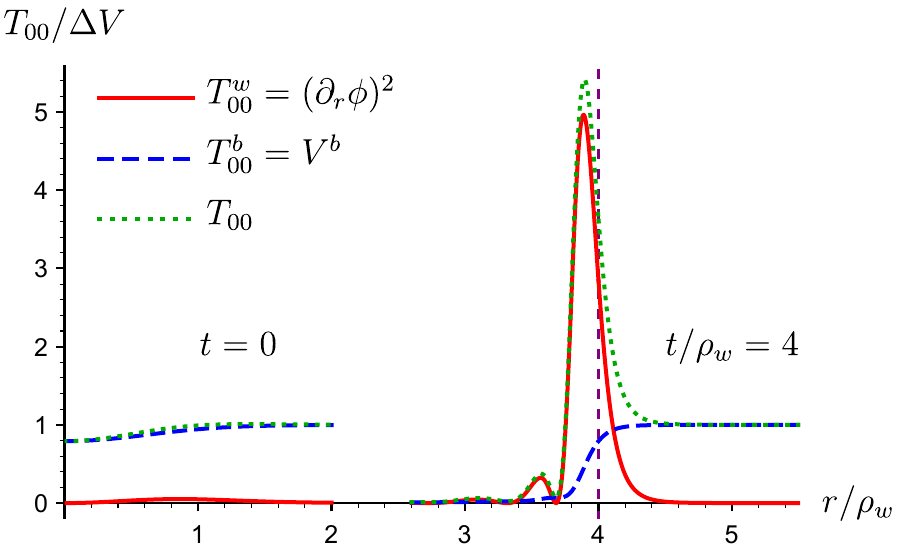}
	\caption{Alternative decompositions of the energy density for the case of Fig.~\ref{fig:potperf2} at two different times. The vertical line indicates the point $r=t$.\label{fig:T002}}
\end{figure}

\begin{figure}[tb]
	\centering
	\includegraphics[height=5cm]{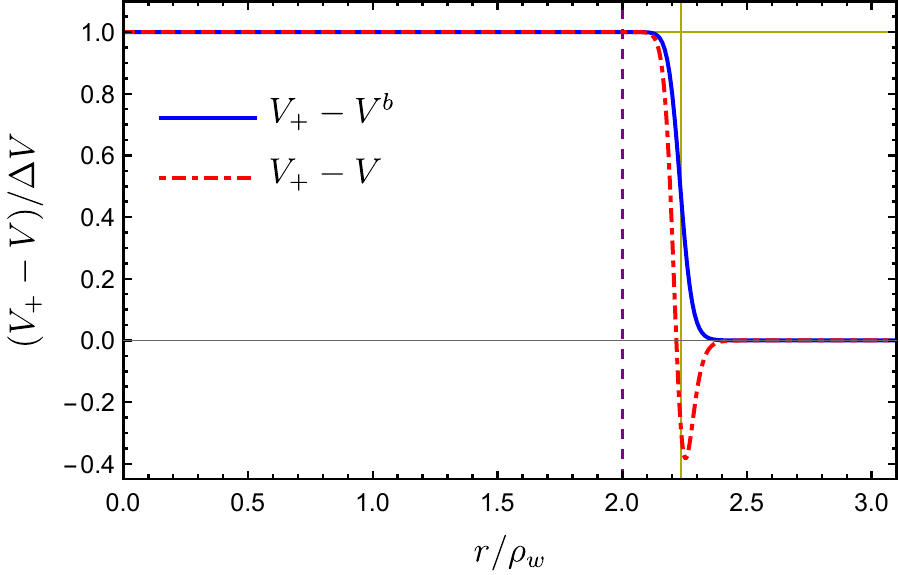}\hfill\includegraphics[height=5cm]{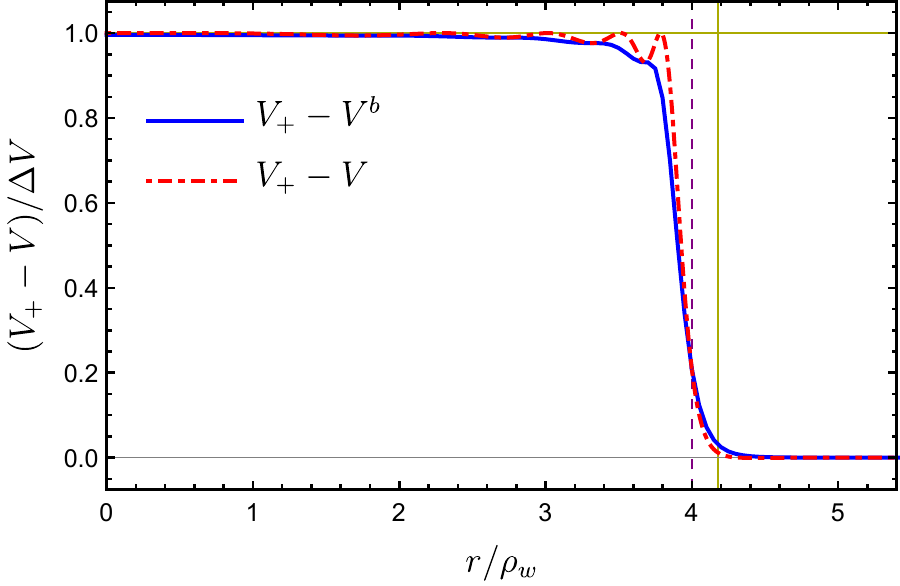}
	\caption{The differences $V_+-V$ and $V_+-V^b$. Left: the case of Fig.~\ref{fig:T001}.
		Right: the case of Fig.~\ref{fig:T002}.
		The vertical lines indicate the points $r=t$ and $r=r_w$.\label{fig:deltav}}
\end{figure}

In Fig.~\ref{fig:T002} we consider our decomposition and the alternative one for the case of Fig.~\ref{fig:potperf2}.
Quantitatively, there is no significant difference.
For a better comparison, we show in Fig~\ref{fig:deltav} the potential  differences $V_+-V$ and $V_+-V^b$ for the cases of Figs.~\ref{fig:potperf1} and \ref{fig:potperf2}.
Notice that $V^b$ is a monotonous function which varies between $V_+$ and $V_-$, while $V$ has oscillations.
In our decomposition, these oscillations are assigned to the wall energy.
The difference is most important for thinner walls, where the potential barrier is higher and its energy is concentrated at the bubble wall.

\subsection{Wall energy}

The metric components in the Gaussian normal coordinates are $\bar{g}_{nn}=-1$, $\bar{g}_{na}=0$, and
$\bar{g}_{ab} =(\rho/\rho_{w})^{2}\gamma_{ab}$, with 
$\gamma_{ab} =\mathrm{diag}\left(\rho_{w}^{2}r_{w}^{-2}, -r_{w}^{2}, -r_{w}^{2}\sin^{2}\theta \right)$
and $\rho=\rho_w+n$, $r_w^2=\rho_w^2+(\xi^{0})^2$, $\theta = \xi^1$.
Thus, Eq.~(\ref{Sabdef}) gives
$\bar{S}_{ab}=\gamma_{ab}\int_{0}^{\infty}d\rho(\rho/\rho_{w})^{7/2}(\partial_{\rho}\phi)^{2}$,
and the replacement $\gamma_{ab}\to h_{\mu\nu}$ gives the complete surface stress-energy tensor $S_{\mu\nu}$.
The tensor $h_{\mu\nu}$ is given by Eq.~(\ref{PmunuColeman}) at $r=r_w,\rho=\rho_w$.
For a very thin wall, we have $\rho\simeq \rho_w$ and $S_{\mu\nu}\simeq \sigma h_{\mu\nu}$.

While $S_{\mu\nu}$ is given by an integral over $n$ with $\xi^a$ constant,
the surface energy density $\varepsilon_w$ is given by an integral over $m$ with $t$ and $\zeta^A$ constant
(hence, $S_{00}$ and $\varepsilon_w$ are different quantities).
As shown in the right panel of Fig.~\ref{fig:cono}, the former is a line integral along a path that remains outside the lightcone,
while the latter is a line integral along a path which enters the lightcone.
In the present case, the wall energy density $T^{w}_{00} = (\partial_{r}\phi)^{2}$ is well defined everywhere, and we can use 
Eq.~(\ref{Ewesf}) for $\varepsilon_w$.
Since we have two different numerical solutions $\phi=\bar{\phi}$, $\phi=\tilde{\phi}$ in different regions, 
we need to separate the integral.
We have $dr=\rho d\rho / r$ for $r>t$ and $dr=-\tau d\tau/r$ for $r<t$,
and we obtain
$E_w=E_{w}^{\mathrm{in}}+E_{w}^{\mathrm{out}}$, where
\begin{equation}
	E_{w}^{\mathrm{in}} = 
	4\pi \int_{0}^{t}d\tau\frac{\left(t^{2}-\tau^{2}\right)^{3/2}}{\tau}\tilde{\phi}'(\tau)^{2},
	\quad
	E_{w}^{\mathrm{out}} =  
	4\pi \int_{0}^{\infty}d\rho\frac{(\rho^{2}+t^{2})^{3/2}}{\rho}\bar{\phi}'(\rho)^{2}.
	\label{Ew}
\end{equation}
If we use the convention that the wall region only extends as far as the Gaussian normal coordinates for $\Sigma$ are well defined, we must consider only $E_{w}^{\mathrm{out}}$.
For an energy density such as that in Fig.~\ref{fig:T001}, $E_w^{\mathrm{in}}$ will be negligible,
whereas it will make a significant contribution to $E_w$ for a case such as that in Fig.~\ref{fig:T002}.

The alternative definition of the wall energy density as $\frac{1}{2}(\partial_t\phi)^2+\frac{1}{2}(\partial_r\phi)^2$
was considered, for instance, in Ref.~\cite{kt93},
where  ${\phi}$ is assumed to depend on $\rho$, with $d\phi/d\rho$ vanishing for $r<t$.
This gives an expression for $E_w$ similar to $E_w^{\mathrm{out}}$ (cf.\ Eq.(9) of Ref.~\cite{kt93}).
Due to energy conservation, 
the total kinetic and gradient energy of the bubble can be computed through the total potential energy.
This fact was exploited, for example, in Ref.~\cite{elnv19}.

\begin{figure}[tb]
	\centering
	\includegraphics[width=0.49\textwidth]{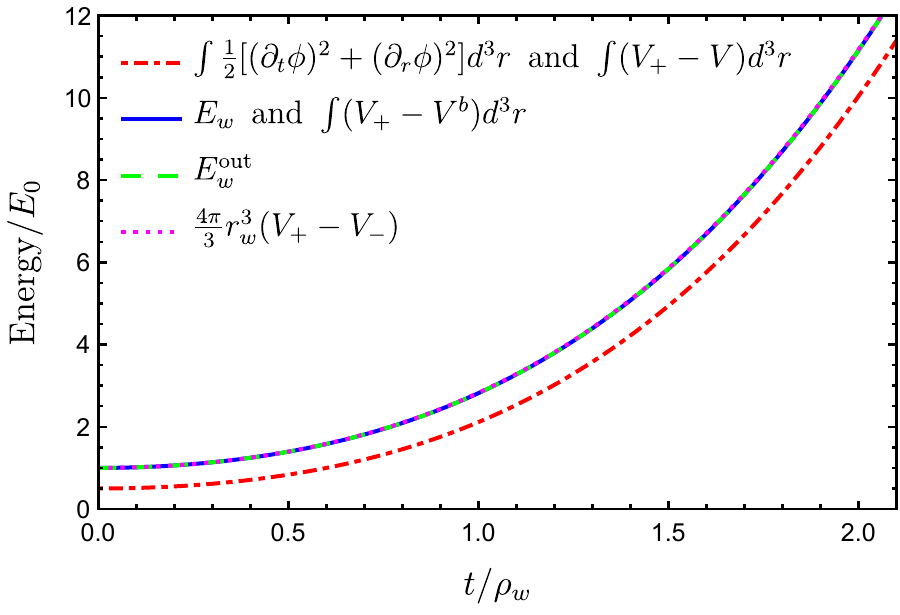}\hfill\includegraphics[width=0.49\textwidth]{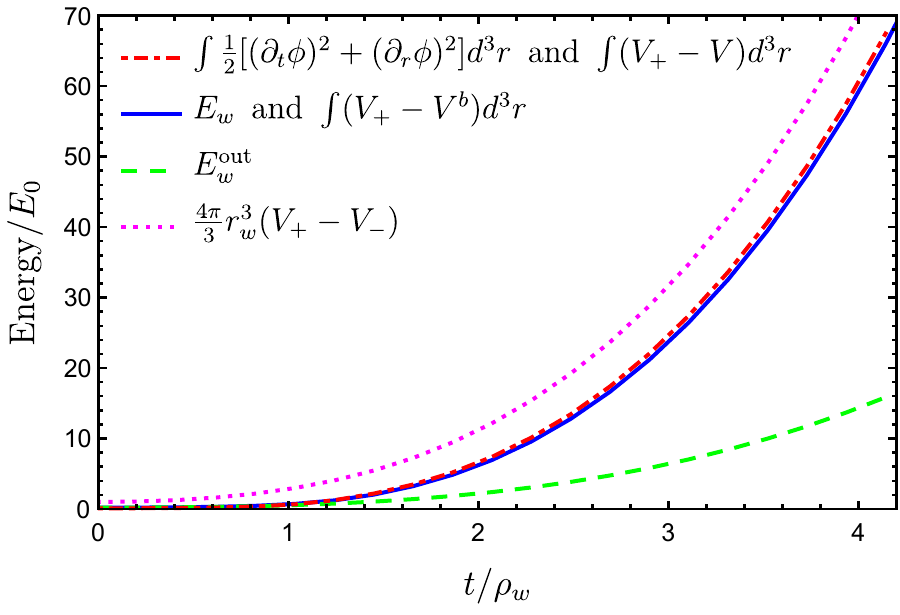}
	\caption{The various energy components for the case of Figs.~\ref{fig:potperf1} and \ref{fig:T001} (left)
		and for the case of Figs.~\ref{fig:potperf2} and \ref{fig:T002} (right).
		Quantities which are equal due to energy conservation are represented by a single curve.
		The energy is scaled with the thin-wall initial value $E_0=\frac{4\pi}{3}\rho_w^3\Delta V$.}\label{fig:energy}
\end{figure}

We compare the various approaches in Fig.~\ref{fig:energy}.
The O(3,1)-invariant solution describes a bubble nucleated at $t=0$ in a vacuum phase transition with vanishing total energy
(see Sec.~\ref{sec:O3}).
Therefore, the integral of $V_+-V$ (dashed-dotted lines in Fig.~\ref{fig:energy})
coincides with the integral of $\frac{1}{2}(\partial_t\phi)^2+\frac{1}{2}(\partial_r\phi)^2$.
Similarly, the integral of $V_+-V^b$  (solid blue lines) coincides with $E_w$.
This equality can be interpreted as the energy released going into the wall.
The case shown in the left panel of Fig.~\ref{fig:energy} is that of Figs.~\ref{fig:potperf1} and \ref{fig:T001},
where we have a significant potential barrier.
Since our definition of the wall energy takes into account this contribution,
it gives a higher value than considering only the gradient and kinetic energy.
As anticipated, in this case we have $E_w\simeq E_w^\mathrm{out}$.
We have included in the comparison the rough thin-wall approximation $\Delta V\times 4\pi r_w^3/3$.
We see that it essentially coincides with the integral of $V_+-V^b$.
This coincidence is also apparent in the left panel of Fig.~\ref{fig:deltav}.

The plot on the right of Fig.~\ref{fig:energy} corresponds to the case of Figs.~\ref{fig:potperf2} and \ref{fig:T002},
where the definition of the bubble wall region is less evident.
The difference between the two alternative definitions of the wall energy is small in this case since the potential barrier is very small.
The integral of $V_+-V$ is a little higher than that of $V_+-V^b$, 
which can also be appreciated in the right panel of Fig.~\ref{fig:deltav}.
In this case, the approximation $\Delta V\times 4\pi r_w^3/3$ overestimates both integrals
because the wall position $r_w$ (indicated by a vertical line in Fig.~\ref{fig:deltav}) is inside the outer profile $\bar{\phi}$.
On the other hand, if we consider the wall to be the part of the profile at $r> t$, then its energy is given by $E_w^\mathrm{out}$, which is much smaller than the energy released. 
The rest of the energy goes into oscillations of the field inside the bubble.

It is worth emphasizing that, although a decomposition of $T_{\mu\nu}$ 
may be irrelevant for a fully numerical computation such as a lattice simulation,
it is relevant for analytic approximations.
Moreover, even in the case of a numerical calculation, it is often of interest to have a proper identification of the wall in order to analyze the results in terms of wall dynamics.
For instance, in Refs.~\cite{jkt19} and \cite{cghw20}, the $\gamma$ factor of the wall is defined through the Lorentz contraction between two points 
corresponding to fixed values of $\phi$ in the bubble profile.
In Ref.~\cite{cghw20}, both reference points are taken outside the lightcone (in the wall region).
However, in Ref.~\cite{jkt19} one of the points lies inside the lightcone (it corresponds to the value $\phi=\phi_-$).
We remark that the ``trajectory'' of such a point is superluminal  (lowest spacelike hyperbola in the left panel of Fig.~\ref{fig:cono}).
The central point of the energy distribution $\frac{1}{2}(\partial_t\phi)^2+\frac{1}{2}(\partial_r\phi)^2$ also has a non-physical trajectory that eventually becomes superluminal.

\section{The thin-wall approximation}
\label{sec:pert_method}

The O(3,1)-symmetric example considered in the previous section is extremely simple because the field depends on a single variable.
The thin-wall approximation also achieves such a simplification by assuming that $\phi$ depends only on the distance $n$ orthogonal to the wall \cite{vs00}. 
The standard procedure uses a few other approximations
to solve for the field profile $\phi(n)$ and then obtain an equation for the wall as a surface.
This method can be generalized to higher orders in the wall width \cite{mm24}.

\subsection{Leading-order approximation}
\label{sec:apbasica}

Assuming that $\phi$ only depends on $n$, Eq.~(\ref{ecfi}) becomes
\begin{equation}
	\partial_{n}^{2}\phi-K\partial_{n}\phi=V'(\phi).
	\label{eq:primerasup}
\end{equation}
It is also a common assumption that $\phi$ reaches asymptotically the values $\phi_\pm$. 
This assumption avoids the problem of considering field dynamics away from the wall,
such as field oscillations at the center of the bubble.
Note that the mean curvature of the wall hypersurface, $K$, depends on the solution $\phi$,
so Eq.~(\ref{eq:primerasup}) is still not trivial.
The usual approach is to assume that the term containing this quantity is negligible,
which is a valid approximation as long as the wall width $l$ is much smaller than the curvature radius $K^{-1}$.
The latter assumption requires $V_{+}=V_{-}$ for consistency \cite{c77}
(see \cite{mm23} for a discussion of these approximations).

To approximate $V$ by a degenerate potential, we write
\begin{equation}
V(\phi)=V_{0}(\phi)+c\phi,
\label{eq:potpert}
\end{equation}
and demand that $V_{0}$ is degenerate.
Therefore, we have $V_{0}=V-c\phi$, whose minima $a_\pm$ are given by the condition $V_0'(a_\pm)=0$,
while the condition $V_0(a_+)=V_0(a_-)$ determines the parameter $c$ (see \cite{mm23} for a generalization of this method).
Thus, we have the equations
\begin{equation}
V'(a_{\pm}) = c, \quad
V(a_{+})-V(a_{-}) = c(a_{+}-a_{-}).
\label{eq:condV0deg}
\end{equation}
For the specific potential (\ref{eq:pot}), we obtain
\begin{equation}
	a_{\pm}=\frac{e}{3\lambda}\left(1\mp\sqrt{3-{9\lambda m^{2}}/{e^{2}}}\right), 
	\quad
	c=\frac{e^{3}}{27\lambda^{2}}\left({9\lambda m^{2}}/{e^{2}}-2\right) ,
\end{equation}
and
\begin{equation}
	V_{0}(\phi)=\frac{\lambda}{4}\left(\phi-a_{+}\right)^{2}\left(\phi-a_{-}\right)^{2} + V_{0+},
	\label{eq:V0}
\end{equation}
where the constant $V_{0+}= V_+ -9\lambda c^{2}/4e^{2}$ is irrelevant for the calculation of the field profile.
For $9\lambda m^{2}/e^{2}= 2$, we have $c= 0$, $a_+=\phi_+= 0$, and $a_-=\phi_-=2e/3\lambda$.

We obtain the basic thin-wall approximation for the field profile, $\phi_0(n)$, by neglecting the second term in Eq.~(\ref{eq:primerasup}) and replacing $V$ by $V_0$,
\begin{equation}
	\phi_{0}''(n)=V_{0}'(\phi_{0}(n)) .
	\label{eq:ecfi0}
\end{equation}
The appropriate boundary conditions are $\phi_{0}(\pm\infty)=a_{\pm}$.
The first integral is 
\begin{equation}
	\phi_{0}^{\prime}(n) = - \sqrt{2[V_{0}(\phi)-V_{0+}]} \equiv -\phi_h(\phi) 
	\label{fih}
\end{equation}
(the sign corresponds to a profile which decreases in $n$ from $a_{-}$ to $a_{+}$),
and we have the implicit solution
\begin{equation}
	n=-\int_{\phi_{*}}^{\phi_{0}}\frac{d\phi}{\phi_h(\phi)}+n_{*} .
	\label{eq:fi0}
\end{equation}
The relation between the integration constants $\phi_*$ and $n_{*}$ depends on the definition of the wall position within the field profile.
We will use the condition (\ref{eq:cero_n}).
For the potential (\ref{eq:V0}), we have $\phi_{h}=\sqrt{\lambda/2}(\phi-a_{+})(a_{-}-\phi)$ and we obtain
\begin{equation}
	\phi_{0}(n)=\phi_{*}-a\tanh\left(\sqrt{\lambda/2}\,a\,n\right),
	\label{eq:phi0}
\end{equation}
with $\phi_{*}=(a_{+}+a_{-})/2$ and $a=(a_{-}-a_{+})/2$.

To obtain an equation of motion for the wall, 
we need to go beyond the approximation $\Delta V=0$.
Therefore, we go back to Eq.~(\ref{eq:primerasup}), this time keeping the second term
but assuming that the quantity $K$ remains approximately constant inside the wall, i.e.,  $K\simeq K|_{n=0}$.
Thus, multiplying Eq.~(\ref{eq:primerasup}) by $\partial_{n}\phi$ and integrating across the wall, the first term vanishes and we obtain $K\simeq -\Delta V/\sigma$.
To calculate the surface tension, we can use the approximation $\phi_0(n)$ for the profile,
\begin{equation}
	\sigma_0=\int_{-\infty}^{+\infty}\phi_0^{\prime}(n)^2dn = \int_{a_+}^{a_-}\phi_h(\phi)d\phi.
\end{equation}
For the potential (\ref{eq:V0}), we have $\sigma_{0}=\frac{4}{3}\sqrt{\lambda/2}\,a^{3}$.
Thus, we have $K = -\Delta V/\sigma_0 $, and using the first relation in Eq.~(\ref{eq:Kcovar})
we obtain an equation for the normal vector $N^\mu$,
which we call $N^\mu_0$ in this approximation,
\begin{equation}
	N^{\mu}_{0;\mu}=\Delta V/\sigma_0 .
	\label{EOM0}
\end{equation}
Using Eq.~(\ref{eq:Nmu}) in Eq.~(\ref{EOM0}) gives an equation for the implicit surface representation function $F$.
A gauge fixing is necessary to solve this equation,
and we will use the explicit form (\ref{eq:Monge}),
$F=x^3-x_w^3(x^a)$ with $a=0,1,2$, which is sometimes called the Monge gauge.
Thus, we have $N_{a}=\partial_{a}x^3_w/s$,
$N_{3}=-s^{-1}$, and
$s^{2}=-g^{33}+2g^{3a}\partial_{a}x_{w}^{3}-g^{ab}\partial_{a}x_{w}^{3}\partial_{b}x_{w}^{3}$.
Inserting in Eq.~(\ref{EOM0}), we obtain an equation for the ``wall position'' $x_w^3$. 
The general expression for this equation of motion can be found in \cite{mm23}.
Below, we consider a specific example.

The field profile $\phi_0(n)$ obtained with the thin-wall approximation is the same for any wall shape.
On the other hand, the function $n(x^\mu)$ depends on the hypersurface $\Sigma$.
For a very thin wall, Eq.~(\ref{eq:n_orden2}) gives
\begin{equation}
	n(x^\mu)=F(x^\mu)/s(x^\mu) = (x^3-x^3_{w0})/s_0,
	\label{n0}
\end{equation} 
where 
the index 0 in $x^3_{w}$ and $s$ indicates that these quantities are calculated using the above 
thin-wall approximations. 
In particular, $x^3_{w0}(x^a)$
is a solution of Eq.~(\ref{EOM0}) and
the quantity $s_0$ is a function of its derivatives $\partial_a x^3_{w0}$.

Since $\partial_a\phi_0=0$,
Eqs.~(\ref{TabNGC2})-(\ref{TnnTanNGC2}) give Eq.~(\ref{eq:Tmunuthin}) in the thin-wall approximation, as already discussed.
The energy-momentum tensor of the wall is given by Eq.~(\ref{eq:Tmunuw}).
For a very thin wall, we have $P_{\mu\nu}\simeq h_{\mu\nu}$, and we obtain
$T_{\mu\nu}^{w}=h_{\mu\nu}{\phi}_{0}^{\prime2}$.
The tensor $h_{\mu\nu}$ is given by Eq.~(\ref{eq:proyector}) and depends on the hypersurface $\Sigma$.
For the surface stress-energy tensor,
both Eqs.~(\ref{Smunudef}) and Eq.~(\ref{Smunuthin}) give the result $S_{\mu\nu}=\sigma_{0}h_{\mu\nu}$ to this order.

For a metric with $g_{00}=1$ and $g_{0i}=0$, the wall energy density is given by Eq.~(\ref{Tw00}),
$T_{00}^{w} =  q^{\prime2} \phi_0^{\prime 2}$, with $q^{\prime 2}=(\nabla n)^2|_{m=0}$.
The surface energy density is given by Eq.~(\ref{eq:Ewdef}),
$\varepsilon_{w} = q^{\prime 2}\int dm {\phi}_{0}^{\prime2}(n)$.
It is convenient to change variables from $m$ to $n$.
Since the integral is along the constant-time geodesic,
we must use Eq.~(\ref{eq:nprin}).
For small $n$, we have $dm=dn/q'$.
Using $q^{\prime 2}|_{m=0} = q^{2}/s^2$, we obtain
\begin{equation}
	\varepsilon_{w} = \sigma_0 q' = \sigma_{0}q/s,
	\label{EwLO}
\end{equation}
with $s^2=-F_{,\mu}F^{,\mu}$ and $q^2=-F_{,i}F^{,i}$.

The energy density in the bulk is given by Eq.~(\ref{defVbulk}). 
In this case, we have $V^{b}(n)=V_{+}+K_{0}I^{(0)}_0(n)$, where $I^{(0)}_0=\int_{n}^{\infty}\phi_{0}^{\prime}(n')^{2}dn'$.
The total energy in the bulk, $E_b=\int d^{3}x(V^{b}-V_{+})$, is given by
$E_b=K_{0}\int d^{2}\zeta\int_{\text{int}}^{\infty}dm\sqrt{\tilde{g}}I_{0}^{(0)}(n)$, where int denotes some point in the bubble interior.
Integrating by parts, we obtain
\begin{equation}
	E_b=K_{0}\int d^{2}\zeta \int_{\text{int}}^{\infty} dm\left(\int_{\text{int}}^{m}dm'\sqrt{\tilde{g}}\right)\phi_{0}'(n)^{2}\partial_{m}n
	=K_0 \mathcal{V} \sigma_0 .
	\label{Ebthin}
\end{equation}
In the last step we have evaluated the function in parenthesis at $m=0$, which is a valid approximation for a very thin wall.
The quantity  $\mathcal{V}=\int d^{2}\zeta \int_{\text{int}}^{0} dm\sqrt{\tilde{g}}$ is the volume of the bubble.
Thus, we obtain the thin-wall result $E_b =\Delta V \mathcal{V}$.

\subsection{Higher orders in the wall width}

We have neglected the terms involving the quantities $\partial_{a}\phi$ and $K$
in Eqs.~(\ref{ecfi})-(\ref{TnnTanNGC2}),
under the assumption that these terms are much smaller than those containing $\partial_n \phi$.
All these terms are of different order in the wall width.
Let $L$ be the length scale associated with the curvature of the hypersurface,
so that we have $K\sim L^{-1}$ and the operator $\partial_a$ is also of order $L^{-1}$,
while $\partial_n$ is of order $l^{-1}$.
On the other hand, $\phi$ is quantitatively of the order of the energy scale of the theory, while the wall width is of the order of the inverse of this scale, so we have $\phi\sim l^{-1}$. 
Besides,
the term $c\phi$ which breaks the degeneracy of the potential in Eq.~(\ref{eq:potpert}) can be assumed to be of order $l/L$ with respect to $V_0$.
To obtain a solution of the field equation at each order in the wall width,
we can consider a formal expansion
$\phi=\phi_{0}+\phi_{1}+\phi_{2}+\cdots$, where the leading-order (LO) solution $\phi_0$ is given by 
Eqs.~(\ref{eq:ecfi0})-(\ref{eq:fi0})
and each term is of order $l/L$ higher than the previous one \cite{mm24}.
Below we briefly review the method.

We need to consider also an expansion $K=K_0+K_1+K_2+\cdots$.
This quantity can also be expanded in powers of $n$, so we have
\begin{equation}
	K= \left(K_0 +  K_1 + \cdots\right)_{n=0} +
	\left(\partial_{n}K_0 +\partial_{n}K_1 + \cdots \right)_{n=0} n +
	\frac{1}{2}\left(\partial_{n}^{2}K_0 + \cdots \right)_{n=0} n^2 +\cdots .
	\label{Kin0}
\end{equation}
Inserting the expansions for $\phi$ and $K$ in Eq.~(\ref{ecfi}) or its first integral (\ref{eq:primeraint}), 
we obtain, order by order in $l/L$, equations for the field corrections $\phi_i$,
which are all of the form
\begin{equation}
	\partial_{n}\phi_{0}\partial_{n}\phi_{i}-\partial_{n}^{2}\phi_{0}\phi_{i}=f_{i},
	\label{eq:primeraintpert}
\end{equation}
where the LO field profile $\phi_0$ is given by Eq.~(\ref{eq:phi0}) and
the source term $f_i$ depends on the previous solutions $\phi_0,\ldots,\phi_{i-1}$. 
We have, e.g.,
\begin{align}
	&f_{1} =  c(\phi_{0}-a_+)  -K_{0}|_{n=0}I_{0}^{(0)}, \label{f1}
	\\
	&f_{2} = \begin{multlined}[t]
	c(\phi_{1}-\phi_{1+}) + 
	{\textstyle\frac{1}{2}}\left[{V_{0}''(\phi_{0})}\phi_{1}^{2} - {V_{0}''(a_{+})} \phi_{1+}^{2} \right]
	\\
	-{\textstyle\frac{1}{2}}{(\partial_n\phi_{1})^{2}} 
	-K_{0}|_{n=0}I_{1}^{(0)}-K_{1}|_{n=0}I_{0}^{(0)}-\partial_{n}K_{0}|_{n=0}I_{0}^{(1)},
	\end{multlined}
	\label{f2}
	\\
	&
		f_{3} =
	c(\phi_{2}-\phi_{2+}) 
	+V_{0}''(\phi_{0})\phi_{1}\phi_{2}-V_{0}''(a_{+})\phi_{1+}\phi_{2+}
	+ {\textstyle\frac{1}{6}}[V_{0}'''(\phi_{0})\phi_{1}^{3}-V_{0}'''(a_+)\phi_{1+}^{3}]
	\label{f3}
	 \\
	 &-\partial_{n}\phi_{1}\partial_{n}\phi_{2} 
	 -K_{0}|_{0}I_{2}^{(0)}-K_{1}|_{0}I_{1}^{(0)}-K_{2}|_{0}I_{0}^{(0)}
	 -\partial_{n}K_{0}|_{0}I_{1}^{(1)}-\partial_{n}K_{1}|_{0}I_{0}^{(1)}
	 -{\textstyle\frac{1}{2}}\partial_{n}^{2}K_{0}|_{0}I_{0}^{(2)}.
	 \nonumber
\end{align}
The terms $K_i|_{n=0}I^{(k)}_j$ come from the expansion of the integrand in Eq.~(\ref{eq:primeraint}),
where
$I^{(k)}_j$ are integrals coming from the expansion of $(\partial_n\phi)^2$,
\begin{equation}
	I_{0}^{(k)}=\int_{n}^{\infty} (\partial_n\phi_{0})^{2}\, n^{\prime k} d n', \quad
	I_{1}^{(k)}=\int_{n}^{\infty} (2 \partial_{n}\phi_{0}\partial_{n}\phi_{1})\, n^{\prime k} dn',
	\, \ldots
	\label{eq:Gi}
\end{equation}
It is convenient to define the wall surface by the condition (\ref{eq:cero_n}), so that the integrals $I_i^{(1)}$ vanish for $n\to -\infty$.
At the lowest orders, 
only the term $K(\partial_{n}\phi)^{2}$ in the integrand in (\ref{eq:primeraint}) contributes to the expansion, since the term $\partial_{n}\phi D_{a}D^{a}\phi$ has an extra factor of $l/L$ and, furthermore, we have $\partial_a\phi_0=0$ and, 
as we shall see, also $\partial_a\phi_1=0$.

The solution of Eq.~(\ref{eq:primeraintpert}) is
\begin{equation}
	\phi_{i}(\xi^{a},n)=\phi_{h}(n)\left[C_{i}(\xi^{a},n)+c_{i}(\xi^{a})\right],
	\label{eq:solfii}
\end{equation}
where
\begin{equation}
	C_{i}(\xi^{a},n)=-\int_{n_{*}}^{n}\frac{f_{i}(n',\xi^{a})}{\phi_h(n')^{2}}dn'
	\label{eq:Ci}
\end{equation}
and $c_i(\xi)$ is determined by the condition (\ref{eq:cero_n}).
The function $f_i$ contains the quantities from the perturbative expansion of $K$ up to the term $i-1$.
These quantities can be obtained from the condition
$f_{i}|_{n=-\infty}=0$, which follows from Eq.~(\ref{eq:primeraintpert}).
Therefore, once $\phi_{i-1}$ is solved, we can solve for $K_{i-1}$ and then for $\phi_i$.
Also, the quantity $K=K_0+\cdots+K_{i}$ gives, through Eq.~(\ref{eq:Kcovar}), the wall EOM to order $i$.

For $i=1$, the condition $f_{1}|_{n=-\infty}=0$ gives $\sigma_0 K_0|_{n=0}=c(a_--a_+)$.
According to Eq.~(\ref{eq:condV0deg}), the right-hand side of this equation is the potential difference $-\Delta V$ to lowest order in $l/L$.
Therefore, we have just re-obtained the leading-order approximation that gives the wall EOM (\ref{EOM0}).
Having determined $K_0|_{n=0}$, we use the function $f_1$ in Eq.~(\ref{eq:Ci}) to obtain $\phi_1$.
Since $f_{1}$ only depends on $n$, $\phi_{1}$ also depends on $n$ alone%
\footnote{As a consequence, the term  $D_{a}D^{a}\phi$ in Eq.~(\ref{ecfi}) is of order $(l/L)^4$ with respect to the term $\partial_n^2\phi$, and does not appear in the field equation at lower orders.}.
For our polynomial potential, we obtain a constant,
\begin{equation}
	\phi_{1}=-c/(2a^{2}\lambda) .
	\label{eq:phi1}
\end{equation}

For $i=2$, the condition $f_2|_{n=-\infty}=0$ gives $K_1|_{n=0}$ 
in terms of the quantities already obtained, namely, $\phi_0$, $\phi_1$, and $K_0|_{n=0}$.
Thus, we can calculate the next-to-leading-order (NLO) value
$K=K_0+K_1$ at $n=0$.
One obtains \cite{mm24} $K|_{n=0} = -\Delta V/\sigma $, where $\sigma=\sigma_{0}+\sigma_{1}$ and
$\sigma_{1} = \int_{-\infty}^{+\infty} 2 \phi_{0}'\phi_{1}'d{n}$.
Since $\phi_0$ and $\phi_1$ only depend on $n$, the correction to the surface tension is a constant. 
Therefore, the NLO EOM has the same form as the LO EOM, namely,
$N^{\mu}_{\ ;\mu}=\Delta V/\sigma$.
For our quartic potential, $\phi_1$ is a constant and
we have $\sigma_1=0$, so this EOM
is exactly the same as the leading-order one.

The function $f_2$ depends on the already obtained quantities $K_0|_{n=0}$ and $K_1|_{n=0}$, but also on the derivative $\partial_{n}K_{0}|_{n=0}$.
According to Eq.~(\ref{eq:Kcovar}), we have
$\partial_{n}K_0|_{n=0}=N_{0;\nu}^{\mu}N_{0;\mu}^{\nu}$, which is readily obtained once the LO EOM, Eq.~(\ref{EOM0}), is solved.
This quantity will generally depend on the variable $\xi^a$, so the source term in the equation for $\phi_2$ is of the form 
$f_{2}={f}_{2a}+f_{2b}\partial_{n}K_{0}|_{n=0}$,
where ${f}_{2a}$ and $f_{2b}$ only depend on $n$.
As a consequence, Eq.~(\ref{eq:solfii}) takes the form 
\begin{equation}
	\phi_2=\phi_{2a}(n)+\phi_{2b}(n)\partial_{n}K_{0}(\xi^a,0).
	\label{phi2}
\end{equation}
For the potential (\ref{eq:V0}), 
we have \cite{mm24} $\phi_{2a}(n)=\phi_h(n)C_{2a}(n)$, $\phi_{2b}(n)=\phi_h(n)C_{2b}(n)$,
where, defining $x= (\phi_{0}-\phi_{*})/a = -\tanh(\sqrt{\lambda/2}\,a\,n)$,
we have
$\phi_{h}=\sqrt{\lambda/2}a^{2}(1-x^{2}) $ and
\begin{align}
	& C_{2a} =\frac{3\sqrt{2}c^{2}}{16\lambda^{5/2}a^{7}}
	\left[\frac{2x}{x^{2}-1}-\log\left(\frac{1+x}{1-x}\right)\right] ,
	\label{C2a}
	\\
	&
	\begin{multlined}[b]
		C_{2b} = -\frac{\sqrt{2}a^{-3}}{48\lambda^{3/2}} 
			\bigg\{ \frac{4x}{(1-x^{2})^{2}} \left[ (1-6\log2)x^{2} + 10\log2 -1 \right]  
			\\
			-2\log(1-x) \bigg[ \frac{x(5x+4)-3}{(1+x)^{2}} + 3\log2 \bigg] 
			+ 2\log(1+x) \bigg[ \frac{x(5x-4)-3}{(1-x)^{2}}+3\log2 \bigg] 
			\\
			+3\log^{2}(1-x)-3\log^{2}(1+x) +6\textrm{Li}_{2}
			\bigg(\frac{1+x}{2}\bigg)-6\textrm{Li}_{2}\bigg(\frac{1-x}{2}\bigg)\bigg\} ,
		\end{multlined}
		\label{C2b}
\end{align}
where
$\textrm{Li}_{2}(z)$
is the dilogarithm function \cite{abramowitz}. 

The correction to the surface tension,
$\sigma_{2} = \int_{-\infty}^{+\infty} 2 \phi_{0}'\partial_n\phi_{2}d{n}$, is of the form
$\sigma_{2}	=\tilde{\sigma}_{2}-\mu_{0}\partial_{n}K_{0}|_{n=0}$,
where 
$\tilde{\sigma}_2$ and $\mu_0$ are constants.
The latter is given by the LO value of the quantity $\mu$ defined in Eq.~(\ref{eq:defl}).
For the potential (\ref{eq:V0}), we have
\begin{equation}
	\tilde{\sigma}_{2}=-\frac{3\sqrt{2}}{4\lambda^{3/2}}\frac{c^{2}}{a^{3}} , \quad \mu_{0}=\frac{\pi^{2}-6}{9}\frac{\sqrt{2}}{\sqrt{\lambda}}\,a .
	\label{sigmamu}
\end{equation}
Defining $\tilde{\sigma}=\sigma_{0}+\sigma_{1}+\tilde{\sigma}_{2}$, we have
\begin{equation}
	\sigma = \tilde{\sigma}-\mu_{0}\partial_{n}K_{0}(\xi^a,0) ,
	\label{sigmatot}
\end{equation}
so the surface tension generally depends on the position on the hypersurface $\Sigma$.

For $i=3$, the condition $f_3|_{n=-\infty}=0$ gives the NNLO correction $K_2$.
The quantity $K=K_0+K_1+K_2$ is given by \cite{mm24}
\begin{equation}
	\sigma K =-\Delta V - (\mu_{0}/2) \partial_{n}^{2}K_{0}|_{n=0} .
	\label{K2}
\end{equation}
There is an extra term with respect to the relation $\sigma K=-\Delta V$ obtained at previous orders.
Besides, the surface tension is no longer a constant along the surface.
Using Eqs.~(\ref{sigmatot}) and (\ref{eq:Kcovar}) in Eq.~(\ref{K2}), 
the NNLO EOM can be written in the form
\begin{equation}
	N^{\mu}_{\ ;\mu}=\frac{\Delta V}{\tilde{\sigma}} + 
	\frac{\mu_0}{\sigma_0} \left( \frac{\Delta V}{\sigma_{0}}N^{\mu}_{0;\nu} N^{\nu}_{0;\mu}
	- N^{\mu}_{0;\nu} N^{\nu}_{0;\rho} N^{\rho}_{0;\mu} \right).
	\label{EOM2}
\end{equation}
The parameter $\mu_0/\sigma_0$ is of NNLO, as can be seen from Eq.~(\ref{eq:defl}), so the quantities $N^{\mu}_{0;\nu}$ can be replaced by $N^{\mu}_{;\nu}$ if convenient, instead of using the LO solution as a source term.

Let us now consider the energy-momentum tensor, Eqs.~(\ref{TabNGC2})-(\ref{TnnTanNGC2}).
Taking into account that $\partial_{a}\phi_{0}=\partial_{a}\phi_{1}=0$
and that $\partial_a\phi_2$ is of order $(l/L)^3$ relative to $\partial_n\phi_0$,
we have, at NNLO, 
\begin{equation}
	T_{\mu\nu} = (\partial_n\phi)^{2}P_{\mu\nu} + V^b g_{\mu\nu} , \label{eq:TmunuNNLO}
\end{equation}
with
\begin{equation}
	(\partial_n \phi)^2 = \phi_{0}^{\prime2}+2\phi_{0}'\phi_{1}'+\phi_{1}^{\prime2}+2\phi_{0}'\phi_{2a}'
	+2\phi_{0}'\phi_{2b}'\partial_{n}K_0|_{n=0} ,
	\label{dfi2NNLO}
\end{equation}
where the factors $\phi_i'$ are the derivatives of $\phi_i(n)$ and do not depend on the specific wall hypersurface, while $\partial_{n}K_0|_{n=0}=N_{0;\nu}^{\mu}N_{0;\mu}^{\nu}$ is a function of $\xi^a$ and depends on the hypersurface. 
The tensor $P_{\mu\nu}=g_{\mu\nu}+n_\mu n_\nu$ can be obtained as an expansion in powers of $n$.%
\footnote{One way to obtain this expansion is using the Taylor series of $g_{\mu\nu}$ and $n_\mu$. 
	From the parallel transport equation $n^{\sigma}\nabla_{\sigma}n_{\mu}=0$, we obtain $\partial_{n}n_{\mu}|_{0}=\Gamma_{\sigma\mu}^{\rho}N^{\sigma}N_{\rho}|_{0}$, and differentiating this equation we obtain the successive derivatives.
	For the expansion of the metric, we write, e.g., $\partial_{n}g_{\mu\nu}|_{0}=n^{\lambda}g_{\mu\nu,\lambda}|_{0}=N^{\lambda}(g_{\mu\rho}\Gamma_{\nu\lambda}^{\rho}+g_{\nu\rho}\Gamma_{\mu\lambda}^{\rho})_{0}$, and so on.
	We obtain $P_{\mu\nu} = A_{\mu\nu} + B_{\mu\nu}n + C_{\mu\nu}n^2 +\cdots$, with
	$A_{\mu\nu}=h_{\mu\nu}$,
	$B_{\mu\nu}=[N^{\lambda}(h_{\mu\rho}\Gamma_{\lambda\nu}^{\rho}+h_{\nu\rho}\Gamma_{\mu\lambda}^{\rho})]$, and 
	$C_{\mu\nu}=\frac{1}{2} N^{\sigma}N^{\lambda} [h_{\mu\rho}(\Gamma_{\lambda\tau}^{\rho}\Gamma_{\nu\sigma}^{\tau}-\Gamma_{\nu\tau}^{\rho}\Gamma_{\sigma\lambda}^{\tau}+\partial_{\lambda}\Gamma_{\nu\sigma}^{\rho})+h_{\nu\rho}(\Gamma_{\tau\lambda}^{\rho}\Gamma_{\mu\sigma}^{\tau}-\Gamma_{\mu\tau}^{\rho}\Gamma_{\lambda\sigma}^{\tau}+\partial_{\lambda}\Gamma_{\mu\sigma}^{\rho})+2h_{\tau\rho}\Gamma_{\mu\lambda}^{\tau}\Gamma_{\nu\sigma}^{\rho}]$.}
In Sec.~\ref{sec:O3} we consider some specific cases.
The expression (\ref{eq:TmunuNNLO}) still has the form of the thin-wall approximation (\ref{eq:Tmunuthin}),
although the field is no longer a function of $n$ alone.
However, at higher order there will be non-vanishing components $\bar{T}_{an}$ in Eqs.~(\ref{TabNGC2})-(\ref{TnnTanNGC2}).

If we use the covariant definition (\ref{Smunuthin}) for the surface stress-energy tensor, we just have $S_{\mu\nu} = \sigma h_{\mu\nu}$, with $\sigma$
given by Eq.~(\ref{sigmatot}).
If we use the more precise definition (\ref{Smunudef}), we must choose a specific coordinate system.
In Gaussian normal coordinates, $\bar{S}_{ab}$ is given by Eq.~(\ref{Sabdef}).
Using the expansion (\ref{gab}) for $\bar{g}_{ab}$, we obtain
\begin{equation}
	\bar{S}_{ab}=(\tilde{\sigma}-\mu_{0}\bar{K}^{cd}\bar{K}_{cd})\gamma_{ab}
	+\mu_{0}[2K\bar{K}_{ab}+{\bar{K}_{a}}^{c}\bar{K}_{cb}+{\textstyle\frac{1}{2}}(K^{2}-\bar{K}^{cd}\bar{K}_{cd})\gamma_{ab}] .
\end{equation}
The parameter  $\mu_0$ is of order $l^2$ relative to $\sigma$, 
so we only need to calculate $\bar{K}_{ab}$ to LO.	

The wall energy density is given by Eq.~(\ref{Tw00}), $T_{00}^{w} =  q^{\prime2} (\partial_n\phi)^2 $,
and the surface energy density $\varepsilon_{w}$ is given by Eq.~(\ref{eq:Ewdef}).
To calculate the integral in Eq.~(\ref{eq:Ewdef}), we must write the integrand as a function of $m$ at fixed $t$ and $\zeta^A$.
The quantity $\sqrt{\tilde g/\tilde{g}|_{m=0}}$ is given by Eq.~(\ref{expangtil}) as a series expansion in powers of $m$
with coefficients depending on $t$ and $\zeta^A$.
The quantity $q^{\prime 2}$ can be written as
\begin{equation}
	q^{\prime2}=q^{\prime2}|_{m=0}+\partial_{m}q^{\prime2}|_{m=0}m+\frac{1}{2}\partial_{m}^{2}q^{\prime2}|_{m=0}m^{2}+\cdots,
\end{equation}
where the notation $|_{m=0}\equiv |_{Y^i(\zeta^A)}$ indicates that here too the coefficients are evaluated at constant $t$ and $\zeta^A$.
In principle, we should also write the functions  $\phi_i'(n)$ appearing in the expansion of $(\partial_n\phi)^2$ in terms of $m$, $\zeta^A$, and $t$.
However, due to the strong dependence of these functions on $n$, 
it is actually convenient to change the integration variable $m$ to $n$ by means of Eq.~(\ref{eq:nprin}). 
Thus, Eq.~(\ref{eq:Ewdef}) takes the form 
\begin{equation}
	\varepsilon_w=\int_{-\infty}^{+\infty} dn \left[A+Bn+Cn^2\right]  (\partial_n\phi)^2 ,
	\label{eq:Ew2o}
\end{equation} 
where the coefficients $A,B,C$ are constant along the integration curve.
The constant $B$ is irrelevant since the integral $\int n(\partial_n\phi)^2$ vanishes 
due to the condition (\ref{eq:cero_n}).
For the term quadratic in $n$, we can replace $(\partial_n\phi)^2$ by its LO contribution $\phi'_0(n)^2$, which gives the parameter $\mu_0$.
On the other hand, the first term gives the NNLO surface tension $\sigma$.
It is straightforward to calculate the quantitites $A$ and $C$, and we obtain
\begin{equation}
\varepsilon_{w} =  \tilde{\sigma} q^{\prime} +
\mu_{0}\left[
\frac{\partial_{m}^{2}q'}{q^{\prime2}} + \frac{\kappa^{2}-\kappa^{ij}\kappa_{ij}}{2q'} 
- \frac{\kappa\partial_{m}q'}{2q^{\prime 2}} 
- \frac{\left(\partial_{m}q'\right)^{2}}{2q^{\prime3}} - \frac{\partial_{m}^{3}n}{2q^{\prime2}} - q^{\prime}N_{;\nu}^{\mu}N_{;\mu}^{\nu}
\right] .
\label{Ewgral}
\end{equation}
All the quantities are evaluated at $S$
and can be written in terms of the vectors $N^\mu$ and $M^i$ and thus
in terms of the covariant derivatives of the function $F$ (see Sec.~\ref{CGN}).
We could also derive an expression for $E_b$ as we did in Eq.~(\ref{Ebthin}), 
but the result is cumbersome and unilluminating, and this quantity is not as useful as $\varepsilon_w$.

\subsection{Spherical bubbles}

For a spherical surface we have $F=r-r_w(t)$ and the covariant derivatives of $N^\mu$ are given by Eq.~(\ref{d2nKesf}).
Thus, the equation of motion for the LO solution $r_{w0}$, Eq.~(\ref{EOM0}), takes the form
\begin{equation}
	\gamma_{w0}^{3}\ddot{r}_{w0}+2{\gamma_{w0}}{r_{w0}^{-1}} ={\Delta V}/{\sigma_{0}} \equiv 3/R_0 .
	\label{eom0esf}
\end{equation}
We see that the wall acceleration is determined by two forces, one due to the pressure difference and the other due to the surface tension.
The NLO EOM is also of the form (\ref{EOM0}),
with $\sigma_0$ replaced by the NLO surface tension.
However, for our polynomial potential, $\sigma$ does not receive corrections at this order, so we have the same EOM.
The NNLO EOM, Eq.~(\ref{EOM2}), gives
\begin{equation}
	\gamma_{w}^{3}\ddot{r}_{w}+\frac{2\gamma_{w}}{r_{w}} =\frac{\Delta V}{\tilde{\sigma}}+2l_{0}^{2}\frac{\gamma_{w0}}{r_{w0}}\left[\left(\frac{\Delta V}{\sigma_{0}}\right)^{2}-3\frac{\Delta V}{\sigma_{0}}\frac{\gamma_{w0}}{r_{w0}}+3\left(\frac{\gamma_{w0}}{r_{w0}}\right)^{2}\right] .
	\label{eom2esf}
\end{equation}
In the last term, we have used the LO equation (\ref{eom0esf}) and the relation $\mu_0=\sigma_0l_0^2$. 
Like in the LO equation, we identify a force due to the pressure difference and another due to the surface tension.
However, this equation includes additional terms proportional to $l_{0}^{2}$.

The energy-momentum tensor of the wall is given by Eq.~(\ref{T00wesf}), $T_{\mu\nu}^{w} =(\partial_{n}\phi)^{2}P_{\mu\nu}$.
In the present case, Eq.~(\ref{dfi2NNLO}) gives
\begin{equation}
	(\partial_{n}\phi)^{2}=
	\phi_{0}^{\prime2}+2\phi_{0}'\phi_{2a}'+6\phi_{0}'\phi_{2b}'\left[R_{0}^{-2}+2\left(R_{0}^{-1}-{\gamma_{w0}}/{r_{w0}}\right)^{2}\right] ,
	\label{dnfi2}
\end{equation}
where the functions $\phi_i^\prime(n)$ are given by Eqs.~(\ref{eq:phi0}), (\ref{eq:phi1}), and (\ref{phi2})-(\ref{C2b}).
Only the last term between brackets depends on $\xi^0$.
Multiplying Eq.~(\ref{eom0esf}) by $r_{w0}^2\dot{r}_{w0}$ and integrating, we obtain the first integral
\begin{equation}
	\left(\gamma_{w0}/r_{w0} - R_{0}^{-1}\right) r_{w0}^{3} = \left(\gamma_{i0}/r_{i0} - R_{0}^{-1}\right) r_{i0}^{3},
	\label{eq:gammaw0primera}	
\end{equation}
where $r_{i0}=r_{w0}(0)$, $\gamma_{i0}=\gamma_{w0}(0)$ are the initial conditions.
Therefore, the last term in Eq.~(\ref{dnfi2}) is proportional to $r_{w0}^{-6}$ and quickly becomes negligible as the bubble grows.

As we have seen, for the spherical case we have $P_{00}=(\partial_r n)^2$.
According to Eq.~(\ref{n0}), at leading order we have $n=\gamma_{w0}(r-r_{w0})$, 
and we obtain
\begin{equation}
		T_{00}^w|_{\mathrm{LO}}=\gamma_{w0}^2\left[\phi_0'\left(\gamma_{w0}(r-r_{w0})\right)\right]^2 .
\end{equation}
This result is a consequence of the Lorentz contraction of the wall profile.
Indeed, for an observer which is instantly at rest with the wall at $n=0$,
$n$ gives the proper distance and
the field profile $\phi_0(n)$ has a wall width $l_0$.
For an observer at rest at the bubble center, the wall width is given by $l_0/\gamma_{w0}$.
We should not expect such a simple relation beyond the LO
since the Gaussian normal coordinates are not an inertial reference frame.
Either from the general expansion (\ref{eq:n_orden2}) or from  Eq.~(\ref{gaussianasO3inv}), we obtain, at NNLO,
\begin{equation}
	n=
	\gamma_{w}(r-r_{w})
	-\dot{r}_{w}^{2}\left(\frac{3}{2R_{0}} - \frac{\gamma_{w}}{r_{w}}\right)\gamma_{w}^{2}(r-r_{w})^{2}
	+ \dot{r}_{w}^{4}\frac{\gamma_{w}}{r_{w}}\left(\frac{\gamma_{w}}{r_{w}}-\frac{1}{R_{0}}\right)\gamma_{w}^{3}
	(r-r_{w})^{3} .
	\label{eq:nesf}
\end{equation}
The wall energy density $T_{00}^w=(\partial_r n)^2(\partial_n\phi)^2$ is readily obtained from Eqs.~(\ref{eq:nesf}) and (\ref{dnfi2}).

If $\phi$ depends only on $n$, we have $T_{00}^w = (\partial_r\phi)^2$.
More generally, we have
\begin{equation}
	T_{00}^w = \left[\partial_r \phi - \partial_r\xi^0 \partial_{\xi^0}\phi \right]^2 .
	\label{T00wdrfi}
\end{equation}
The derivative $\partial_{\xi^0}\phi$ vanishes at LO and NLO. 
At NNLO we have $\partial_{\xi^0}\phi = \phi_{2b}\partial_{\xi^0}\partial_{n}K_{0}$, where $\partial_n K_0$ is given by Eqs.~(\ref{eq:Kcovar}) and (\ref{d2nKesf}).
We can keep the quantities that multiply $\phi_{2b}$ to the leading order.
In particular, from Eq.~(\ref{gaussianasO3inv}) we have $\partial_r\xi^0 = -\gamma_{w}^{2}\dot{r}_{w}$.
Using also Eq.~(\ref{eq:gammaw0primera}), we obtain
\begin{equation}
	\partial_{r}\xi^{0}\partial_{\xi^{0}}\phi
	=36\left(\gamma_{i0}-\frac{r_{i0}}{R_{0}}\right)^{2}\left[\frac{r_{i0}}{R_{0}}+\left(\gamma_{i0}-\frac{r_{i0}}{R_{0}}\right)\frac{r_{i0}^{3}}{r_{w0}^{3}}\right]^{2}\dot{r}_{w0}^{2}\frac{r_{i0}^{2}}{r_{w0}^{5}}\phi_{2b}(n) .
	\label{difTdrfi}
\end{equation}
This term is initially suppressed by the factor $\dot{r}_{w0}^2$ and decreases as $r_{w0}^{-5}$ as the bubble grows.
As a consequence, this contribution is even smaller than the NNLO terms in Eq.~(\ref{dnfi2}),
and we can use the approximation $T_{00}^w = (\partial_r\phi)^2$.
We have checked that the term (\ref{difTdrfi}) is indeed negligible for 
the specific examples of the next section, 
both in the thin-wall approximation and in the numerical solutions of Eq.~(\ref{ecfiO3}).

The surface energy density is given by Eq.~(\ref{EwLO}) at LO and Eq.~(\ref{Ewgral}) at NNLO, with 
$q'=\partial_r n$ for the spherical case.
Thus, at LO we obtain the well-known result
\begin{equation}
	\varepsilon_{w0}=\sigma_{0}\gamma_{w0},
	\label{Ew0}
\end{equation} 
while at NNLO we have
\begin{equation}
	\varepsilon_{w}=\gamma_{w}\left\{\tilde{\sigma}
	-\frac{\sigma_{0}l_{0}^{2}}{R_{0}^{2}}
	\left[\frac{9}{2}
	\left(3-\frac{2}{\gamma_{w}^{2}}+\frac{1}{\gamma_{w}^{4}}\right)
	-3\left(5-\frac{3}{\gamma_{w}^{2}}+\frac{2}{\gamma_{w}^{4}}\right)\frac{R_{0}\gamma_{w}}{r_{w}}+5\frac{R_{0}^{2}\gamma_{w}^{2}}{r_{w}^{2}}\right]\right\}.
	\label{Ew2}
\end{equation}
(the solutions $r_w$ and $r_{w0}$ are interchangeable in the term proportional to $l_0^2$).
At leading order, the bulk energy is given by Eq.~(\ref{Ebthin}), $E_b=\frac{4\pi}{3}r_w^3 \Delta V$.
Therefore, the total energy of the bubble is given by
\begin{equation}
	E_w+E_b=\sigma_{0}\gamma_{w0}\, 4\pi r_{w0}^{2}- \Delta V\frac{4\pi}{3} r_{w0}^{3} .
	\label{eq:gammaw0}	
\end{equation}
To check that this is a conserved quantity, notice that this expression is just the left-hand side of Eq.~(\ref{eq:gammaw0primera}) multiplied by $4\pi$.
As already mentioned, obtaining $E_b$ at higher orders is rather cumbersome.
On the other hand, the first integral of Eq.~(\ref{eom2esf}) can be obtained analytically \cite{mm24}.
Multiplying this conserved quantity by $4\pi$, we obtain the NNLO total energy,
\begin{equation}
	E_w+E_b = \tilde{\sigma}\gamma_{w}r_{w}^{2}
	-\frac{\Delta V}{3}r_{w}^{3}
	-\frac{\mu_{0}}{R_{0}^{3}}r_{w}^{3}\left[3 - 3\frac{R_{0}\gamma_{w}}{r_{w}} + 3\left(\frac{R_{0}\gamma_{w}}{r_{w}}\right)^{2} - \left(\frac{R_{0}\gamma_{w}}{r_{w}}\right)^{3}\right] .
	\label{eq:primeraintesf}
\end{equation}

\section{Specific cases}
\label{sec:O3}

Our general EOM, either the LO version (\ref{EOM0}) or the NNLO version (\ref{EOM2}), describes the evolution of a wall surface $S$ from a given initial condition.
For example, the initial surface may represent part of a bubble wall after a bubble collision or after undergoing some deformation process.
In order to make a comparison with the numerical solution of the field equation, we are going to consider the simpler case of a spherical bubble.

\subsection{Vacuum bubbles}
\label{subsec:O31}

For the O(3,1)-symmetric solution discussed in Sec.~\ref{sec:O31}, the field equation takes the form (\ref{eq:ecfiColeman}) for $r>t$ and (\ref{eq:ecfiColemanin}) for $r<t$, which give the solutions $\bar{\phi}(\rho)$ and $\tilde{\phi}(\tau)$, respectively.
In particular, for $r>t$ we have $\phi(r,t)=\bar{\phi}(\sqrt{r^2-t^2})$, so the initial bubble configuration is given by this solution, $\phi(r,0)=\bar{\phi}(r)$.
This is consistent with the bubble nucleation being governed by the O(4)-invariant bounce instanton in a vacuum phase transition \cite{c77}.
The bounce configuration is the analytic continuation of the O(3,1)-symmetric solution $\bar{\phi}$ to imaginary time.
As in Sec.~\ref{sec:O31}, we define the wall position $\rho_w$ from  $\bar{\phi}(\rho)$ through Eq.~(\ref{rhow}).
The condition $\rho=\rho_w$ gives the bubble radius at time $t$, $r_w(t) = \sqrt{\rho_w^2+t^2}$.

To compare the numerical solution with the thin-wall approximation, we could use the numerical value $\rho_w$ as an initial condition in the LO equation (\ref{eom0esf}) or the NNLO equation (\ref{eom2esf}).
However, we can also obtain thin-wall approximations for the initial condition, so we do not need to use a numerical result in the approximation.
Indeed, using the standard thin-wall approximation directly in Eq.~(\ref{eq:ecfiColeman}) instead of the general field equation
gives $\rho_w =R_{0}=3\sigma_{0}/\Delta V$ \cite{c77}.
Similarly, using our perturbative method in Eq.~(\ref{eq:ecfiColeman}) we can obtain the NNLO approximation \cite{mm24}.
We discuss this approach in App.~\ref{app:instanton}%
\footnote{Alternative attempts to improve the thin wall approximation for the bounce instanton can be found in Refs.~\cite{mr00,b16,mpw17}.}.
Here we simply note that the total energy of the nucleated bubble vanishes for a vacuum phase transition, so we can obtain the nucleation radius by imposing this condition on the approximations (\ref{eq:gammaw0})-(\ref{eq:primeraintesf}).
We readily obtain the LO value $R_0$ and the NNLO value%
\footnote{The condition $E_w+E_b=0$ for $\gamma_w$=1 gives a cubic equation for $r_w$, but we can write this equation in the form $r_w= 3\tilde{\sigma}/\Delta V+\cdots$ and replace this expression recursively in the NNLO terms.}
$R=3\tilde{\sigma}/\Delta V - 2l_0^2/R_0$. 

We will use the initial condition $r_{i0}=R_0$ in Eq.~(\ref{eom0esf}) and $r_i=R$ in Eq.~(\ref{eom2esf}).
In this case, the LO solution has the same form as the exact solution, $r_{w0}^{2}=R_{0}^{2}+t^{2}$.
We compare the approximations for $r_w$ in App.~\ref{app:instanton}.
Here we focus on the energy-momentum tensor and the wall energy.

\begin{figure}[tb]
	\centering
	\includegraphics[width=0.48\textwidth]{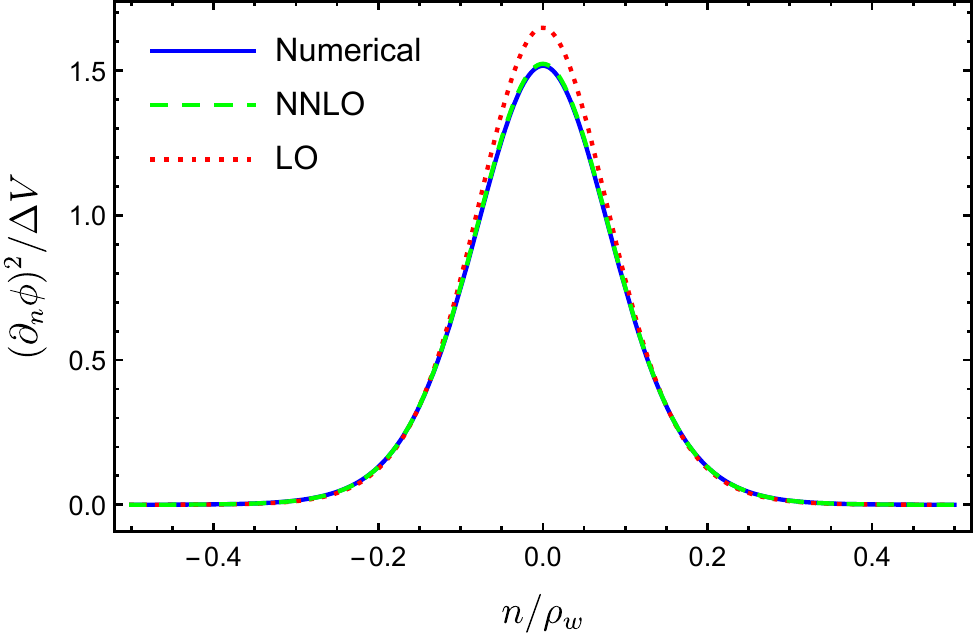}\hfill\includegraphics[width=0.48\textwidth]{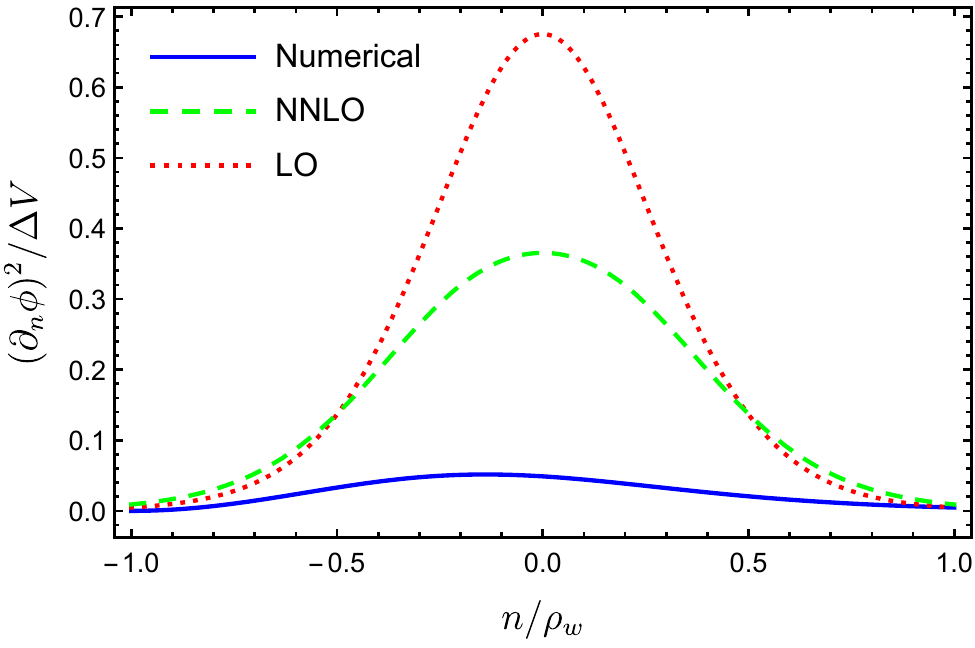}
	\caption{The quantity $(\partial_n\phi)^2$ as a function of $n$ for the two example potentials discussed in Sec.~\ref{sec:O31}.
		Left: the case of Fig.~\ref{fig:potperf1}. Right:  the case of Fig.~\ref{fig:potperf2}.}
	\label{fig:fipri2ap}
\end{figure}

All the components of the wall energy-momentum tensor (\ref{eq:Tmunuw})
are proportional to the quantity $(\partial_n \phi)^2$.
In this case, the quantity (\ref{eq:gammaw0primera}) vanishes
and the expansion (\ref{dnfi2}) depends only on $n$, as the exact solution.
In Fig.~\ref{fig:fipri2ap} we compare the approximations for $(\partial_n \phi)^2$.
We see that the NNLO approximation is very good if the wall is not too thick (left panel).
For a very thick wall (right panel), the approximation fails.
We remark that the field as a function of $n$ only describes a part of the bubble profile,
where the Gaussian normal coordinates are valid
(the part to the right of the crosses in Fig.~\ref{fig:potperf2}).
In contrast, the thin-wall approximation assumes a complete profile where the field varies between the values $\phi_-$ 
and $\phi_+$.
Writing $n$ as a function of $r$ and $t$, this approximation will give a better description of the bubble profile at later times, once the field has reached the minimum $\phi_-$ inside the bubble (see below).

\begin{figure}[tb]
	\centering
	\includegraphics[width=0.48\textwidth]{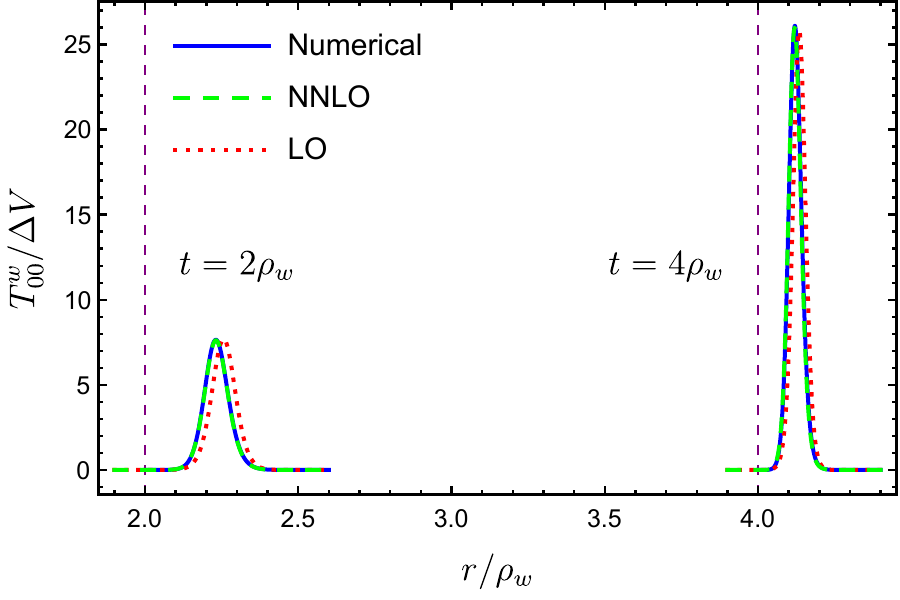}\hfill\includegraphics[width=0.48\textwidth]{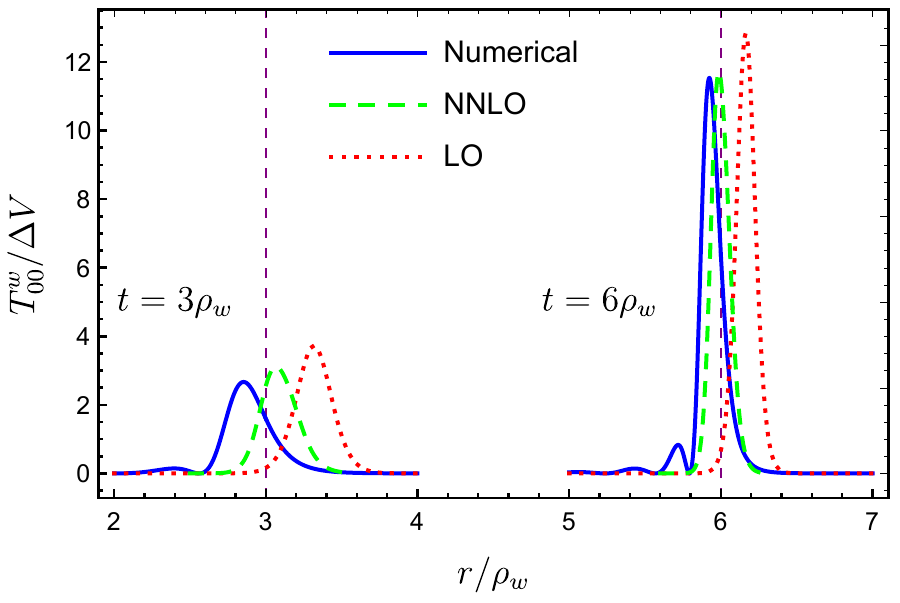}
	\caption{Wall energy density at different times for the two example potentials.\label{fig:Twallap}}
\end{figure}

The wall energy density is given by
$T_{00}^w = (\partial_r n)^2(\partial_n\phi)^2 = (\partial_r\phi)^2$.
For the numerical solution, the latter expression can be used beyond the limit of the normal variable $n$,
where the solution is given by the function $\tilde\phi(\tau)$.
Initially, though, we have $n=r-r_i$ and $T_{00}^w = (\partial_n\phi)^2=(\partial_\rho\bar{\phi})^2$, 
so the wall energy density has the profile shown in Fig.~\ref{fig:fipri2ap}.
At subsequent times, this profile becomes thinner and taller due to the Lorentz contraction,
as shown in Fig.~\ref{fig:Twallap}.
In particular, at the point $r=t$ separating the inner and outer solutions, we have $\partial_r\phi=\frac14 t V'(\phi_i)$.
This point is indicated by vertical dashed lines in Fig.~\ref{fig:Twallap}.
We see that the approximations for the shape of the profile are generally quite good, 
but there is also an error in the wall position $r_w(t)$, which decreases with time.
If the wall is not too thick, the thin wall approximation is very good even at leading order.

\begin{figure}[tb]
	\centering
	\includegraphics[width=0.48\textwidth]{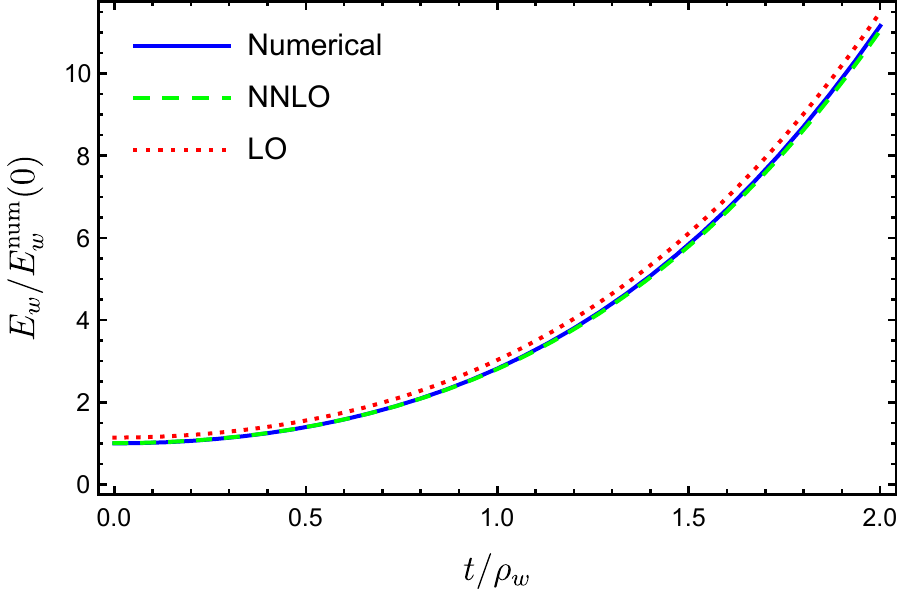}\hfill\includegraphics[width=0.48\textwidth]{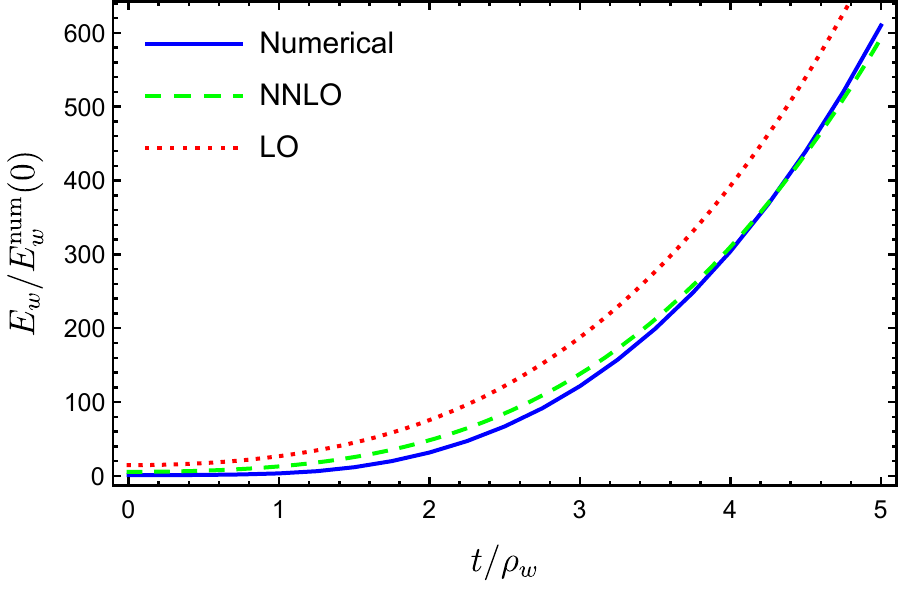}
	\caption{Wall energy as a function of time for the two example potentials.
		The wall energy is scaled with the initial value for the numerical solution.}\label{fig:Ewap}
\end{figure}

In Fig.~\ref{fig:Ewap} we show the wall energy $E_w$, which is given by Eq.~(\ref{Ewesf})
for the numerical solution (cf.\ Fig.~\ref{fig:energy})
and by Eqs.~(\ref{Ew0}) and (\ref{Ew2}) for the approximations.
We see that the NNLO approximation can significantly improve the LO approximation.

\subsection{Near-critical bubbles}
\label{subsec:O3}

If the bubble has O(3) symmetry but not O(3,1) symmetry, its evolution is given by the more general equation (\ref{ecfiO3}) and we need to consider an initial condition.
We will consider a bubble nucleated by thermal activation, where our toy-model potential (\ref{eq:pot}) represents the finite-temperature effective potential at a given nucleation temperature. 
Neglecting the presence of a plasma for the subsequent evolution is a good approximation in a very supercooled phase transition.
In this case, the relevant instanton is a saddlepoint of the three-dimensional action
\begin{equation}
	S_3 = 4\pi \int_0^\infty r^2dr\left[\frac{1}{2}\left(\frac{d\phi}{dr}\right)^2+V\right].
	\label{S3}
\end{equation}
Extremizing $S_3$ we obtain the equation
\begin{equation}
	\frac{d^2\phi}{dr^2} + \frac{2}{r} \frac{d\phi}{dr} =\frac{dV}{d\phi} 
	\label{ecinstO3}
\end{equation}
with the boundary conditions $d\phi/dr=0$ at $r=0$ and $\phi\to 0$ for $r\to\infty$.
This equation is very similar to Eq.~(\ref{eq:ecfiColeman}), and its solution  $\phi_c(r)$ can also be easily obtained using the shooting method.
This bubble is in unstable equilibrium between expansion and collapse.
Indeed, notice that Eq.~(\ref{S3}) is also the expression for the energy of a static field configuration.

\begin{figure}[tb]
	\centering
	\includegraphics[width=0.48\textwidth]{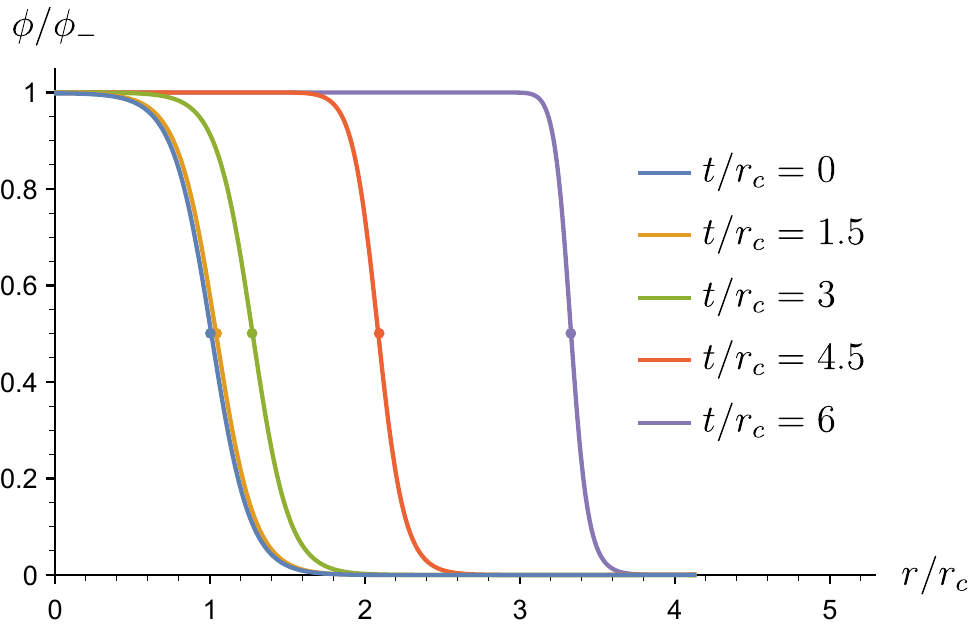}
	\hfill\includegraphics[width=0.48\textwidth]{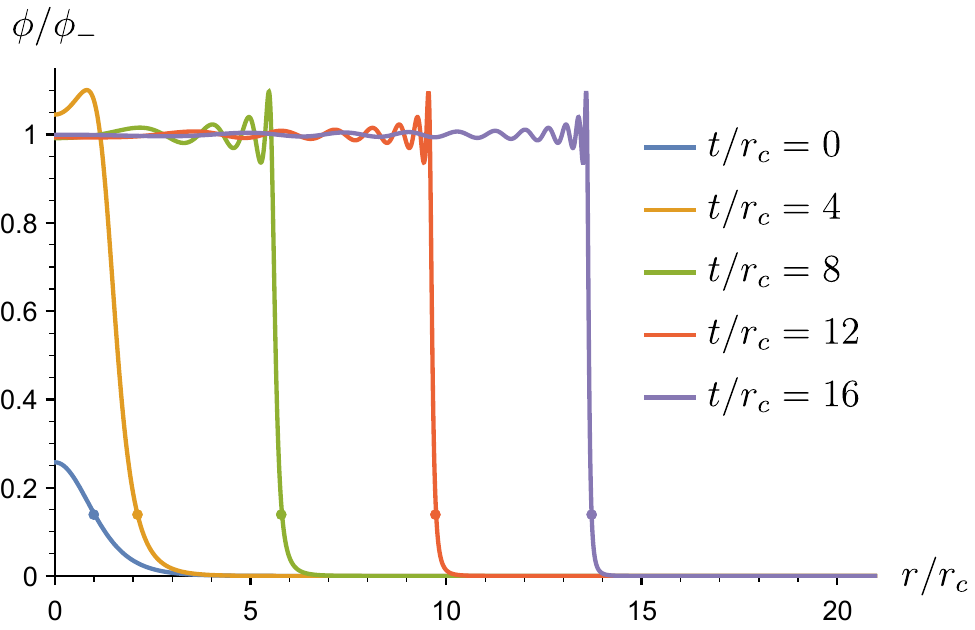}
	\caption{Evolution of the bubble profile of a nearly-critical bubble ($\epsilon=10^{-2}$)
		for the potential of Fig.~\ref{fig:potperf1} (left) and that of Fig.~\ref{fig:potperf2} (right).}\label{fig:profO3}
\end{figure}

We will consider an initial bubble profile that is slightly displaced from the unstable configuration,
$\phi(0,r)=\phi_c(r-\epsilon r_c)$,
where we define the radius of the critical bubble as
\begin{equation}
	r_c = \frac{\int_0^\infty r [\phi_c'(r)]^2 dr}{\int_0^\infty  [\phi_c'(r)]^2 dr} .
	\label{rc}
\end{equation}
As an example, in Fig.~\ref{fig:profO3} we consider the case $\epsilon=10^{-2}$ for our two example potentials.
The field profiles are qualitatively similar to those of the O(3,1)-symmetric case.
In the left panel we have the case of a potential where the height of the barrier is of the same order as the energy difference between the minima.
Although not exactly thin, the bubble wall is not extremely thick in this case
(we have $l/r_c\simeq 0.14$).
The case on the right panel corresponds to a potential with a very small barrier compared to the energy difference between the minima.
In this case, the wall is initially very thick (we have $l/r_c\simeq 0.47$) and the field inside the bubble is far from the stable minimum.
As the bubble grows, the wall becomes thin and the field inside the bubble oscillates around the minimum.

In previous sections, we used the condition (\ref{eq:cero_n}) as the definition of the wall position.
In the present case, this condition gives $r_w(t)$ as an average similar to that of Eq.~(\ref{rc}), but with the range of integration in $r$ limited to the range of the Gaussian normal coordinates for the hypersurface $\Sigma$.
The advantage is that this range excludes the field oscillations in the right panel of Fig.~\ref{fig:profO3}.
However, this is a recursive definition of $r_w(t)$ since
the Gaussian normal coordinates depend on the hypersurface generated by the evolution of $r_w(t)$.
This is highly impractical to apply to the numerically obtained $\phi(t,r)$.
Therefore, we will use a slightly different definition.
Once the critical radius $r_c$ is calculated, we take the value $\phi_w = \phi_c(r_c)$
and define the bubble radius by $\phi(t,r_w(t))=\phi_w$.
The result is indicated by a dot on each curve in Fig.~\ref{fig:profO3}.
For cases where $(\partial_r\phi)^2$ falls virtually to zero within the range of the Gaussian normal coordinates,
we have verified that this definition numerically coincides with the average of $r$ weighted with this quantity.

The construction of the Gaussian normal coordinates is shown in the right panel of Fig.~\ref{fig:curved}
for the potential of Fig.~\ref{fig:potperf1} and the initial condition with $\varepsilon=0.1$.
The hypersurface $\Sigma$ is represented by the curve of $r_w(t)$.
The hypersurfaces $\Sigma_1$ and $\Sigma_2$ correspond to $n=\pm 5 r_c$ and are obtained by varying $\xi^0$ in Eq.~(\ref{gaussianasO3}).
The normal geodesics are obtained by varying $n$ in Eq.~(\ref{gaussianasO3}) for $\xi^0/r_c=0.8,1.6,2.4,3.2$.
The dotted line indicates the lightcone that is the boundary of these coordinates.
Its position is determined numerically as the asymptote of the curve of $r_w(t)$.

Like in the previous case, we can obtain the thin-wall approximation for the nucleated bubble
by applying our perturbative method to the ordinary differential equation (\ref{ecinstO3}).
We can also obtain the critical radius from the thin-wall bubble energy, by
maximizing the LO expression (\ref{eq:gammaw0}) or the NNLO expression (\ref{eq:primeraintesf}).
Since this bubble is in unstable equilibrium between expansion and collapse,
it is even more straightforward to obtain its critical radius
by imposing the conditions $\ddot{r}_{w}=0$, $\dot{r}_{w}=0$ in the LO EOM (\ref{eom0esf}) or the NNLO EOM (\ref{eom2esf}).
We compare these approaches in App.~\ref{app:instanton}.
We obtain the LO value $R_{c0}=2\sigma_0/\Delta V$ and the NNLO value
$R_c=2\left(\Delta V/\tilde{\sigma} +2l_0^2/R_{c0}^3 \right)^{-1}$.
We will use the near-critical initial condition $r_{i0}=(1+\epsilon)R_{c0}$ for 
the LO EOM and $r_i=(1+\epsilon)R_c$ for the NNLO EOM.
In App.~\ref{app:instanton} we compare these approximations with the numerical evolution.

\begin{figure}[tb]
	\centering
	\includegraphics[height=5.cm]{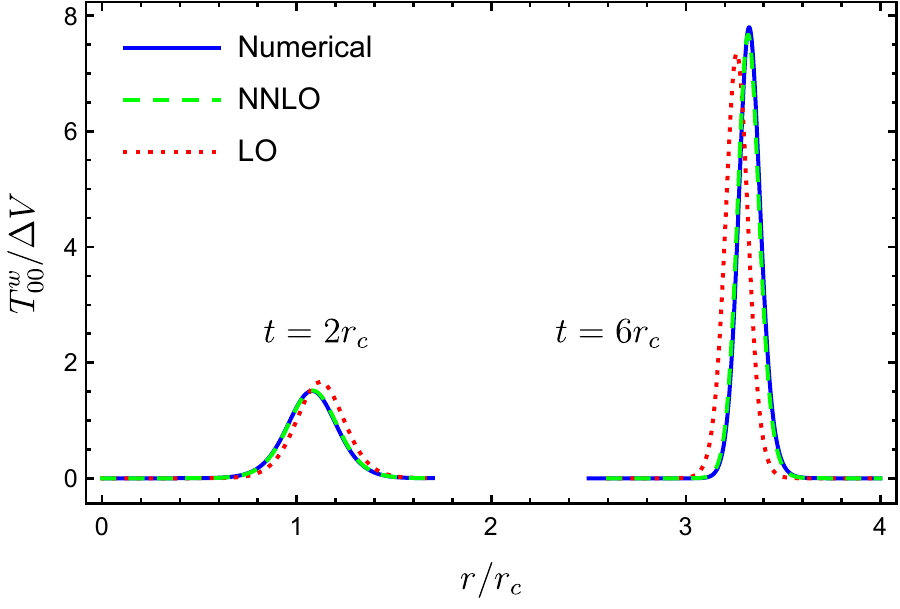}\hfill\includegraphics[height=5.cm]{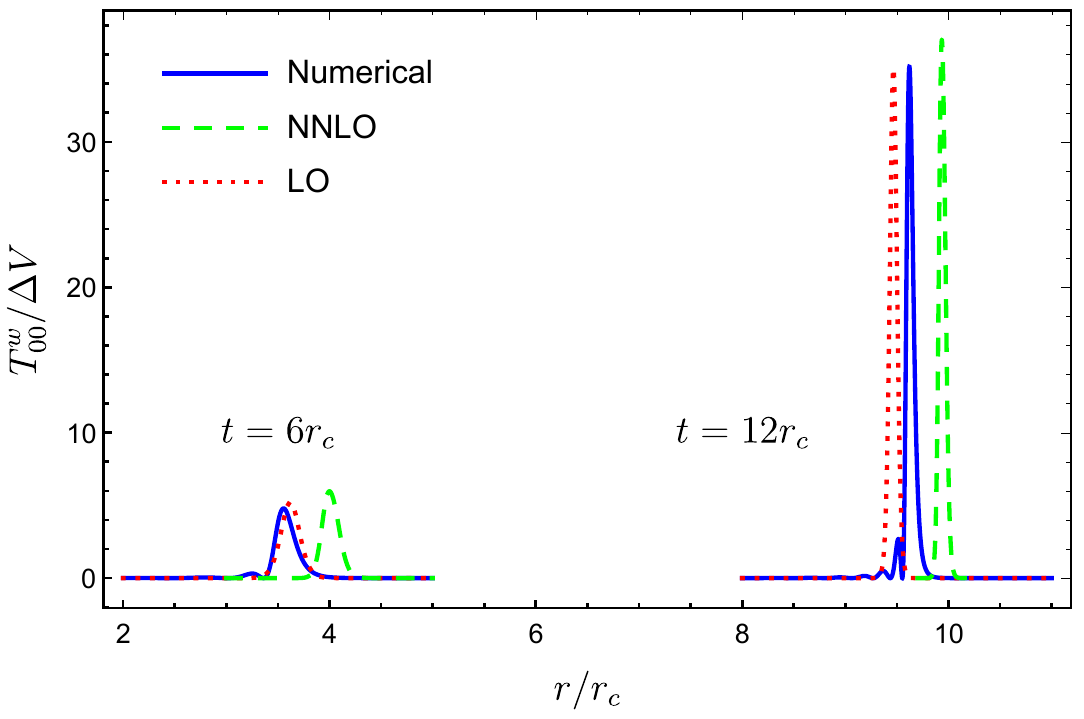}
	\caption{Wall energy density at different times for the two example potentials,
		for the initial condition $\epsilon=10^{-2}$.\label{fig:TwallapO3}}
\end{figure}

The energy density of the wall is given by Eq.~(\ref{T00wesf}), $T_{00}^w=(\partial_r n)^2(\partial_n\phi)^2$.
For the thin-wall approximation, this quantity can be readily calculated from Eqs.~(\ref{dnfi2}) and (\ref{eq:nesf})
as a function of $r$ and $t$.
For the numerical solution $\phi(t,r)$, the calculation of this quantity is rather cumbersome, as explained below Eq.~(\ref{T00wesf}).
As a consequence, $T_{00}^w$ is difficult to obtain with good numerical precision.
Nevertheless, we can use the approximation $T_{00}^w = (\partial_r\phi)^2$, since the difference between these quantities is beyond the NNLO, as discussed below Eq.~(\ref{T00wdrfi}).
Furthermore, this expression can be easily extended inside the lightcone.
In Fig.~\ref{fig:TwallapO3} we compare the different approximations.
When the wall is not very thick (left panel), the LO approximation is quite good and the NNLO approximation is much better.
The second example shows how the thin-wall approximation breaks down if the wall is too thick.
In this case, the NNLO approximation is worse than the LO one.
However, the LO approximation is not bad.
In particular, the approximation for the profile is actually quite good.
Furthermore, the relative error in the position becomes insignificant at higher times.

\begin{figure}[tb]
	\centering
	\includegraphics[width=0.48\textwidth]{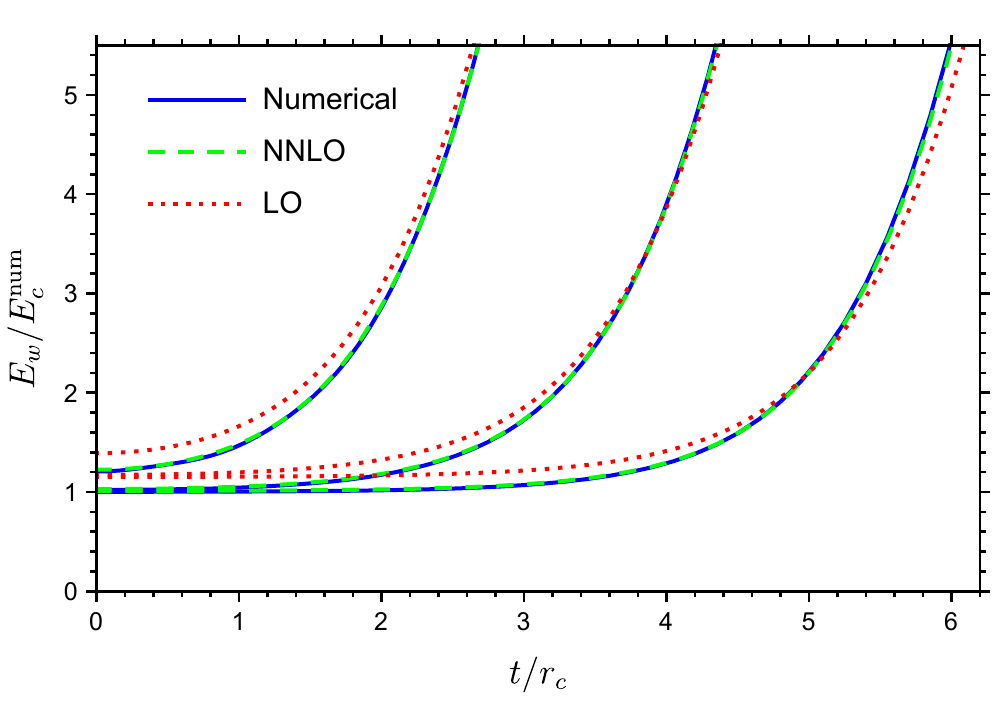}\hfill\includegraphics[width=0.48\textwidth]{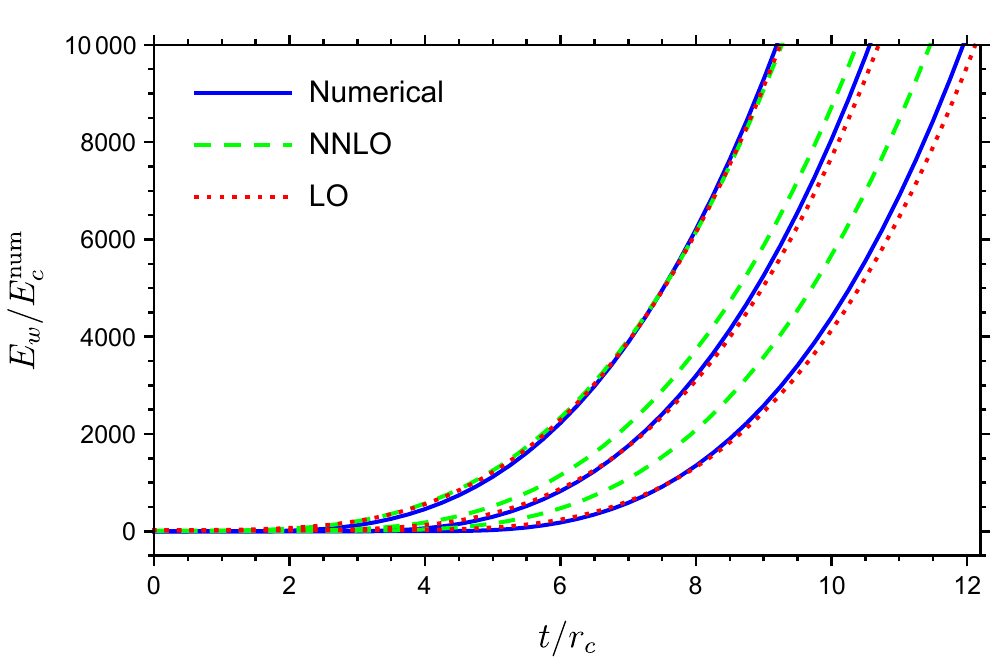}
	\caption{Wall energy as a function of time for the two example potentials
		and, from top to bottom, for the cases $\epsilon=10^{-1}$, $10^{-2}$, and $10^{-3}$.
		The numerical value for the critical configuration is used as the reference scale.}\label{fig:EwapO3}
\end{figure}

\begin{figure}[tb]
	\centering
	\includegraphics[height=5.cm]{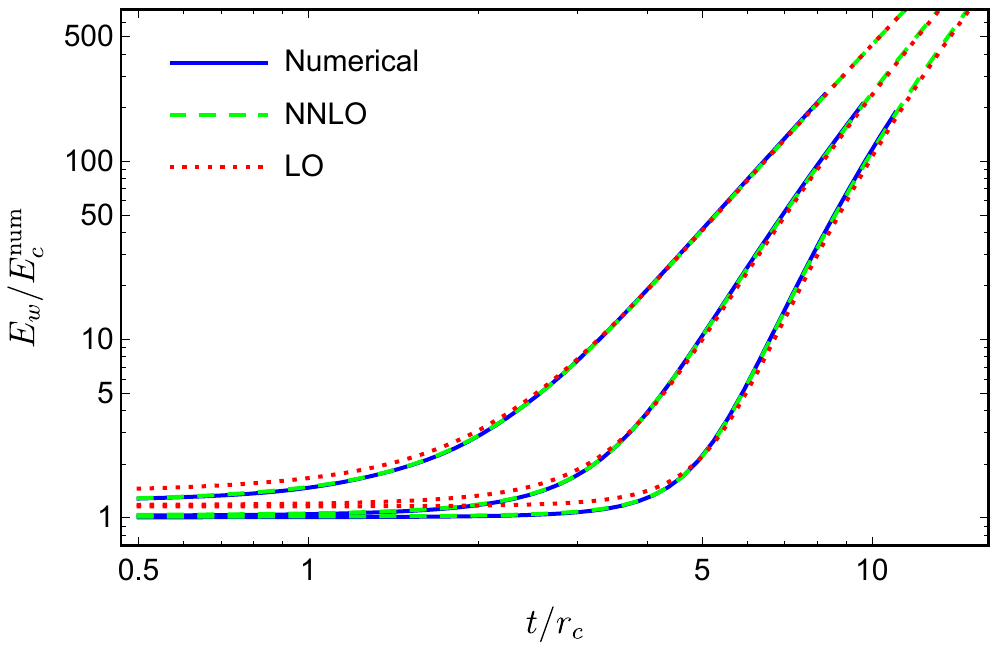}\hfill\includegraphics[height=5.cm]{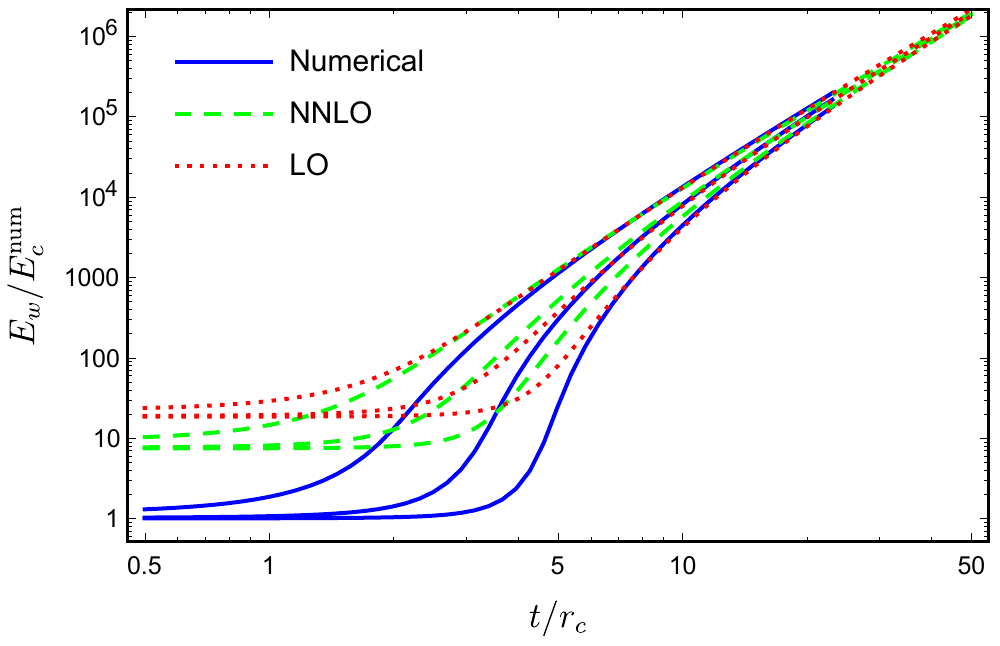}
	\caption{Like Fig.~\ref{fig:EwapO3}, for a wider time range and in log scale.}\label{fig:EwapO3log}
\end{figure}

In Fig.~\ref{fig:EwapO3} we show the wall energy for the numerical solution and the approximations.
As expected, the NNLO approximation is very good for the first example potential.
For the second example, the perturbative expansion breaks down, but the LO
approximation is not bad, as might be expected from the profiles of Fig.~\ref{fig:TwallapO3}.
Actually, in this case the thin-wall approximation fails at the beginning of the bubble evolution.
In Fig.~\ref{fig:EwapO3log} we show the same curves in logarithmic scale, which shows in more detail
the initial values of the wall energy.
For the case on the right, the approximation departs from the numerical curve for $t\lesssim 5 r_c$. 
This is the time it takes the field inside the bubble to reach the value $\phi_-$ 
(see Figs.~\ref{fig:profO3} and \ref{fig:bbini}).
At high $t$, all the log curves approach the asymptote given by $E_w=\Delta V \frac{4\pi}{3}t^3$, 
as the bubble radius approaches the behavior $r_w=t$.

The log-log curves in Fig.~\ref{fig:EwapO3log} are similar to those plotted in Fig.~2 of Ref.~\cite{elnv19}.
In that work, a hybrid approximation was proposed for the wall energy.
In this approximation, the LO thin-wall approximation is used to calculate $r_w(t)$,
but using as initial condition the value $r_{i0}=[3E_{0,V}/4\pi \Delta V]^{1/3}$, where $E_{0,V}$
is the potential energy of the actual initial bubble (obtained numerically).
Then, the wall energy is calculated as
$E_w= E_0+\frac{4\pi}{3}r_w(t)^3\Delta V(t)$, where $E_0$ is the total energy of the initial bubble 
and $\Delta V(t)=V_+-V(\phi(t,r=0))$.
To obtain $\phi$ as a function of time at the center of the bubble, the partial differential equation must be solved numerically, since there is currently no general analytical approximation for the evolution of the field inside the bubble%
\footnote{For the O(3,1)-symmetric solution, where we have a function of a single variable, $\tilde{\phi}(\tau)$,
	approximating the potential near $\phi_-$ by a parabola gives
	the damped oscillations around the minimum analytically \cite{mm23}.}.
Our perturbative thin-wall approximation does not require such a numerical input, although it breaks down when $\phi$ departs from $\phi_-$ at $r=0$.
Comparing Fig.~\ref{fig:EwapO3log} with Fig.~2 of Ref.~\cite{elnv19},
it is apparent that our approximation is much better beyond $t\sim r_c$.
Indeed, the approximation of Ref.~\cite{elnv19} seems to work well only at the very beginning of the wall evolution.

\section{Conclusions}
\label{sec:conclu}

This work is the third of a series where we study the dynamics of a bubble wall beyond the 
usual approximations of an infinitely-thin wall and a spherical bubble \cite{mm23,mm24}.
Here, we have focused on the energy-momentum tensor of the wall beyond the thin-wall approximation.
We neglected the presence of fluid in the evolution of the bubble wall,
but the generalization of our method is straightforward and we will consider the plasma in a forthcoming paper. 

Ambiguities in the definition of the wall position and width are inevitable for a wall that is not infinitely thin,
and the wall energy may inherit these ambiguities.
We have proposed a decomposition of the energy-momentum tensor in which the wall contribution is unambiguously identified
and does not depend on an arbitrary definition of the wall boundaries.
In contrast to a decomposition often used in the literature, where 
the kinetic and gradient energies are assigned to the wall and the potential energy to the bubble,
our decomposition includes the potential barrier in the wall component.
Furthermore, our treatment is based on the Gaussian normal coordinates, so that the energy density profile is a function of the proper distance along the normal geodesics to the wall hypersurface.
These coordinates are valid on a neighborhood of the surface where the geodesics do not cross, which is a natural boundary if the width of the wall is of the same order as its radius of curvature.

We remark that the definition of the wall component of $T_{\mu\nu}$ is not just a matter of terminology, since
some approximations rely on this decomposition.
In particular, the energy stored in the bubble walls is relevant to the generation of a gravitational wave background by bubble collisions.
In the bubble collision mechanism, the walls are treated as infinitely thin shells.
Our decomposition is suitable for the thin-wall approximation, where it gives a clearer separation of the energy into volume and surface contributions.
We have calculated the energy-momentum tensor of the wall and the surface energy density at next-to-next-to-leading-order in the wall width.
We have found analytical expressions for an arbitrary wall shape.

We have considered the specific case of a spherical bubble to compare the approximations with a numerical calculation.
It is well known that the thin-wall approximation is valid for a potential that is nearly degenerate, i.e., 
where the potential difference between minima, $\Delta V$, is much smaller than the height of the potential barrier, $V_{\max}$.
To test our approximations, we have considered two example potentials that deviate from this condition,
one with $\Delta V\gtrsim V_{\max}$ and one with $\Delta V\gg V_{\max}$.
In the first case, the NNLO approximation is a significant improvement over the LO approximation.
In the second case, the perturbative expansion is no longer reliable
(for a vacuum bubble it improves the LO approximation, but for a near-critical bubble the LO approximation is better).
However, the LO thin-wall approximation is still better than other approaches considered in the literature.

\section*{Acknowledgements}

This work was supported by UNMdP grant EXA1202/24.

\appendix

\section{Initial bubble profile and bubble radius}
\label{app:instanton}

The initial bubble after nucleation is given by the O(3)-symmetric equation (\ref{ecinstO3}) for thermal nucleation and by
the O(4)-symmetric equation (\ref{eq:ecfiColeman}) for vacuum nucleation.
Both equations are of the form
\begin{equation}
	\phi''(\rho) + (j-1)\rho^{-1} \phi'(\rho) =V'(\phi) ,
	\label{ecinst}
\end{equation}
where $j=4$ for the O(4)-symmetric case and $j=3$ for the O(3)-symmetric case
(with $\rho=r$ in this case).
The boundary conditions are the same, so we can study the two cases together.
In Ref.~\cite{mm24} we used our perturbative method to obtain the bubble profile and radius
at NNLO in the wall width for the case
$j=4$.
The generalization is straightforward and we only need to make the replacement $4\to j$ in the results.
Below, we briefly describe the derivation.

We define the wall position as
$\rho_w = \sigma^{-1}\int_{-\infty}^{+\infty}\phi^{\prime 2}(\rho) \rho\, d\rho $,
with 
$\sigma = \int_{-\infty}^{+\infty}\phi^{\prime 2}(\rho)d\rho$.
If we shift the variable $\rho$ to $n=\rho-\rho_w$, the definition of $\rho_w$ is equivalent to the condition $\int_{-\infty}^{+\infty}\phi^{\prime 2}(n) ndn=0$.
Physically, the variable $n$ is the Gaussian normal coordinate to the $(j-1)$-dimensional surface. 
In the three-dimensional case, $\rho_w$ and $n$ represent the quantities $r_w$ and $m=r-r_w$, respectively.
In terms of the variable $n$, the first integral of Eq.~(\ref{ecinst}) gives
\begin{equation}
	\frac{1}{2}\phi^{\prime 2}(n) +
	\int_{\infty}^{n}\frac{(j-1)}{\rho_w+n'}\phi^{\prime 2}(n') dn' =V(\phi)-V_+ .
	\label{primeraintinst}
\end{equation}
In particular, at $n=-\infty$, we have
\begin{equation}
	\int_{-\infty}^{+\infty}\frac{(j-1)}{\rho_w+n}\phi^{\prime 2}(n) dn = \Delta V .
	\label{ecparedinst}
\end{equation}
The quantity $K= -(j-1)(\rho_w+n)^{-1}$ is the mean curvature of the surface.

The basic thin-wall approximation consists in neglecting $K$ in Eq.~(\ref{primeraintinst}) and, for consistency, approximating $V$ by a degenerate potential $V_0$.
This leads to the solution $\phi_0(n)$ already obtained in Sec.~\ref{sec:pert_method}.
Inserting this solution in Eq.~(\ref{ecparedinst}), we obtain the LO approximation for $\rho_w$, 
\begin{equation}
	\rho_0=(j-1)\sigma_0/\Delta V.
	\label{eqap:rho0}
\end{equation}
To obtain the wall profile $\phi$ and the radius $\rho_w$ to higher order, we expand $K$ in powers of $n$ and assume the expansions $\rho_w=\rho_0+\rho_1+\cdots$,
$\phi=\phi_0+\phi_1+\cdots$.

Inserting these expansions in Eq.~(\ref{primeraintinst}), we obtain the equation
(\ref{eq:primeraintpert}) for $\phi_i$, whose solution is given by Eqs.~(\ref{eq:solfii})-(\ref{eq:Ci}).
The source terms $f_i$ are given by Eqs.~(\ref{f1})-(\ref{f3}) but, in this case,
the parameters from the expansion of $K$ are given by
\begin{align}
	&K_{0}|_{0} =-(j-1)/\rho_{0}, \ \ K_{1}|_{0}=(j-1)\rho_{1}/\rho_{0}^{2}, \ \ K_{2}|_{0}=(j-1)\rho_{2}/\rho_{0}^{2}-(j-1)\rho_{1}^{2}/\rho_{0}^{3}, \nonumber
	\\
	&\partial_{n}K_{0}|_{0}  = (j-1)/\rho_{0}^{2}, \ \ \partial_{n}K_{1}|_{0}=-2(j-1)\rho_{1}/\rho_{0}^{3}, \ \ \partial_{n}^{2}K_{0}|_{0} = -2(j-1)/\rho_{0}^{3}.
	\label{expanKinst}
\end{align}
For our specific potential (\ref{eq:pot}), the $\phi_i$ are given by Eqs.~(\ref{eq:phi1})-(\ref{C2b}).
The essential difference with the general method of Sec.~\ref{sec:pert_method} is that the quantities (\ref{expanKinst})
are constant in this case.

Expanding Eq.~(\ref{ecparedinst}) in a similar way, we obtain equations for $\rho_i$,
\begin{equation}
	\frac{\sigma_{0}}{\rho_{0}}=\frac{\Delta V_{1}}{(j-1)}, 
	\quad
	\frac{\sigma_{1}}{\rho_{0}} - \frac{\rho_{1}}{\rho_{0}^{2}}\sigma_{0}=\frac{\Delta V_{2}}{(j-1)}, 
	\quad
	\frac{\sigma_{2}}{\rho_{0}}-\frac{\rho_{1}\sigma_{1}}{\rho_{0}^{2}}
	-\frac{\rho_{2}\sigma_{0}}{\rho_{0}^{2}} + \frac{\rho_{1}^{2}\sigma_{0}}{\rho_{0}^{3}} 
	+ \frac{\mu_{0}}{\rho_{0}^{3}} = \frac{\Delta V_{3}}{(j-1)},
\end{equation}
where $\Delta V_i$ are the terms of the expansion of $\Delta V$.
Adding the first two of these equations, we obtain the NLO radius
$\rho_w = (j-1)\sigma/\Delta V$.
For our potential, the NLO correction to $\sigma$ vanishes and we have $\rho_w=\rho_0$.
Adding the three equations, we obtain the NNLO value
$\rho_w = (j-1)(\sigma + \mu_0/\rho_0^2)/\Delta V$.
The NNLO correction to $\sigma$ is given by 
$\sigma_{2}	=\tilde{\sigma}_{2}-\mu_{0}\partial_{n}K_{0}|_{0}$,
with $\tilde{\sigma}_2$ and $\mu_0$ given by Eq.~(\ref{sigmamu}).
Therefore, we obtain
\begin{equation}
\rho_{w}=\frac{(j-1)}{\Delta V}\left[\tilde{\sigma}-(j-2)\frac{\mu_{0}}{\rho_{0}^{2}}\right]=\frac{(j-1)}{\Delta V}\tilde{\sigma}-(j-2)\frac{l_{0}^{2}}{\rho_{0}} .
\label{eqap:rhow}
\end{equation}

\begin{figure}[tb]
	\centering
	\includegraphics[height=5.2cm]{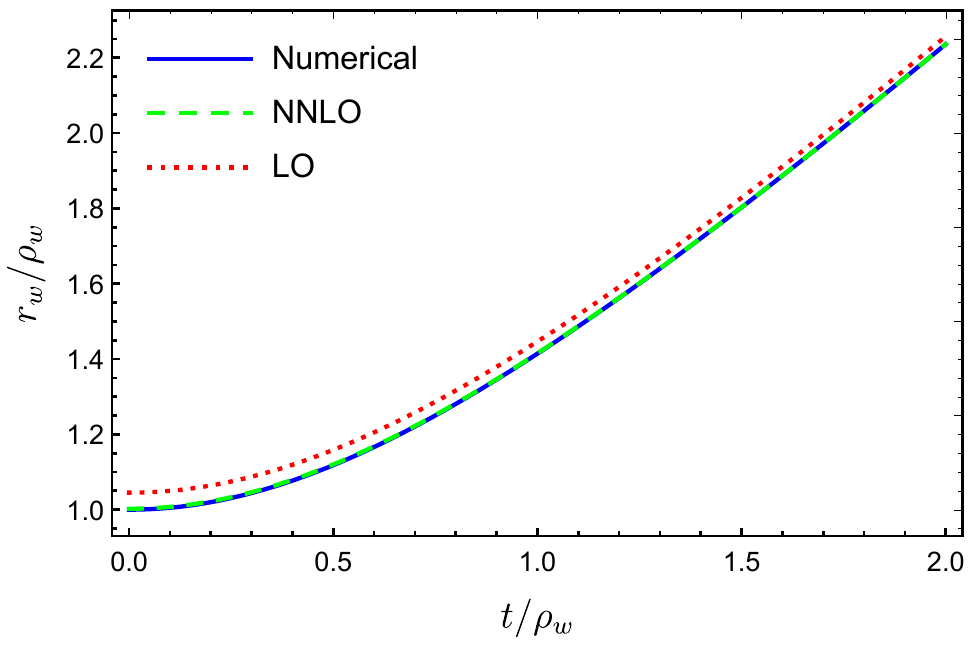}\hfill\includegraphics[height=5.2cm]{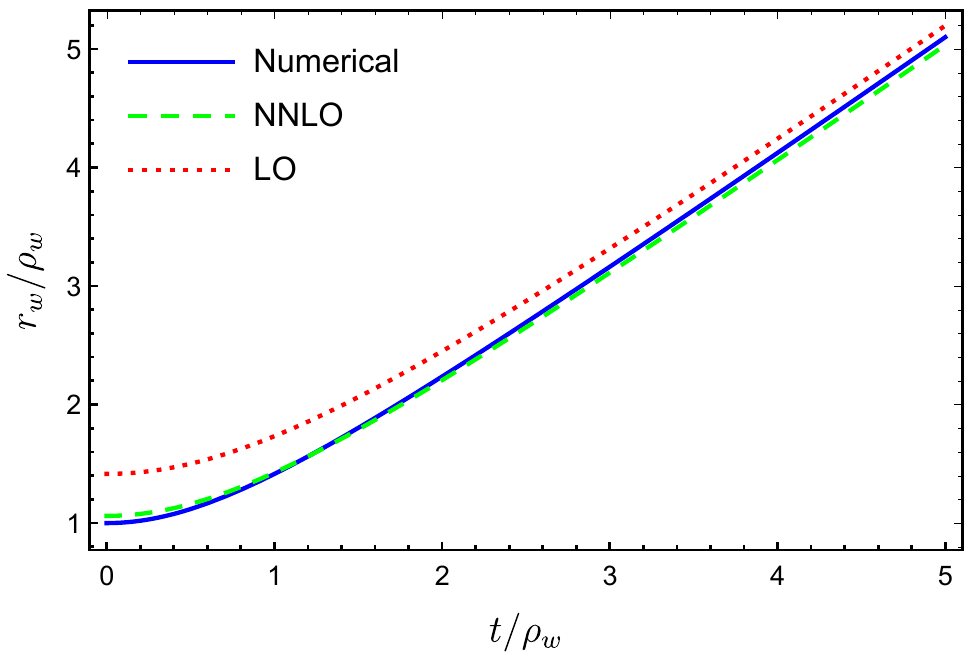}
	\caption{The bubble radius of the O(3,1)-symmetric bubble as a function of time for the potential of Fig.~\ref{fig:potperf1} (left) and that of Fig.~\ref{fig:potperf2} (right).}\label{fig:rw}
\end{figure}

For $j=4$, Eq.~(\ref{eqap:rho0}) gives $\rho_0=3\sigma_0/\Delta V\equiv R_0$, 
which is the well known value for the instanton radius in the thin-wall approximation.
On the other hand, Eq.~(\ref{eqap:rhow}) gives $\rho_{w}=3\tilde{\sigma}/{\Delta V}-2{l_{0}^{2}}/{R_{0}}\equiv R$, 
which coincides with the value obtained in Sec.~\ref{subsec:O31} from the requirement that the total bubble energy 
(\ref{eq:primeraintesf}) vanishes.
In Ref.~\cite{mm24} we discussed in detail these analytic approximations for $\rho_w$ and the field profile.
In Fig.~\ref{fig:rw} we compare the evolution of $r_w$ using the LO EOM with the LO nucleation radius and the NNLO EOM with the NNLO nucleation radius.
We see that the approximation improves significantly at the NNLO.
For the case on the left, the LO approximation gives $R_0\simeq 1.05 \rho_w$, while the NNLO approximation gives $R\simeq 1.002 \rho_w$.
For the case on the right, we have $R_0\simeq 1.4\rho_w$ and $R\simeq 1.06\rho_w$.
At high $t$, the difference between the approximation for $r_w$ and the numerical result approaches a constant value,
which is of the same order for the LO and the NNLO approximation.

For $j=3$, Eq.~(\ref{eqap:rho0}) gives $\rho_0=2\sigma_0/\Delta V\equiv R_{c0}$.
This is the well-known value for the thermal instanton radius in the thin-wall approximation.
It
coincides with the value obtained in  Sec.~\ref{subsec:O3} with the unstable equilibrium condition $\ddot{r}_{w0}=0$,
and it maximizes the LO bubble energy (\ref{eq:gammaw0})
for $\gamma_{w}=1$.
On the other hand, Eq.~(\ref{eqap:rhow}) gives the NNLO value
$\rho_{w} = R_c $, where
\begin{equation}
	R_c = 2\tilde{\sigma}/{\Delta V}-{l_{0}^{2}}/{R_{c0}} .
	\label{eqap:Rc}
\end{equation}
We obtain the same result by maximizing the NNLO bubble energy (\ref{eq:primeraintesf}).
It is also easy to check that the value of $R_c$ obtained 
with the unstable equilibrium condition $\ddot{r}_w=0$
is in agreement with Eq.~(\ref{eqap:Rc}). 
Indeed, we obtain
\begin{equation}
	R_c=2\left(\Delta V/\tilde{\sigma} +2l_0^2/R_{c0}^3 \right)^{-1}= 2\tilde{\sigma}/{\Delta V}-{l_{0}^{2}}/{R_{c0}}+\mathcal{O}(l_0^3) .
	\label{eq:Rcap}
\end{equation}
Since near-critical bubbles are very sensitive to the initial conditions,
the small difference between these estimates can be relevant.
For calculating the initial condition to use in the equation of motion (\ref{eom2esf}), 
the first expression in (\ref{eq:Rcap}) is more consistent,
since it implies exactly $\ddot{r}_w=0$ for $r_w=R_c$ for this EOM.

\begin{figure}[tb]
	\centering
	\hspace*{\fill}\includegraphics[width=0.48\textwidth]{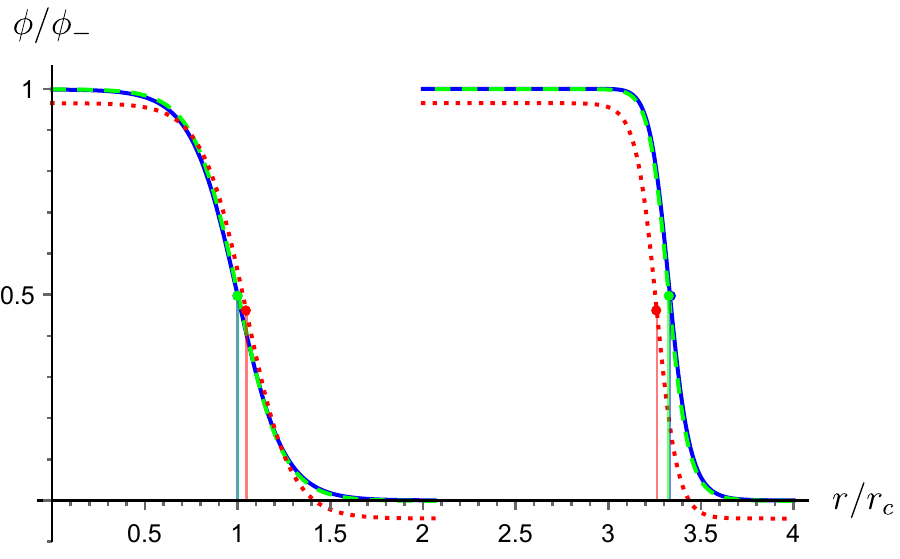}%
	\hfill%
	\includegraphics[width=0.48\textwidth]{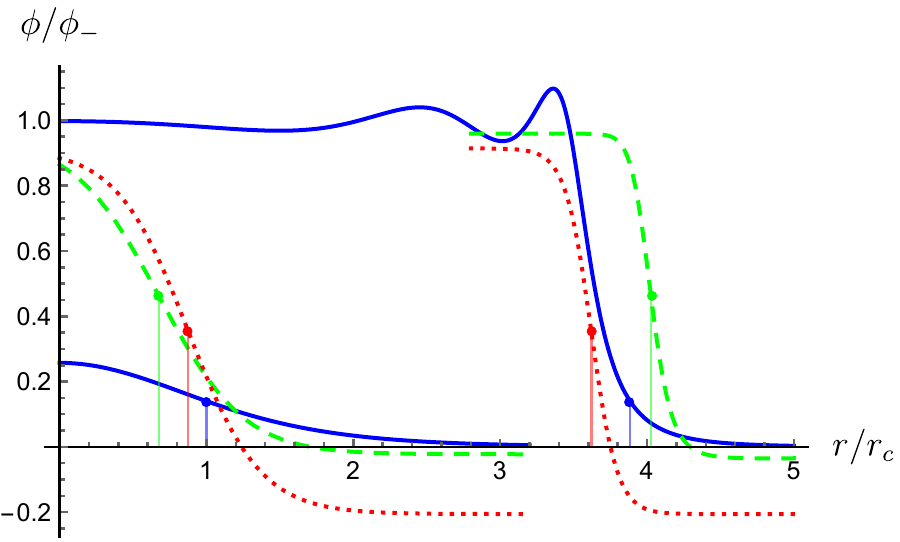}%
	\hspace*{\fill}
	\caption{The profile of the critical bubble and the bubble profile at $t=6r_c$ for the case $\varepsilon=10^{-2}$ for the two example potentials.}%
	\label{fig:bbini}
\end{figure}

In Fig.~\ref{fig:bbini} we compare the approximations for the critical bubble with the numerical solution for the two example potentials considered in this article.
The figure also shows the profile at a later time.
For the approximations, the field profile does not approach the asymptotic values $\phi_\pm$ but the minima of the corresponding approximation for $V(\phi)$.
In particular, the LO field approaches the minima of the degenerate potential $V_0(\phi)$.
For the case on the left, where the bubble wall is not very thin but not extremely thick, 
we see that the NNLO approximation significantly improves the LO approximation.
For the case on the right, it is apparent that the thin-wall approximation breaks down.
In particular, the NNLO value for the radius does not improve the LO value.
The NNLO profile does improve on the interpolation between the values $\phi_\pm$.
However, this is not relevant for the energy density since the shape of the LO profile already gives a reasonable approximation.

\begin{figure}[tb]
	\centering
	\includegraphics[height=5.5cm]{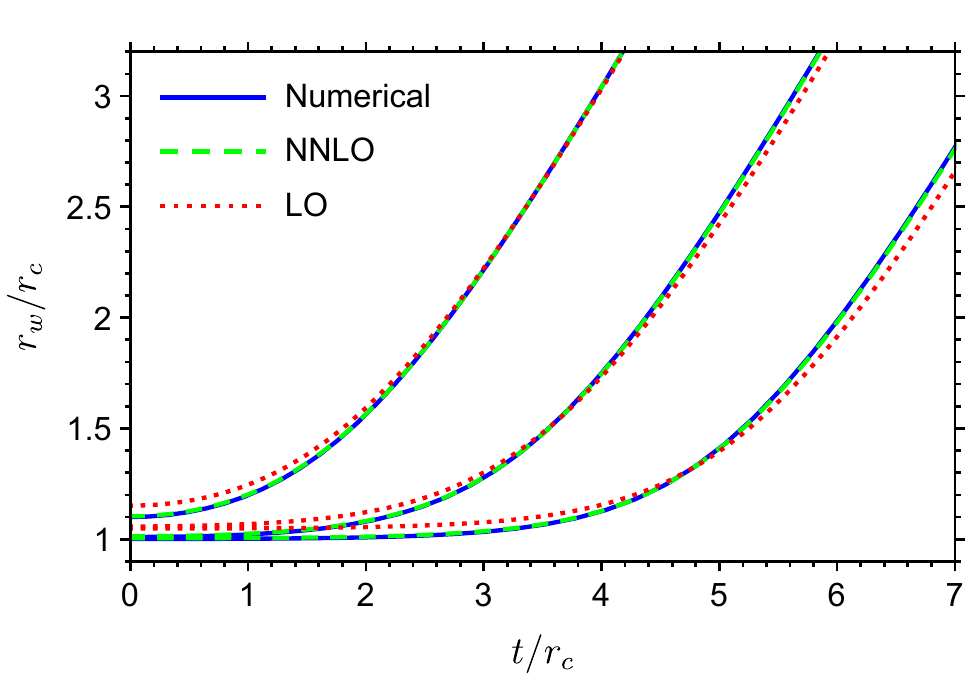}\hfill\includegraphics[height=5.5cm]{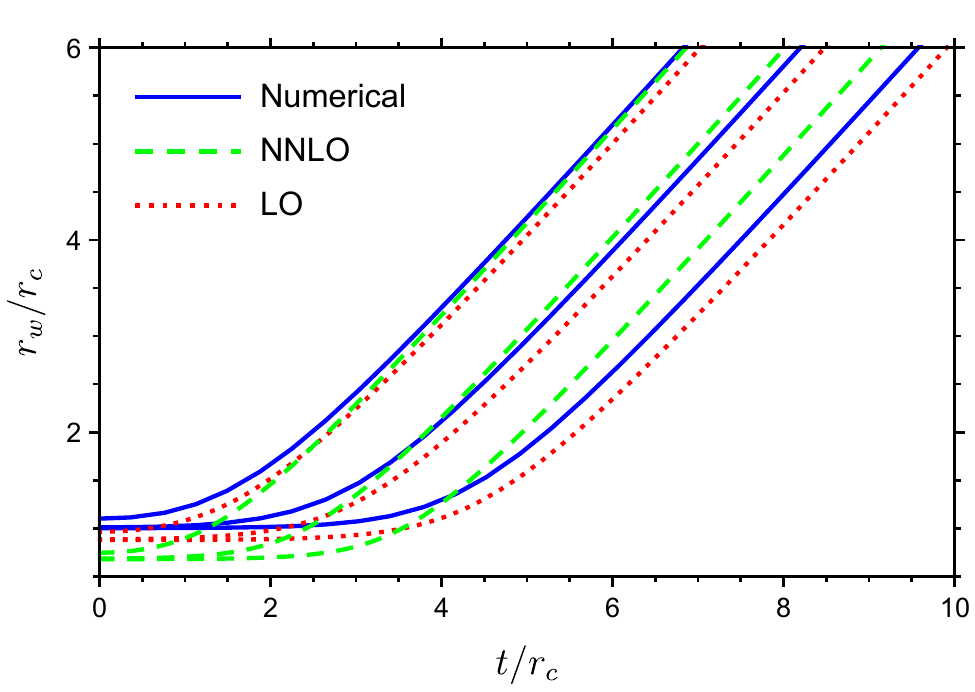}
	\caption{The bubble radius as a function of time for the two example potentials
		and for three different initial conditions.
		From top to bottom we have $\epsilon=10^{-1}$, $10^{-2}$, and $10^{-3}$.}\label{fig:rwo3}
\end{figure}

In Fig.~\ref{fig:rwo3} we show the bubble radius as a function of time.
We consider the same two potentials and different initial bubbles.
For the case on the left, the NNLO approximation is again very good.
The LO approximation becomes worse as the initial bubble gets too close to the critical one
and the evolution becomes too sensitive to the initial condition.
For the case on the right, we see that both approximations give a reasonable description of the
evolution of the bubble radius at later times.
Although the NNLO approximation appears to be better in some regions, 
these curves are somewhat misleading in the present case
because the wall position $r_w$ represents points that are not equivalent 
in the numerical and thin-wall profiles (see Fig.~\ref{fig:bbini}).

\bibliographystyle{jhep}
\bibliography{papers}
\end{document}